\documentclass[twoside,12pt]{article}

\usepackage{graphicx}
\usepackage{amsmath}
\usepackage{amssymb}
\usepackage{array}
\usepackage{wrapfig}
\usepackage{units}
\usepackage{caption}
\usepackage{slashed}
\usepackage{xcolor, hyperref, float}
\usepackage{doi}
\usepackage{epsfig}
\usepackage[caption=false]{subfig}

\newcommand{\be}{\begin{equation}}
\newcommand{\ee}{\end{equation}}
\newcommand{\bea}{\begin{eqnarray}}
\newcommand{\eea}{\end{eqnarray}}

\DeclareMathOperator{\msun}{{\rm M}_\odot}

\begin{document}

\title{\vspace{1cm} 
The Milky Way, Coming into Focus: \\
Precision Astrometry 
Probes its Evolution \\ and its Dark Matter
}
\author{Susan Gardner$^{1}$, Samuel D.\ McDermott$^2$,  and Brian Yanny$^2$ 
\\
\\
${}^{1}$Department of Physics and Astronomy, University of Kentucky, \\
Lexington, KY 40506-0055, USA 
\\
${}^2$Fermi National Accelerator Laboratory, \\
Batavia, IL 60510, USA\\}
\maketitle

\begin{abstract} 
The growing trove of precision astrometric observations
from the {\it Gaia} space telescope and other surveys
is revealing 
the structure and dynamics of the Milky 
Way in ever more exquisite detail. We summarize
the current status of our understanding of the structure and the characteristics of the Milky Way, and we review the 
emerging picture: the Milky Way is evolving through
interactions with the massive satellite galaxies that
stud its volume, with evidence pointing to 
a cataclysmic past. 
It is also woven with stellar streams, and  
observations of streams, satellites, and field stars 
offer new constraints on its dark matter,
both on its spatial distribution and its fundamental 
nature. The recent years have brought much focus to the
study of dwarf galaxies found within our Galaxy's halo 
and their internal matter distributions. 
In this review,  we focus on
the predictions of the cold dark matter 
paradigm 
at small mass scales through 
precision astrometric measurements,
and we summarize the modern consensus on the extent to which small-scale probes are consistent with this paradigm.
We note the discovery prospects of these studies, 
and also how they 
intertwine with probes of 
the dynamics and evolution of
the Milky Way in various and distinct ways. 
\end{abstract}

\eject
\tableofcontents

\section{Introduction} 
 It has long been recognized that detailed observations of our
Milky Way (MW) galaxy and its stars could lead to a 
{\it near-field} cosmology. This is distinct from 
far-field studies of 
the large-scale structure of the 
Universe~\cite{freeman2002revnewgalaxy}, which are
realized through observations of 
the cosmic microwave background (CMB)~\cite{hu2002reviewcmb,Planck2020legacy} or of the clustering of galaxies~\cite{bahcall1988reviewcluster,alam2017sdssclustering}.
Although small-scale structure probes can also be
made at far field,  
through, e.g., 
gravitational lensing~\cite{huterer2010reviewweaklens,massey2010reviewgravlens}
or from the imprint of neutral hydrogen 
in the intergalactic medium on the spectra of 
distant quasars~\cite{rauch1998reviewlymanalphaforest,viel2013warmdmlymanalpha},
in this review we focus on the  small-scale tests 
possible in 
the nearest of fields: the MW itself.
Studies within the MW offer 
probes of dark matter with a precision not currently possible through any other means. 
In contrast, all terrestrial experiments in particle, nuclear, or atomic physics 
that search for 
dark matter (DM) 
require that the DM particles have non-gravitational interactions
with the particles of the Standard Model (SM). 
This may change --- advances in sensing beyond 
the quantum limit~\cite{Caves1981QSL}, stimulated by the sensitivity
needs of the LIGO gravitational wave 
detector~\cite{LIGO2013squeezed}, have spurred
other DM 
studies~\cite{graham2016accelerometer,Chaudhuri2018dmquantum,Carney2020gravdirectdetection,sikivie2021RvMPinvisibleaxions} --- and 
gravitational detection of a DM candidate may yet be realized~\cite{Carney2020gravdirectdetection}. 
Nevertheless, the unique 
precision with which we are able
to probe 
the structure and dynamics of our own galaxy 
extends our ability to hunt for DM beyond the
reach of terrestrial experiments. 
In this review we consider how sharpening observations of 
MW stars drive new and reinvigorate old investigations
of the Galaxy, to probe its structure and evolution,
including all of its DM and 
its own matter distribution and the fundamental properties of its constitutive elements.
Probes of MW structure are hence probes of the nature of dark matter. This follows a long tradition: all of the positive evidence for DM thus far comes from astronomical observations within and beyond our own Galaxy~\cite{bertone2005DMreview, buckley2018gravitational}. 

\begin{figure}[t!]
\centering
\includegraphics[scale=0.6]{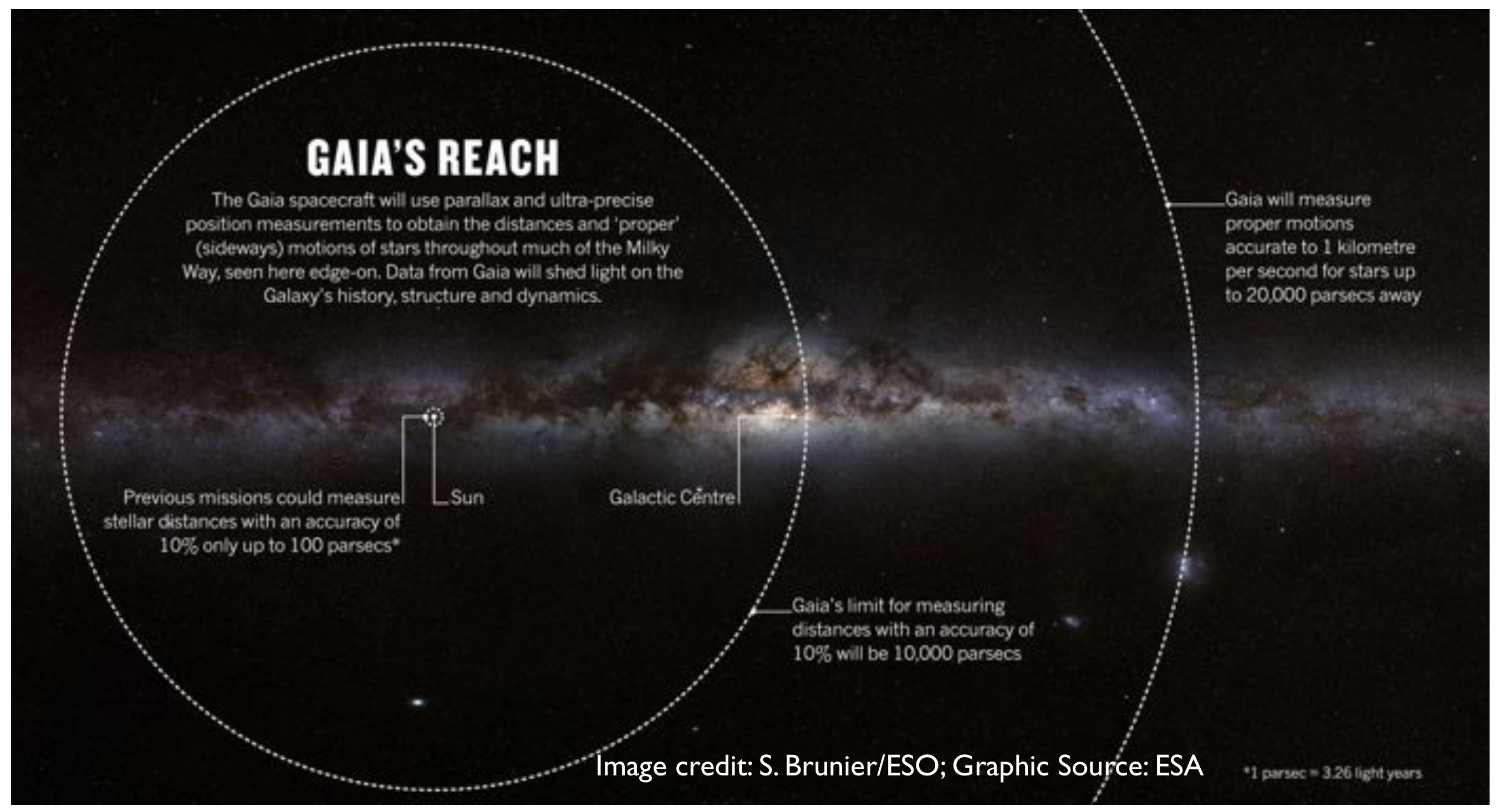}
\begin{minipage}[t]{16.5 cm}
\caption{
An illustration of {\it Gaia}'s reach --- note that
``previous missions'' refer to 
Hipparcos~\cite{hipparcos1997catalog}. 
From \cite{gaia2013naturenews}, reprinted by
permission from Springer Nature. 
}
\label{fig:Gaiareach}
\end{minipage}
\end{figure}
\begin{figure}[t!]
\centering
\includegraphics[scale=0.6]{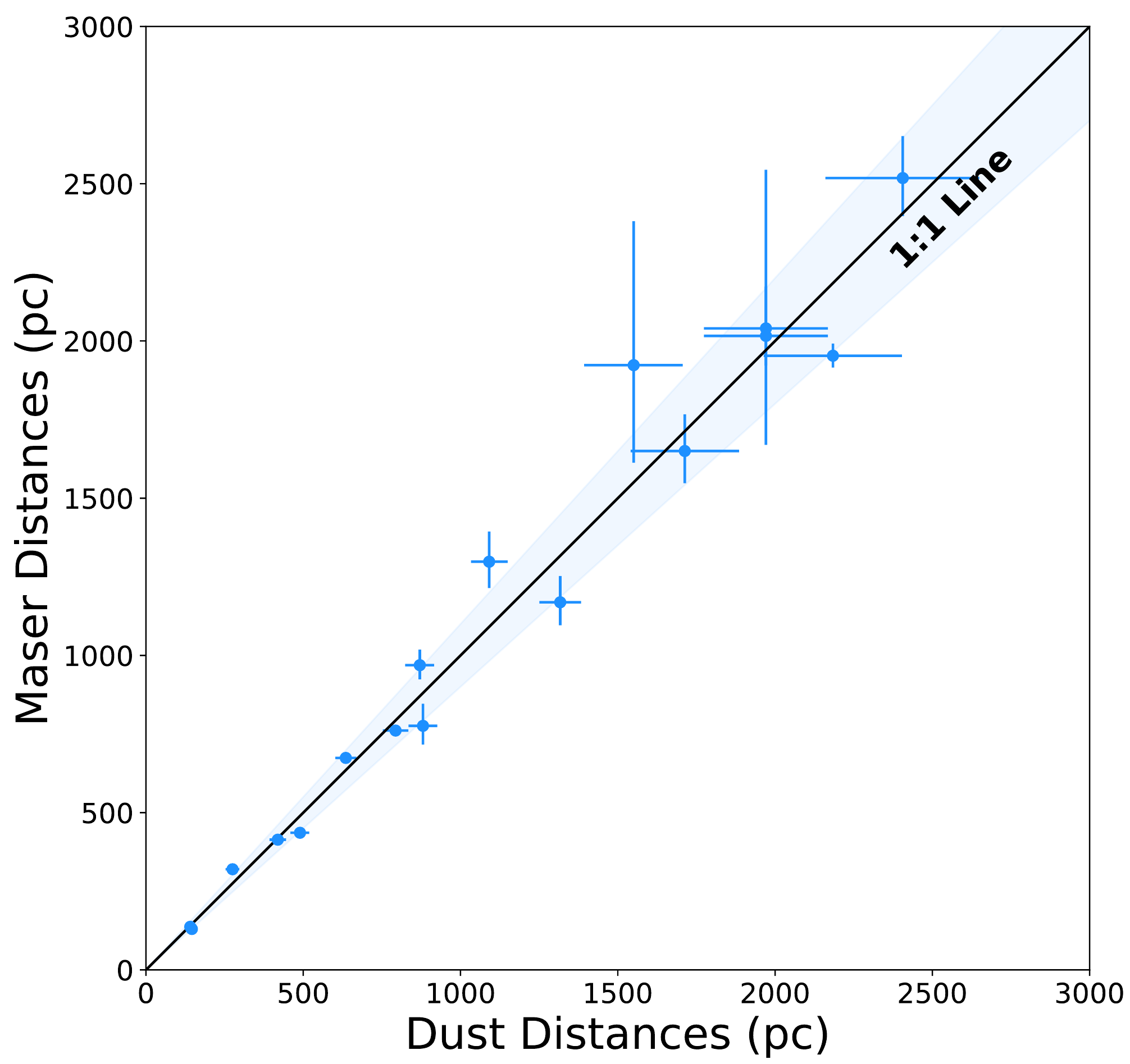}
\begin{minipage}[t]{16.5 cm}
\caption{
Two methods of assessing distances to nearby 
dust clumps compared: one method uses VLBI parallaxes, 
mainly to masers, and
the other employs {\it Gaia} DR2 data. 
The distance assessments for the 
two methods agree within $\le 10\%$ for distances 
ranging from 100 pc to 2.5 kpc with a negligible 
offset. From \cite{Zucker2020Gaiavsmaser}, reproduced
with permission $\copyright$ESO. 
}
\label{fig:Gaiavsmaser}
\end{minipage}
\end{figure}

All of this has been made possible through 
the rise of large-scale astronomical surveys
over the last decades, beginning with 
the Sloan Digital Sky Survey (SDSS) \cite{york2000sdss}, 
and continuing with 
2MASS \cite{skrutskie2006mass2}, Pan-Starrs1 \cite{chambers2016panstarrs}, and DES \cite{Abbott2018DES}. The sensitivity and reach of 
these studies have been greatly enhanced through the
advent of data releases from 
the {\it Gaia} space telescope~\cite{prusti2016gaia,lindegren2018gaia,Gaia2020EDR3}.
We show an estimated 
footprint of the 
{\it Gaia} data in Fig.~\ref{fig:Gaiareach}.
The {\it Gaia} 
mission enormously extends the reach and precision of
the astrometry from 
the Hipparcos mission of the early 1990's. 
{\it Gaia} data account for the parallaxes of more 
than 1.3 billion objects\footnote{More precisely, the {\it Gaia} data enable a 5-parameter
astrometric solution corresponding to the sky location, parallax and
proper motions of each object.} within 10 kpc in {\it Gaia} Data Release 2 (DR2)~\cite{brown2018gaia,lindegren2018gaia}. This is roughly 1\% of the MW's stars,
and this trove greatly expands on our knowledge of the MW compared to the parallaxes for 2.5 million objects 
within 200 pc 
measured by Hipparcos~\cite{hipparcos1997catalog}.
Astrometric parallaxes give an enormous improvement
in the quality of distance 
assessments over ground-based surveys, which are 
largely forced to rely on photometric methods, and 
the precision of the distance assessments~\cite{lindegren2018gaia}, 
and the completeness of the stellar samples~\cite{arenou2018gaia}, 
with {\it Gaia} DR2
are extremely high. It is possible, e.g., 
to select a data sample of some 14 million stars
within 3 kpc of the Sun's location with an 
average relative parallax error of less than 
10\%~\cite{HGY20}. As a separate example, 
we compare two distance assessment 
methods to local dust clouds --- 
one uses {\it Gaia} parallaxes and the 
other uses parallaxes of masers measured by the VLBI ---
and show the result in Fig.~\ref{fig:Gaiavsmaser}. 

Through these studies our perspective of
the MW has shifted: its mass distribution is 
neither that of an isolated system, nor is it 
in steady state. 
These outcomes are at odds with 
long-standing theoretical assumptions. 
Although the time
scales associated with these changes are long, 
their impacts are appreciable nonetheless, giving
us the opportunity
to study the MW 
as it responds to external forces, 
and yielding new probes 
of its particle dark matter. 
To set the stage for later discussions, 
we open in Sec.~\ref{sec:Prologue}
with a perspective on the 
theoretical framework for the 
matter distribution in our Galaxy prior to
these discoveries. We review the 
distribution function formalism for
both visible and dark matter, noting the so-called 
standard halo model (SHM)~\cite{drukier1986SHM} employed 
in dark 
matter direct detection experiments 
and
observational evidence for its limitations, 
before turning to recap
the cold dark matter (CDM) paradigm, noting 
its 
long-standing 
small scale problems~\cite{Weinberg2015CDMcontroversiessmall,bullock2017challengeLCDM,buckley2018gravitational} and limited relaxation 
mechanisms. 

We then turn, in Sec.~\ref{sec:Parameters},  
to a brief summary of the gross
features of the MW as they are 
known thus far. We provide current best values of the MW mass, size and shape; and 
we list important components of the MW, survey its known rotation curve, and provide insights into the 
environment in which it is situated. 
We refer the interested reader to earlier, 
extensive reviews~\cite{freeman2002revnewgalaxy,bland-hawthorne2016galaxy} for extended discussion of earlier results
and historical context. 

Beginning in 
Sec.~\ref{sec:smallscaleprobes} we turn to 
smaller scale features of the MW. We describe 
the Magellanic clouds, 
satellite galaxies, stellar streams, and 
patterns in MW stars 
and how they probe the structure of the MW. The purpose of Sec.~\ref{sec:smallscaleprobes} is
to give a current inventory of the constituent parts of the MW, and to describe how
to harness their unique characters to extract information about the nature of the MW halo.
In Sec.~\ref{sec:dmcand} we describe 
how, and in what manner, 
such systems constrain
DM particle properties.
We start this section with an overview of the implications of the hierarchical nature of the assembly
of DM halos for the MW. We use this as 
an entry point to discuss specific models of dark matter in more detail.
We start with dark sectors that are characterized by their kinematics rather than by other microphysical
aspects. We then move to a handful of other dark sectors, each of which has a qualitatively different
galactic-scale phenomenology.
Finally in Sec.~\ref{sec:Change} 
we look at these same systems for the
manner in which they reveal 
non-isolating and non-steady-state
effects. In Sec.~\ref{sec:dm_phasespace}, we consider 
how these newly established effects 
impact the assessment of the local DM density 
and velocity distribution, 
important to DM direct detection experiments, 
and we conclude our review in Sec.~\ref{sec:Summary}, 
offering an assessment of future discovery prospects.

\section{Past as Prologue}
\label{sec:Prologue}

In this section we note how 
the theory of the matter
distribution in the MW emerges from kinetic
theory, along with its commonly employed assumptions,
to set the
stage for our discussion of recent discoveries and
their implications. 
The
one-body density of a system of $N$ particles 
in six-dimensional phase space is determined by~\cite{kardar2007SM}
\begin{equation}
    f_1(\mathbf{p},\mathbf{q}, t) 
    = \left\langle \sum_{i=1}^N 
    \delta^{(3)} (\mathbf{p} - \mathbf{p}_i)
    \delta^{(3)} (\mathbf{q} - \mathbf{q}_i)
    \right\rangle \,,
\end{equation}
where the average is over the full 
phase space density 
$\rho(\mathbf{p}_1,\mathbf{p}_2,\dots, \mathbf{p}_N, 
\mathbf{q}_1,\mathbf{q}_2,\dots, \mathbf{q}_N, t)$.
Liouville's theorem tells us that
the full phase-space density behaves as an 
incompressible fluid, 
so that 
\begin{equation}
\frac{d\rho}{dt} = \frac{\partial \rho}{\partial t} + 
\{ \rho, {\cal H} \} =0 \,, 
\end{equation}
where the curly brackets denote a Poisson bracket.
Upon adopting a Hamiltonian
${\cal H}$ with pairwise forces, this yields the 
Bogoliubov, Born, Green, Kirkwood, and Yvon (BBGKY) 
hierarchy, relating the time-evolution of the $s$-body 
density to the $(s+1)$-body density. If $a$ is the range
of the two-body force and $n=N/V$, the BBGKY hierarchy 
collapses to a equation in $f_1$ only if either 
the system is dilute and/or has short-range forces, 
$n a^3 \ll 1$, or it is dense and/or has
long-range forces, $n a^3 \gg 1$. The former limit 
yields the Boltzmann equation and can be used
to model nuclear heavy-ion 
collisions~\cite{Bertsch1984boltzmann,Kruse1985vlasov}.
The latter limit, if we let $f_s \propto (f_1)^s$ so that
the particles are uncorrelated, yields the Vlasov, 
or collisionless Boltzmann, equation 
\begin{equation}
    \left[ \frac{\partial}{\partial t} +
    \frac{\mathbf{p}}{m}\cdot \frac{\partial}
    {\partial \mathbf{q}} - 
    \frac{\partial {\cal U}_{\rm eff}}{\partial \mathbf{q}}\cdot 
    \frac{\partial}{\partial \mathbf{p}}
    \right]f_1(\mathbf{p},\mathbf{q}, t) =0
    \,,
\end{equation}
where ${\cal U}_{\rm eff}$ is the effective potential,  
and is apropos to galactic dynamics~\cite{binney2008GD}.

In this review we focus 
on the mass distribution of
the MW, considering both its visible and
dark matter.
We particularly focus on the component of
its visible matter in stars, its dominant
component away from the Galactic 
midplane\footnote{At the Galactic midplane, the
volume density of the interstellar medium, comprised of 
atomic and molecular hydrogen, ionized gas, and dust, 
is thought to exceed that of stars by about 
50\%~\cite{binney2008GD}, but 
gas and dust are very much localized to the mid-plane region.
}.
To go from $f_1(\mathbf{p},\mathbf{q}, t)$ for a $N$
particle system in the $na^3 \gg 1$ limit to a description
of the Galactic matter distribution requires 
further assumptions~\cite{binney2008GD}: 
(i) that the long-range nature of the gravitational
forces allows us to segue from a $N$-particle system
to a fluid with some total mass and (ii) that the birth
and death of stars have negligible impact. 
We have already neglected both collisions, which
is supported by estimated stellar relaxation times that 
exceed the age of the Universe, and correlations. 
Neglecting the finite stellar lifetime also incurs some errors, because
the force on a star from 
neighboring stars is attractive, but this force is far less than that 
from a distant but much more massive matter distribution,
because of the long-range
nature of the gravitational force: in a system 
of uniform mass density
the most distant members of an
ensemble dominates the gravitational force on any given point \cite{binney2008GD}.
Thus the physics
that allows us to replace a collection of stars
with a 
smooth mass distribution 
also allows us to neglect correlations. 
With this we replace 
$f_1(\mathbf{p},\mathbf{q}, t)$ with the smooth
distribution function $f(\mathbf{v},\mathbf{x}, t)$.
Introducing $\Phi$ as the 
gravitational potential per unit mass, the 
Vlasov equation takes the form~\cite{binney2008GD}
\begin{equation}
    \left[ \frac{\partial}{\partial t} +
    \mathbf{v}\cdot \frac{\partial}
    {\partial \mathbf{x}} - 
    \frac{\partial \Phi }{\partial \mathbf{x}}\cdot 
    \frac{\partial}{\partial \mathbf{v}}
    \right]f(\mathbf{v},\mathbf{x}, t) =0
    \,.
\end{equation}
This is to be solved simultaneously with 
Poisson's equation, relating the 
gravitational potential per mass to the mass density:
\begin{equation}
    \nabla^2 \Phi = 4 \pi G \rho \,,
\end{equation}
where $\rho(\mathbf{x},t) = M\int d^3\mathbf{v} 
f(\mathbf{v},\mathbf{x}, t)$ and $M$ is the total 
mass of the system. Isolated, steady-state systems
have distribution functions that correspond to 
equilibrium solutions of this system of equations. 
In what follows we develop this connection explicitly. 

We emphasize that we have chosen a nonrelativistic normalization such that $f(\mathbf{v},\mathbf{x}, t)$ has mass dimension 3 in the ``natural units'' familiar to 
a high-energy particle physicist. At risk of ambiguity, but following convention, we will use a similar notation for the distribution function after we have projected out the spatial component, such that $f(\mathbf{v}, t) = \int d^3 \mathbf{x} f(\mathbf{v},\mathbf{x}, t)$ is dimensionless in natural units, or has units ${\rm (velocity)}^{-3}$ more generally. We show explicit examples of steady-state $f({\bf v})$
in the MW
in Sec.~\ref{sec:dm_phasespace}.

\subsection{Mass Distribution of a Static, Isolated Galaxy --- and its Symmetries}
\label{sec:df}

We expect an isolated system in steady state to 
be characterized by 
integrals of motion; we also suppose it to be self-gravitating
and of finite mass. An integral of motion $I$ of
interest to us is referenced to 
stellar orbits, 
so that $I(\mathbf{x}(t), \mathbf{v}(t))$ is 
constant for a given orbit. We thus focus on {\it isolating}
integrals~\cite{binney2008GD}. 
As long established, if $I$ is such an integral of motion, it 
is also a solution of the steady-state Vlasov 
equation, and indeed --- as per Jeans theorem --- any 
steady-state solution of that equation can be 
expressed in terms of the system's 
integrals of motion, or functions thereof~\cite{jeans1915}. 
The particular integrals that appear depend on the 
symmetries of the system. Noether's theorem guarantees the existence
of a conserved quantity for each continuous 
symmetry~\cite{noether1918}. 
For example, if ${\cal H}$ is time-independent, making 
it invariant under translations in time, then 
the system's energy $E$ will be an integral of motion
and $f(E)$. Similarly if the system is also spherically 
symmetric, so that it is invariant under rotations, 
then the orbital angular momentum 
$\mathbf{L}$
is also an integral of motion and $f(E, \mathbf{L})$; if it is axially symmetric about the $\hat z$-axis,
$L_z$ is an integral of motion and
$f(E, L_z)$ instead.

There has been much discussion of whether 
the converse of Noether's theorem could also hold, with 
explicit counterexamples in well-known textbooks of 
classical mechanics~\cite{goldstein1980CM,arnold1989CM}
showing that it does not. These counterexamples concern 
systems in which 
invariant quantities exist for which there
are no associated continuous symmetries. 
Our interest here, however, is in 
integrals of motions that are invariant along
stellar orbits. In this case, Noether's theorem
and its converse are both guaranteed, where we
refer to Theorem 5.58 of Olver~\cite{olver2012applications}.  
Thus if $L_z$ is an integral of motion, then the 
associated matter distribution is 
axially symmetric; also if $\mathbf{L}$ is an 
integral of motion, then the associated matter
distribution is spherically symmetric. Thus
a failure of axial symmetry speaks to failure of
$L_z$ as an integral of motion~\cite{GHY20}. 
Along related lines, we note that, as an extension
of Lichtenstein's theorem~\cite{lindblom1992symmetries} in fluid mechanics, an 
isolated, static, self-gravitating, 
ergodic\footnote{An ergodic distribution function 
$f(E)$ uniformly samples its energy surface
in phase space~\cite{binney2008GD}.}
stellar systems has been shown to be spherically symmetric~\cite{binney2008GD}, though this  
can also be shown without reference to fluid mechanics, where we refer to \cite{an2017reflection} for an extended discussion
and further references. If the density distribution 
associated with $f(E)$ is spherical, then 
$f(E)$ is non-negative as well, as expected on 
physical grounds, though the 
Eddington formula for $f(E)$ does not in itself 
guarantee it~\cite{binney2008GD}. 
As a further consequence, 
Noether's theorem says that $\mathbf{L}$ must
be an integral of motion as well, yielding $f(E,\mathbf{L})$. 
Consequently $f(E,\mathbf{L})$ is non-negative as well. 
Finally, if the distribution function of an isolated,
static system is 
axially symmetric, so that it takes the
form $f(E,L_z)$, then it is also reflection or 
north-south symmetric~\cite{an2017reflection}, 
 where we note 
\cite{Schulz2013grav} as well for a slightly less 
general proof, so that 
it is symmetric under $z \to 2z_0 -z$
with $z_0$ the center of the galactic mid-plane. 
We discuss observational
probes of these symmetries and 
the implications of the pattern of their
breaking in Sec.~\ref{sec:Change}.

\begin{figure}[t!]
\centering
\includegraphics[width=1.0\textwidth]{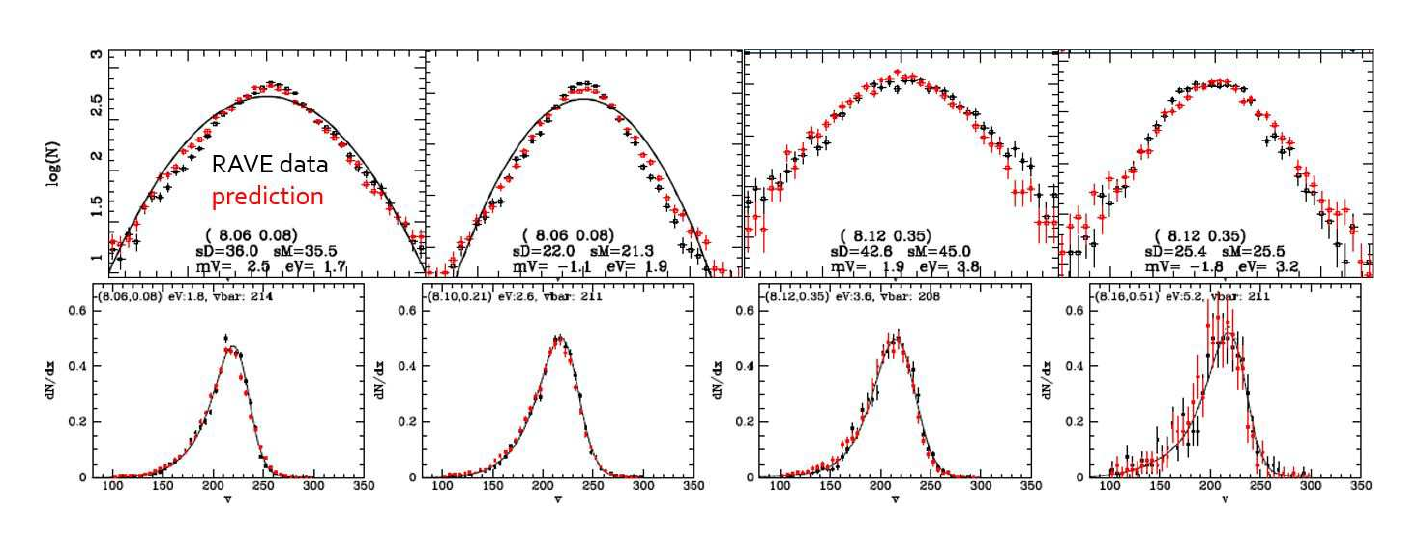}
\begin{minipage}[t]{16.5 cm}
\caption{
Comparisons of the velocity distributions from 
$f(\mathbf{J})$ modeling fitted to data from the 
CGS survey (red points)
with those of 
red dwarfs from the RAVE survey (black points). The
top row shows velocity components 
$V_R$, $V_z$ near the Sun's location
and the Galactic mid-plane (left two panels) and at 
$|z| \sim 0.35\,\hbox{kpc}$. The lower row shows 
$V_\phi$ distributions at increasing heights from 
the plane, from left to right, with the furthest bin centered
on $0.51\,\hbox{kpc}$.
From \cite{binney2019modelling}. 
}
\label{fig:cfRAVE}
\end{minipage}
\end{figure}

The modeling of the Galaxy is commonly 
realized through a superposition 
of its 
disk, bulge, bar, and halo 
components~\cite{robin2003synthetic,robin2012stellar}, with 
each component modelled by a distribution function 
$f_i$~\cite{binney2008GD} in steady state. 
Although $E$ is an integral of motion, it is more
useful to choose the action integrals $J_R$, $J_\phi$, and 
$J_z$ as its arguments~\cite{binney2012actions}, 
where $R$, $\phi$, and $z$ 
are the (in-plane) radial, azimuthal, and vertical coordinates 
with respect to the plane of the Galactic disk. 
For reference $J_\phi$ is 
the angular momentum about the symmetry axis 
$\hat{z}$ of an axisymmetric disk. We note
that $f(\mathbf{J})$ modeling
gives a very good description of the
velocity distributions observed by the RAVE survey
\cite{binney2014galactic,binney2019modelling}, 
and we show the comparison of data versus 
model in Fig.~\ref{fig:cfRAVE}.
Although we see that 
$f({\mathbf J})$ modeling can work very well, it
is a Jeans-theorem-based analysis, and it contains
all of the underlying assumptions we have noted. We
discuss evidence for the breaking of these assumptions,
noting the context in which they occur, 
in Sec.~\ref{sec:Change}.

\subsection{The Galactic Dark Matter Halo}
\label{sec:dmhalo}

A galactic dark matter halo can be described
within the distribution function framework we have
just detailed. Its distribution function 
is poorly known, even in our 
own Galaxy, simply because the only established
dark matter interactions are 
gravitational. Yet 
the possibility of the direct detection of particle
dark matter through its elastic scattering with 
nuclear targets~\cite{goodman1984detect} 
in highly sensitive, low-background 
underground experiments~\cite{gaitskell2004direct,hooper2008strategiesreview}
has stimulated keen interest in the 
physics and characteristics 
of the Galactic data matter halo. 
Indeed, such information 
is essential 
to translating the outcomes of a 
dark matter direct detection 
experiment to limits on a dark matter candidate's
mass and nuclear cross section~\cite{green2012dmastroreview,green2017dmdirectreview}.
In what follows we consider the particular Galactic inputs
needed at the Sun's location, 
noting that the local effects from dark matter capture on 
the solar system have been determined to be  
small~\cite{Peter2009DMSolarSysI,Peter2009DMSolarSysIII}. 

In a typical dark matter direct detection 
experiment the nuclear
recoil energy $E$ and possibly 
its direction $\mathbf{\Omega}$, in some coordinate system, would be detected. 
This is possible if the candidate particle's 
mass is in the range of roughly 
10-100 GeV, as expected for the long popular weakly-interacting massive particle, or WIMP, 
dark matter candidate~\cite{feng2010dmcandidates}.
More recently, the idea of measuring electronic
recoils to probe dark matter candidates in 
the sub-GeV mass 
range ~\cite{Essig2011subGeVDM,Essig2015DMeSiconcept, Hochberg2016DMescat1, Hochberg2016DMescat2, Hochberg2017DMDirac1, Hochberg2018DMDirac2, Hochberg2019DMnano, Coskuner2019DManiso, Kurinsky2019DMdiamond, Blanco2020DMaromatic1, Blanco2021DMaromatic2, Tiffenberg2017SENSEI,Barak2020SENSEI}
in scattering experiments has
been developed; the interpretation of such 
experiments also requires astrophysical 
information on dark matter \cite{Radick2021DMevelocity}. 
In what follows we focus on dark matter-nuclear
interactions, though our considerations 
largely generalize to the case of electronic
recoils as well --- we refer to 
\cite{Catena2019DMdirectatomicresponse} for
a discussion of new features resulting from atomic
excitations. 
We also note \cite{Gardner2013entrain} for a review 
of nuclear and hadron physics issues in 
the evaluation of the cross 
section for dark-matter-nuclear scattering
and to 
\cite{Gluscevic2015DMdirectdetectIDtheory} for
a review of how these experiments, taken en masse, 
constrain dark matter models. We also note
\cite{Lee2014FocusDMdirectdetectAnnualMod}
for a discussion of astrophysical issues pertinent
to the interpretation of an annual modulation signal. 

If the scattering of dark matter and nuclear
target were simply elastic, as usually 
assumed, then information on the 
local dark matter density and 
velocity distribution would suffice 
to interpret the results~\cite{green2017dmdirectreview}.
In particular, an integral over
the {\it lab frame} dark-matter velocity distribution 
$f_{\rm lab}( \mathbf{v}, t)$ involving
\begin{equation}
    v_{\rm min}= \left( 
    \frac{E M_A}{ 2 \mu_{\chi A}^2}\right)^{1/2}
\end{equation}
would be required, where $M_A$ and $\mu_{\chi A}$ are
the nuclear and reduced masses, respectively; and  
the integral is 
computed 
up to the escape velocity. 
Potentially, too, the directional information 
such an experiment can provide is not
only a sensitive discriminant of dark matter 
models~\cite{Mayet2016reviewreachDMdirectdetect}, but it 
can also 
yield constraints on the dark matter velocity 
distribution~\cite{Lee2012constrainDMvdistrib}. 
In the event that the recoil direction is not 
detected, the integral is simply some function of 
$v_{\rm min}$, but it is also time dependent, 
because the Earth's velocity with respect to 
the Galactic rest frame is time dependent. 
This last velocity is relatively 
well-known, and, assuming
that the local standard of rest coincides with
the rotational standard of rest, it is fixed
by the vector sum of the local circular speed, 
the peculiar velocity of the Sun with respect
to the rest frame of Galactic rotation, and the 
velocity of the Earth about the Sun.
Nevertheless, the underlying dark matter
velocity distribution is not well-known, nor
is its needed integral, which we term 
$g(v_{\rm min},t)$. 
We note in passing if the 
scattering were {\it inelastic}~\cite{TuckerSmith2001inelasticDM,Fan2013dissipativeDM}, 
as possible if the particle were 
composite~\cite{Khlopov2005compositeDM,gardner2008dmgyrofaraday,Alves2009compositeinelastic,Kribs2009quirkycompositeDM}, though this is not required, 
then the nuclear response to the dark matter
probe is also involved~\cite{Fitzpatrick2012EFTDMdirectdetect,Fitzpatrick2012modindDMdirectdetect}, 
complicating the connection between the 
experimental outcomes, the 
dark matter astrophysical inputs,  and the
desired dark matter constraints. 
The use of effective field theory for 
dark matter-nuclear scattering 
shows that additional currents could 
also mediate the effect~\cite{Fan2010eftdmdirectdetect,Fitzpatrick2012EFTDMdirectdetect,Fitzpatrick2012modindDMdirectdetect}, impacting the 
determination of DM parameters~\cite{Anand2013dmelastnuclearresponse}.

There has been much discussion of strategies 
to eliminate the 
ill-known function $g(v_{\rm min})$,
given studies with different nuclear targets
and an assumption of elastic scattering~\cite{Fox2010integratingout}, or
simply of how a combined analysis 
could be made~\cite{Peter2010DMdirectdetectplusastro}, 
though the former appears to 
require that a
signal is observed in one nuclear 
target
first~\cite{Fox2010integratingout}.

Given these uncertainties, 
all direct
detection experiments assume the SHM \cite{drukier1986SHM}
 in order to put
 their results on the same footing. 
 Thus the assumed 
 distribution function 
 is in steady-state and 
 is that of an isotropic, isothermal sphere, so that
  its velocity distribution is of 
  Maxwell-Boltzmann form\footnote{In a galaxy, a velocity distribution may be of Maxwell-Boltzmann form, but this does {\it not} imply that equipartition (or the usual results of equilibrium statistical mechanics) apply~\cite{lynden-bell1967violentrelax}.}
  \begin{equation} \label{eq:isothermal-MBfv}
      f(v) \propto \exp\left(-\frac{v^2}{\sigma_v^2}\right).
  \end{equation}
If the DM density has a radial profile $\rho(r) \propto 1/r^2$,
then the circular speed of the DM has a radial dependence
$v_{\rm c} \sim 1/\sqrt{r}$\cite{binney2008GD}.
Extensive evidence now exists to suggest that the SHM 
is not realistic on 
several counts. Chief among the ways that the DM halo is 
believed to depart from isotropy and isothermality are that: (i) its shape is not
spherical; (ii) its velocity-distribution is 
somewhat modified by these shape effects, and its 
high velocity tail, pertinent to searches for 
lighter mass WIMP candidates, may be modified
by Galactic evolution, such as 
debris flow effects~\cite{Lisanti2011debrisflow,Kuhlen2012debrisflow};
(iii) its mass distribution, particularly in 
smaller mass halos, is a matter of debate; and
(iv) the Galaxy itself is not in steady state. 
We address the first two points briefly here, 
consider point (iii) in the next section, and 
reserve that of (iv) and its broader consequences to 
Sec.~\ref{sec:Change}. We delve into the implications of 
these refinements for the local dark matter density and
velocity distribution 
in Sec.~\ref{sec:dm_phasespace}.

Fully accounting for all of these noted effects 
would take us beyond the framework we have outlined
in Sec.~\ref{sec:df}, though it is worth emphasizing that 
the SHM 
can already be probed and refined 
within its scope
in a data-driven way. It has, after all, 
a number of simplifying assumptions. 
We refer to Green~\cite{green2017dmdirectreview} for
an extended discussion. 
For example, if spherical symmetry 
is assumed, the structure of the dark halo can 
be determined from the stars alone~\cite{binney2015darkhalo}, 
in that 
the circular speed with Galactocentric radius $r$ 
inferred from the effective 
Galactic potential, reconstructed 
from astrometric 
measurements of stars 
with {\it Gaia} DR2 data, 
is
compatible with the circular speed directly measured with
Cepheids~\cite{mroz2019rotation}.

We now turn to a discussion of 
the points we have outlined. 
The shape of the Galactic halo is not well-known, 
but it can be constrained 
though observations of stars and/or HI gas~\cite{olling2000darkhaloshape}.
There is considerable evidence, of long standing, for distortions 
in the Galactic disk, as it is both 
warped and flared in HI gas \cite{levine2006vertical, kalberla2007dark} and 
in stars \cite{alard2000flaring, ferguson2017milky}. 
Striking evidence for the latter has emerged
recently from three-dimensional maps of samples of 1339 and 2431 Cepheids, respectively \cite{chen2019intuitive,skowron2019three}. 
Galactic warps have been thought to have a dynamical origin, appearing and disappearing on time scales short
compared to the age of the universe, due to interactions with the halo and its satellites 
\cite{nelson1995damping, shen2006galactic}, though it has 
also been suggested 
that the warp in HI gas is due to the presence of the Large Magellanic Cloud (LMC)~\cite{weinberg2006magellanic}. 
We refer to Sec.~\ref{sec:MWsize} for further
discussion of the current status 
and to Sec.~\ref{sec:Change} for a broader discussion
of non-steady-state effects in the MW and its origins.

The velocity ellipsoid \cite{hagen2019tilt} and DM
distribution \cite{posti2019mass} are not spherical either, with the evidence favoring a prolate 
matter distribution. Studies of flaring HI gas in the outer galaxy also support a prolate DM 
distribution \cite{banerjee2011progressively}; these authors note that a prolate halo can support
long-lived warps \cite{ideta2000time}, which would help to explain why they are commonly seen  \cite{banerjee2011progressively}. 
It has also been suggested that some of these features could arise from a dynamically active disk \cite{chequers2018bending} in isolation. 

The Galactic velocity distribution can also be
impacted by the tidal disruption of 
dark matter subhaloes\footnote{A halo that orbits inside
a larger halo is a {\it subhalo}. Subhalos are a qualitative prediction of the hierarchical nature of galaxy formation, as we discuss in more detail in Sec.~\ref{sec:smallscaleprobes}.}, and  
the resulting debris flow
can also impact the high velocity tail of the dark matter  
distribution, as studied in the context of the via Lactea II simulation~\cite{Lisanti2011debrisflow,Kuhlen2012debrisflow}.
A broader issue is the impact of Galactic evolution
on the survival and evolution of CDM 
substructure~\cite{vandenbosch2018dmdisruption} itself, 
and we turn to this in the next section. 

\subsection{The Cold Dark Matter 
Paradigm}
\label{subsec:CDMpara}
The extreme uniformity of the observed cosmic 
microwave background suggests that the early universe 
was nearly homogeneous and isotropic.
This initial state 
can be 
realized through a yet earlier inflationary epoch \cite{Guth1980inflation1}. 
The quantum field sourcing this inflationary epoch was beset by fluctuations, as all such fields are \cite{Guth1982inflation2, Albrecht1982inflationGUT, Bardeen1983inflationscalefree}.
These fluctuations manifested as perturbations to the overall density field $\rho(\mathbf{x})$, which grew with time into large-scale structures, such as galaxies. 

Defining the overdensity $\delta(\mathbf{x})$ as the
fractional density difference from the mean $\rho_0$, 
determined over a volume for which the universe appears
homogeneous, 
it is apparent that the overdensities at two nearby
points may be correlated. The power spectrum ${\cal P}(\mathbf{k})$ of these correlations is the Fourier transform of
that correlation function. With
$\delta_{\mathbf{k}} =\int_V d^3\mathbf{x}\, \delta({\mathbf{x}})
\exp(-i \mathbf{k}\cdot \mathbf{x})$, we have
${\cal P}(\mathbf{k}) \equiv \langle |\delta_{\mathbf{k}}|^2 \rangle /V $. Then
${\cal P}$ depends only on the scalar wavenumber
$k$, because the universe is isotropic.
In the CDM 
model~\cite{Peebles1982colddarkmatter}, the supposed power
spectrum ${\cal P}(k) \propto k$, 
due to Harrison~\cite{Harrison1970fluctations} and 
Zeldovich~\cite{Zeldovich1972hypo}, is scale 
invariant, meaning that the gravitational potential 
associated with a root-mean-square (RMS) 
fluctuation at scale $k$ is 
independent of $k$~\cite{binney2008GD}, thus 
sidestepping the problem that the fluctuations 
might prove to be too large at either large {\it or} small scales~\cite{Zeldovich1972hypo}. 

It was quickly realized that this model would engender
an inside-out growth of structure~\cite{Press1974formationselfsimilar,white1978corecondensation}, as consistent with the observation
of galaxies at large $z$, with 
interesting implications at galactic scales~\cite{blumenthal1984colddarkmatter}. 
It had also been realized that 
the size and mass of spiral 
galaxies should be much larger than once 
thought~\cite{Ostriker1974darkmatter,Einasto1974darkmatter,Tremaine1999comments} 
and that the bulk of that mass would be dark. Thus, $N$-body numerical 
simulations should prove powerful probes of the CDM
distribution and 
evolution~\cite{davis1985colddarkmatter}. 
Moreover, in order to test the 
CDM paradigm, 
it is essential to test whether this supposed
scale-free hierarchy is actually reflected
in the data. The evolution of structure with scale $k$ and redshift $z$ is 
encapsulated by the transfer function 
\begin{equation}
    T^2(k) \equiv \frac{{\cal R}(k,z=0) }{{\cal R}(0,z=0)} \,,
\end{equation}
where 
${\cal R}(k,z) \equiv \langle \delta_k^2\rangle|_{z}/
\langle \delta_k^2\rangle|_{z\to\infty}$
and the power spectrum in the linear regime is 
given by 
\begin{equation}
    {\cal P}(k) \propto 
    T^2(k) {\cal P}_{\rm prim} (k) 
\end{equation}
where the primordial power spectrum 
${\cal P}_{\rm prim} (k)$ is also taken 
to be scale invariant, and thus of Harrison-Zeldovich
form.

\begin{figure}[t!]
\centering
\includegraphics[scale=0.9]{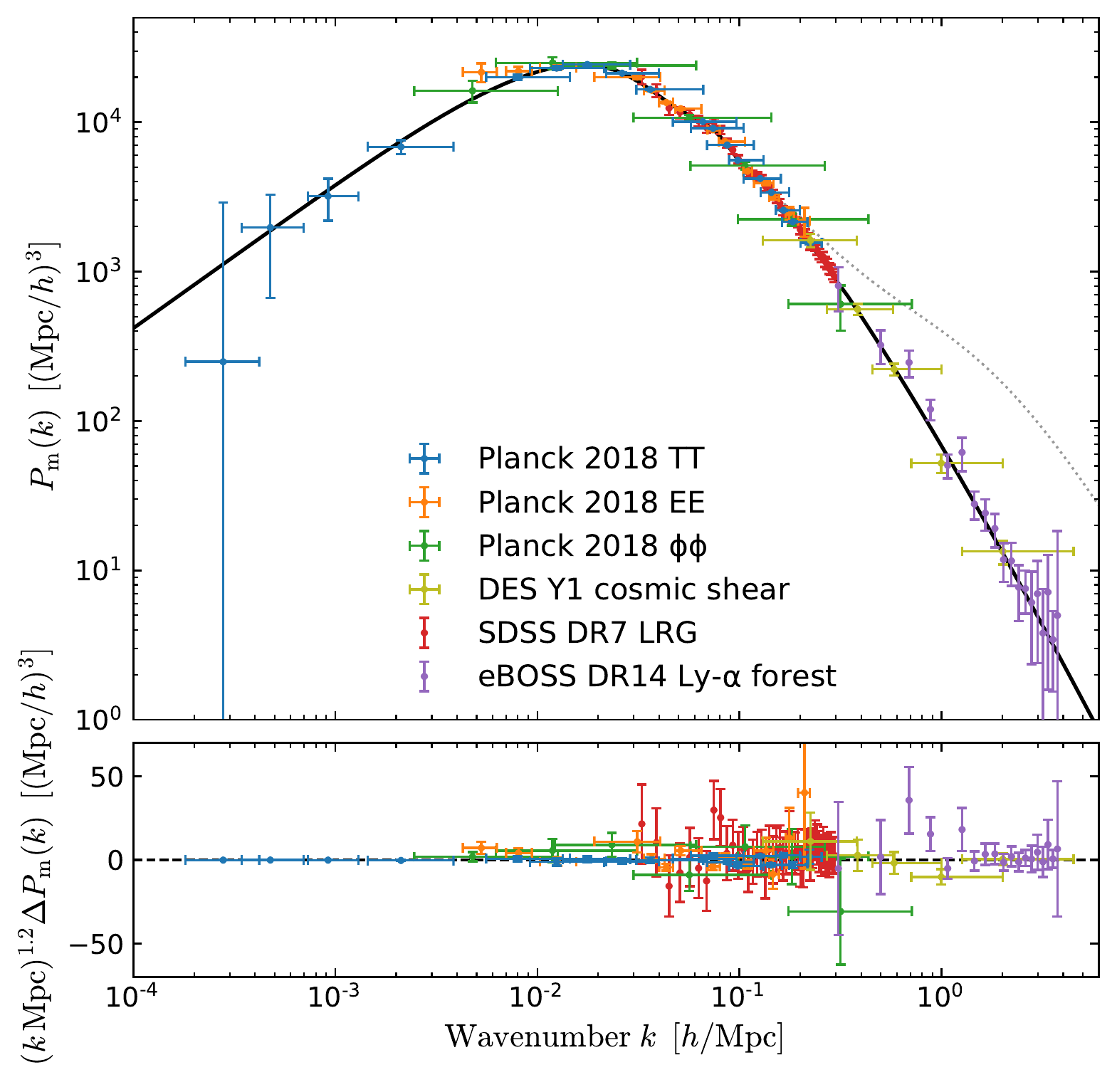}
\begin{minipage}[t]{16.5 cm}
\caption{Upper panel: The measured matter power spectrum compared to the 
prediction of the best-fit Planck 2018 cosmology \cite{Planck2020legacy} (solid)
and an example prediction of a nonlinear power spectrum (dashed). Large wavenumbers correspond to small physical distances, $L \simeq 2\pi/k$. Lower panel: Deviation of the matter power spectrum $\Delta P$ from the Planck model, scaled by $k^{1.2}$.
The MW probes the nonlinear, high-$k$ regime.
 From \cite{Chabanier2019mps}. 
}
\label{fig:linearpowerCDM}
\end{minipage}
\end{figure}

A comparison across a wide range of physical scales between the linear structure formation prediction assuming a 
CDM model with inflation 
and recent observational data is shown in Fig.~\ref{fig:linearpowerCDM}.  
The data in this figure is evidently well-described by the linear theory prediction derived from the best-fit Planck 2018 parameters \cite{Planck2020legacy}. This is true even at high-$k$, corresponding to (relatively) small length
scales, since even the low-$z$, high-$k$ measurements of the Ly-$\alpha$ forest probe the possibility of overdensities in underdensities \cite{rauch1998reviewlymanalphaforest}. However, a departure from linearity 
is necessarily anticipated at high matter densities and large wavenumbers \cite{Peebles1982colddarkmatter, Peebles1980book, Jain1994nonlinear, Carlson2009perturbationtheory}. 
{\it Ab initio} cosmological calculations in the effective theory of large scale structure can probe mildly nonlinear scales \cite{Carrasco2012eft}, but remain limited to scales larger than roughly 20 Mpc \cite{Lewandowski2017eft1, Lewandowski2018eft2}.
Thus, the MW is part of the small-scale frontier in which we hunt departures from the CDM 
paradigm. Studies of the MW are squarely in the nonlinear regime, 
and comparisons to numerical simulations of cosmological structure necessarily play a key role. 

For this reason, we turn to the study of the cold matter
structure from $N$-body simulations, and comparison to observations, to determine
whether indeed there is 
small scale structure without end. This issue
has incurred much discussion. Some $N$-body numerical 
simulations show that large fractions of 
dark matter subhaloes undergo complete 
disruption, prompting the question as to 
whether the origin of this effect is physical, 
arising from tidal heating and stripping, or
is a numerical artifact. 
This issue has been recently been studied
carefully by van den Bosch 
and collaborators~\cite{vandenbosch2018dmdisruption,vandenbosch2018dmdisruptionII}, and they find that the 
destruction of CDM 
subhaloes in the
absence of baryonic effects is extremely rare. 
They identify a number
of numerical effects that could yield 
numerical overmerging, driven by 
resolution limitations and finite system size. 
Moreover, it appears that dark matter
substructure survives Galactic evolution
up to the current epoch~\cite{green2019tidalevolutiondmI,green2021tidalevolutiondmII}. 
This puts in context the current lack of
consensus as to why subhalo 
destruction happens at all. Certainly, however, 
some measure of disruption, from tidal interactions, 
generating gravitational heating, is expected, 
and this should be 
revealed by the formation of tidal tails 
in self-gravitating subhaloes and in 
their perturbations.

A separate point of interest is the matter 
distribution, which reflects 
$d^3\mathbf{v}\, f$ in the DF framework. 
Thus we are considering a steady-state configuration --- 
the result of the relaxation processes we describe later
in this section --- and we term this a halo. 
Studies of the luminosity distribution of 
elliptical galaxies
are well-described by a two-power density model~\cite{binney2008GD}, 
and it is natural to consider a 
similar model of the DM distribution as well. 
Indeed, the well-known 
Navarro, Frenk, and White (NFW) model~\cite{Navarro1995assembly}
\begin{equation}
    \rho(r) = \frac{\rho_0}{(r/a) (1+r/a)^2 }
    \label{eq:NFWform}
\end{equation}
is a specific case of just such a form. 
Although $\rho_0$ and $a$ would appear to be free parameters, 
the numerical simulations of 
\cite{Navarro1996structurecdmhalo} reveal them 
to be strongly correlated, so that 
Eq.~(\ref{eq:NFWform}) can be regarded as 
a one-parameter family of shapes. 
That remaining parameter can be fixed by choosing a 
maximum radius. A conventional choice is 
$r_{200}$, the radius at which the mean density
is 200 times the {\it critical density} 
$\rho_{\rm c}$\footnote{So-called because
in a flat universe, 
$\Omega (t)\equiv \rho(t)/\rho_c(t) =1 \,\forall\, t$.},
where $\rho_{\rm c}(t) = 3H^2(t)/8\pi G$; we discuss this choice in more detail below.
The concentration $c$ of the halo is given by 
$c=r_{200}/a$, and the
simulations of \cite{Navarro1996structurecdmhalo}
show little sensitivity to its value --- 
the total mass enclosed within $r_{200}$ can 
vary by powers of ten, yet $c$ changes by no more than a
factor of a few. 
The NFW model is ``cuspy'' because $\rho(r)$ diverges in the $r\to 0$; it also potentially suffers from 
a logarithmic divergence of its total mass at large radii, though this divergence is in practice cutoff by choice of a maximum radius such as $r_{200}$ mentioned above. 
Remarkably, Navarro, Frenk, and White have 
determined that the form of 
Eq.~(\ref{eq:NFWform}), if applied to a
primary halo, appears to be 
universal~\cite{navarro1997universal,Navarro2004cdmhaloIII},
with few exceptions~\cite{navarro2010diversitycdmhalo}, 
describing systems differing by 
over 20 orders of magnitude in mass~\cite{Wang2019NFWuniversal20}. 

The density profiles of subhaloes, 
in contrast, 
are more
typically described by a single-power-law form, 
with a cutoff at larger distance, engendered by tidal 
effects from the host 
halo~\cite{kazantzidis2004subhaloprofile}. 
A diversity of subhalo profiles have been 
observed, as reviewed by \cite{Tulin2017reviewtulinyu}, with 
evidence for both cuspy and cored profiles and much 
corresponding debate as to their differing origin. 

Returning to the profiles of primary haloes, 
we note that the origin of the observed universal
behavior is not well understood. For example, 
numerical studies have shown that NFW profiles
emerge even if the initial power spectrum 
${\cal P}(k)$ is set to zero above some $k=K$~\cite{huss1999universaldarkhalo,moore1999coldcollapse}.
Thus a cuspy profile in this case cannot arise
as a relic of an initial condition; rather,
dynamics must insure the effect. 
A halo is the outcome of the violent relaxation of 
a phase of initial gravitational collapse, to yield
a system to which the virial theorem applies; this
process is sometimes called virialization --- and
thus this is what must act, regardless of initial
condition, to yield the NFW form. We note that 
a virial analysis suggests that $r_{200}$ is crudely
the radius over which the halo is in virial equilibrium, making it the {\it virial radius}\footnote{The mass associated with the volume enclosed by the virial radius 
is the {\it halo mass}. It is apparent that only a rough assessment of a halo mass is possible.}, 
with the mass beyond that radius being apparently in 
first infall~\cite{binney2008GD}. 
This rationalizes
its two power-law form. We refer to 
\cite{dalal2010origin} for further thoughts on
its origin.

\subsubsection{Small-Scale Challenges}

Although the CDM 
paradigm has been 
enormously successful 
in explaining the large-scale structure of the 
Universe, small-scale challenges to it 
have slowly emerged as  well~\cite{Weinberg2015CDMcontroversiessmall,bullock2017challengeLCDM}. These are potentially entrained
with numerical challenges in simulating
the number and structure of 
CDM subhaloes~\cite{vandenbosch2018dmdisruption,vandenbosch2018dmdisruptionII,green2019tidalevolutiondmI,green2021tidalevolutiondmII}. 
There are observational 
challenges as well, though the ability to identify 
faint subhaloes in the MW has greatly improved in 
recent years. In addition, the role of baryons 
in determining the structure of subhaloes is still being
clarified. 
Nevertheless, we may yet
establish the limits of the CDM paradigm 
by (i) determining whether there is indeed 
a 
deficient number of observed satellites with respect to the 
number expected, (ii) determining whether the structure of 
subhaloes is consistent with their intrinsic luminosity, or by (iii) 
resolving whether the cores of subhaloes are 
cored or cusped. These are known as
the ``missing satellites'', ``too big to fail'', and ``core-versus-cusp'' problems, respectively.
It has been suggested that the missing satellites
problem has been 
solved~\cite{kim2018nomissingsatellites}, 
yet a complete consensus has not been reached 
--- see, e.g., the ``Dark Matter'' 
review of \cite{PDG2020} --- and we refer 
to the review of \cite{buckley2018gravitational} for a detailed discussion. In Sec.~\ref{sec:smallscaleprobes} we 
consider small-scale DM probes that would seem less sensitive
to baryon effects. 

\subsubsection{Relaxation Mechanisms}

``Relaxation'' encompasses the 
processes by which a system can approach 
equilibrium --- that is, how it can approach 
a steady state. We emphasize that a galaxy 
can attain dynamical equilibrium, but not
thermodynamic equilibrium, because there is no 
maximum entropy state in this case~\cite{binney2008GD}. 
We have already 
noted how the collisionless Boltzmann 
equation, along with the Poisson equation, 
governs the construction of the static  
galactic distribution function. This is
possible, in part, because the stellar relaxation
time, mediated by the diffusion of a star through 
two-body collisions --- in contrast to its evolution
in the smooth mass field of the Galaxy --- 
is exceptionally long. That is, the 
time scale for the star 
to change its speed $v$ by that same amount is 
estimated to be~\cite{binney2008GD}
\begin{equation}
    t_{\rm relax} \sim \frac{0.1\,N}{{\rm ln} N} t_{\rm cross} \,,
\end{equation}
where the 
time for a star of speed $v$ to cross the Galaxy
is $t_{\rm cross} \sim R_G/v$, 
and $R_G$ is its radius. Thus in our Galaxy with 
$N\sim 10^{11}$ and of a few hundred $t_{\rm cross}$
in age~\cite{binney2008GD}, 
the two-body relaxation time is 
absurdly long, as noted long ago by 
Zwicky~\cite{zwicky1939timescale,lynden-bell1967violentrelax}, and we must look to other
processes to determine how a system 
with gravitational interactions can evolve
with time. Otherwise we would have the 
conundrum of having to explain how a galaxy 
might form very close to the state in which 
we observe it today.

We note three basic 
mechanisms by which a 
$N$-body gravitational system can evolve to a 
steady state~\cite{binney2008GD,mo2010galaxyformation,thorne2017MCP,binney2008GD}: phase mixing, violent relaxation, 
and chaotic mixing. Only the last leads 
to irreversibility, through its extreme sensitivity to 
the system's initial conditions. In all cases the
collisionless Boltzmann equation applies, so that 
Liouville's theorem holds --- but only if we resolve
the fine-grained phase-space structure and 
consider populated
orbits. In violent relaxation, the 
potential is time-dependent, as in the example of gravitational 
collapse, so that the energy is not a constant of motion. 
Thus the distribution function is not static, 
but $df/dt =0$ still applies. We contrast chaotic mixing
and phase mixing in that the orbits in the former
case are stochastic 
rather than regular, so that over time two orbits
that were initially close in phase space separate
exponentially 
with time. Nevertheless, 
the mechanism by which 
the $N$-body
gravitational system can 
relax 
is common 
in all cases. That is, the population of orbits
in phase space spreads out with time, even if 
Liouville's theorem requires that the total volume of 
phase space remains constant. Viewed broadly, 
after some time, a fixed region of phase space
will contain both occupied and unoccupied regions. 
If we define a coarse-grained distribution
function, $\bar{f}$, blurring the occupied and unoccupied
regions, we can realize $\bar{f}< f$ and thus relax
to a higher entropy state. Even in the case of regular
orbits, the time scale of this process can be 
rather smaller than the age of the Galaxy, allowing 
it to evolve from its initial state. Whether 
visible and dark matter evolve to a similarly 
coarse-grained distribution function is a 
matter of assumption~\cite{thorne2017MCP}.

We conclude this section by emphasizing that the 
paths by which a $N$-body gravitational system can 
achieve steady-state are limited. This stands
in 
stark contrast to 
systems in which inelastic or dissipative 
processes 
are present. We refer to 
\cite{Rosenberg2017coolingdissipativeDM}
for just such an exemplar dark matter study. 
More generally, studies of the global 
population of dark matter in phase space, 
to identify, e.g.,  
a dark disk
in the MW~\cite{Fan2013dddm,Fan2013ddu}, 
can serve to anchor novel features of the 
dark universe. We consider this in greater
detail in Sec.~\ref{sec:dmcand}.

\section{Targeted Review of Parameters of the Milky Way \label{sec:Parameters}}

Early parameterizations of our home MW were simplified due to the limitations of the available data.  Infrared and microwave studies able to penetrate the dust obscuration at low latitudes in the disk and toward the Galactic center in the 1990s improved our overview significantly \cite{hauser2001cobemission}. Most recently, the {\it Gaia} dataset with accurate parallax-based distance and proper motion information has again enormously improved the breadth and depth of our knowledge of the MW.  Studies of motions and densities of stellar streams, satellite dwarf galaxies and globular clusters in the Galactic halo have served as probes of the Galactic potential,
presumably dominated by a DM component.

The overall picture of our MW remains one of a pair of stellar disks, thin and thick \cite{Gilmore1983thinthick}, surrounded by a massive dark halo of uncertain extent, shape, orientation, and clumpiness.  Many details, however, are beginning to be filled in: as one looks more closely, these reveal more structure on many scales \cite{bland2016galaxy}.
Moreover, systematic studies of the spatial distribution of 
stellar metallicities, i.e., a chemical cartography, particularly in the ratio of the alpha
chain elements to Fe \cite{Hayden2015ChemCartography,Bland-Hawthorn2019dissectdisc}, 
shows dissection by chemistry to be key to 
distinguishing the two disk components~\cite{Bland-Hawthorn2019dissectdisc}.
We discuss 
small-scale structures in the ensuing sections, and focus here on the gross properties of the MW itself.

\subsection{Milky Way Mass\label{sec:MWmass}}

The total mass of the MW, including its dark halo, 
is roughly $10^{12}$ times the mass of the Sun, denoted $\msun$, enclosed within a virial radius of $\approx 200$ kpc \cite{Wang2020massMWreview}.  Our best measures come from studies of the orbits of satellites such as the LMC/SMC system \cite{besla2015magcloudmasses} as well as studies of the distribution of standard candles such as blue horizontal branch (BHB) stars and assuming they have come into approximate equilibrium under the Jeans approximation \cite{xue2008massmilkyway}. 
That assumption of relaxed equilibrium is not true in detail, and so overall the estimates for mass still come with rather hefty error bars of 30\% to 50\% \cite{Wang2020massMWreview}.

There are models for the distribution of non-dispersive DM on many scales from N-body simulations.
These generally find DM halos of galaxies well fit by a NFW \cite{navarro1997universal} profile with inner slope $\rho(r) \sim r^{-1}$ and outer slope $\rho(r) \sim r^{-3}$, plus a central density and a concentration scale indicating a transition from inner to outer slope, as in Eq.~(\ref{eq:NFWform}).  Even simpler isothermal models of DM density $\rho(r) \sim r^{-2}$ are reasonable fits to many halo simulations over a wide range of scales, but, of course, the total integrated mass of an isothermal halo tends to infinity at large radius and so must be cut off by a steeper falloff at large $r$.   Studies of the stellar distribution in the outer parts of the Galaxy show a steeper-than-NFW outer slope falloff, with $\rho(r) \sim r^{-4.5}$ for stars \cite{stringer2020rrlyraesinhalo}, and it is possible that the DM falls off more quickly than the NFW profile predicts as well.  In the inner parts of galaxies, while the NFW model and all non-dissipative models of DM predict cuspy ($\alpha < 0$) density spikes at small $r$ where $\rho(r) \sim r^{\alpha}$, in fact, there is little observational evidence for any central cusps with slope as steep as $\alpha = -1$ in stellar density or in DM.   Studies of velocities and densities of stars in the central regions of the MW's largest nearby satellites Fornax and Sculptor \cite{walker2011corecusp} show $0 > \alpha > -0.5$.   Researchers \cite{Read2005cusptocore} have shown how baryonic dissipation can flatten out cuspy spikes in the center of galaxies and help understand the relative paucity of observed satellites compared to numbers of DM clumps predicted in computer simulations \cite{Brooks2013toofewsatellites}.  We note that the observations of the centers of galaxies and clusters to determine the profile slope remains an exceedingly difficult measurement.  In clusters with a central black hole, \cite{Bahcall1976blackholecuspslope} showed that the expected density profile near a central black hole would approach $\alpha \sim -1.75$. In the dense cluster M15, thought to have an intermediate mass black hole, \cite{vanderMarel2002M15blackhole}  have shown evidence for some cuspiness in the stellar profile, though we stress that the 
number of stars in the very central region 
is extremely small leading to large Poisson error on the inner density slope.

\subsection{Size and Shape \label{sec:MWsize}}

Where does the MW end? To a large extent, the answer to this question is a matter of definition. Various proposals for the end of a dark matter halo include: comparisons against the background density at the time that the halo separated from the Hubble flow \cite{Tinker2008hmf}, comparisons against the background density at a floating redshift \cite{Bryan1998virialoverdensity}, and dynamical measures, such as the splashback radius \cite{More2015splashback}. The total mass of a virialized NFW halo depends logarithmically with the maximum distance, noting Eq.~(\ref{eq:NFWform}), is 
\begin{equation}
    M_{\rm NFW} = \int_0^{R_{\rm max}} d^3r \rho_{\rm NFW} = 4 \pi \rho_0 a^3 \left[ \ln(1+R_{\rm max}/a) - \frac{R_{\rm max}}{a+ R_{\rm max}} \right].
\end{equation}
Thus, $d M_{\rm NFW} / d R_{\rm max} \propto R_{\rm max}  /(R_{\rm max} + a)^2$, 
so that for $R_{\rm max} \simeq 10 a$ choices of how to truncate the Milky Way halo can affect its estimated mass at the $\sim \mathcal O(10)\%$ level.

The DM halo of our MW extends to at least 50 kpc as the orbits of the LMC/SMC are clearly strongly affected by it. The stars and gas of the LMC/SMC and its associated DM appear to be on their first orbital pass around the MW \cite{besla2007lmcsmcfirstpassage, kallivayalil2006lmc}. The details of the orbit and distribution of gaseous and stellar tidal tails shows strong evidence for tidal friction and an alteration in the semi-major axis of the combined MW --- LMC/SMC system orbit, which provides evidence for drag from the DM halos of the two systems slowly drawing closer together \cite{besla2007lmcsmcfirstpassage}.  Beyond 50 kpc, tracers are rare, though the dwarf galaxy Draco at 80 kpc from the MW does show a flattened stellar distribution, which could be a sign of either tidal influence of the MW at 80 kpc or evidence for having its own DM halo.  Recent work suggests the latter, as no evidence of tidal stripping of Draco stars is seen \cite{segall2007dracohasnotidaltails}.  In contrast, nearly every dwarf satellite companion that approaches within 20 kpc of the MW center appears to have strong evidence for tidal distortion and in many cases stars from the satellite object are stretched into an elongated stream by tidal interaction with the MW and its DM halo \cite{shipp2018streamsfromdes}.

We conclude that the influence of DM is strong in the halo of the MW out to at least 50 kpc, but appears to diminish significantly at radii greater than 80 to 100 kpc.  Our nearest large spiral neighbor, Andromeda, which is $\gtrsim700$ kpc distant from the MW \cite{vanderMarel2008distM31}, has its own complex system of tidal streams and associated dwarfs, and has its own dark halo of uncertain extent \cite{Tollerud2012M31dwarfs}. Both the MW and Andromeda's dark halos likely extend out to beyond 200 kpc from their respective centers with density decreasing at the NFW outer slope of $-3$ or steeper.  The extent to which the halos overlap in between or can be said to be a common halo is uncertain.  The currently favored CDM 
hierarchical structure formation scenario --- with build up of structure from smaller clumps densities to larger --- suggests that the halos began well separated. Their overlap is increasing over time and they will completely merge a few billion years in the future. There are no known sufficiently luminous stars or other tracers in between the two large spiral systems to map the DM distribution between them in detail.   A problem with using luminous tracers far out in the halo of the MW, beyond 100 kpc, is that the timescales for completing an orbit or responding to dynamical friction effects approach or exceed the Hubble time, $\tau_H \equiv cH_0^{-1} \simeq T_{\rm Univ} \simeq 10^{10}$ yr. For this reason it becomes difficult to distinguish systems in equilibrium or which have relaxed from those that are just forming or interacting for the first time.

\subsection{Components \label{sec:MWcomp}}

The shape of the MW thin and thick stellar disks is clear. The more massive thin disk has an exponential distribution in radius with scale length of  over 3 kpc and an exponential falloff in vertical ($|Z|$) scale height of 300 pc. 
These quantities refer to the $1/e$ fall-off in the directions parallel and perpendicular 
to the plane of the disk, respectively.
It is strongly dissipated and rotates at about 
$220~{\rm km/s}$
at a radius of about $8 \rm kpc$
from the Galactic Center, the radius at which the Sun orbits (another estimate of mass). We discuss precision determinations of these parameters later. 
Such thin disks are unstable to clumping and strong bar formation \cite{mo2010galaxyformation}, and thus the existence of only a weak bar in the MW (and other so-called Grand Design spiral galaxies) was early evidence for a supporting massive dark halo \cite{ostriker1973instabilityofdiskwithnohalo} surrounding each spiral galaxy.  In addition to the thin disk, the MW has a chemically distinct thick disk, which consists of stars with alpha peak element metallicity about 2x higher ($[\alpha/{\rm Fe}] \approx 0.3$ on a log base 10 scale) than in the thin disk
\cite{Gilmore1989disks}.  This population of stars presumably comes from an environment that has been enriched by the debris of remnants of exploded high-mass progenitor (type II) supernovae.   

Recent findings made possible in large part by {\it Gaia} satellite measurements point to a major, approximately face-on (as opposed to a prograde or retrograde infalling) merger of a massive ($>20\%$ of the MW's mass) dwarf galaxy. This has been called the {\it Gaia}-Enceladus/Sausage event, and is estimated to have occurred between 10 and 12 Gyr in the past \cite{helmi2018merger,belokurov2018co}.  This event was gas rich and is thought to have provided impetus for a significant epoch of higher-mass star formation. This may have led to the high abundance of alpha-element rich stars of the thick disk.  In addition, that event disturbed and dynamically heated, primarily vertically, an existing proto-thin disk, explaining both the larger scale height (0.8 kpc) and shorter 
scale length (2 kpc)
of the thick disk compared with today's thin disk \cite{vincenzo2019sausagemetals,grand2020sausagemashup}. 
Recent studies of the $[\alpha/{\rm Fe}]$ composition of stars at a variety of Galactocentric radii confirm the picture of a short scale length for the high-$[\alpha/{\rm Fe}]$ thick disk, essentially ending just beyond $R_0$, and additionally find support for an extended, low-$[\alpha/{\rm Fe}]$ thin disk, flared beyond $R_0$, with on-going star formation \cite{Bensby2011flaringalphapoor,Bland-Hawthorn2019dissectdisc}.

The so-called asymmetric drift or lag of thick disk stars, which complete a rotation around the galaxy at a slower pace than thin disk stars at similar radii from the Galactic center, can also be understood as the result of this ancient massive merger. 
Mergers can lead to a partial cancellation of the existing disk angular momentum, with some of the circular velocity of rotation of the thick disk becoming an up-and-down vertical component of motion \cite{binney2008GD}. 

While the {\it Gaia}-Enceladus/Sausage event is accepted as the most significant merger that our Galaxy has had in the past 10 Gyr, there is evidence for other 
substantial merger events, possibly in the much
more recent past~\cite{Donlon2020shells}, and for on-going interactions with our satellite neighbors.  
Determining the predicted kinematic signature in the Galaxy's disk or halo resulting from a close interaction with a massive satellite such as Sagittarius \cite{poggio2020sagmodeling,grionfilho2020overviewmodelsgr} 
is a subject of ongoing theoretical study, even if 
Sagittarius is less massive \cite{blandhawthorn2021snail}
than once suggested \cite{purcell2011sagittarius}.
Moreover, the LMC  \cite{erkal2019LMCandorphan, GHY20,conroy2021halogiantasym}, with or without Sagittarius \cite{Vasiliev2021Tangoforthree},
can influence the Milky Way, and studying their impacts
remains an active area of research.  Also see Section \ref{sec:intruders} below.

\begin{figure}[t]
\begin{center}
\includegraphics[width=0.85\textwidth]{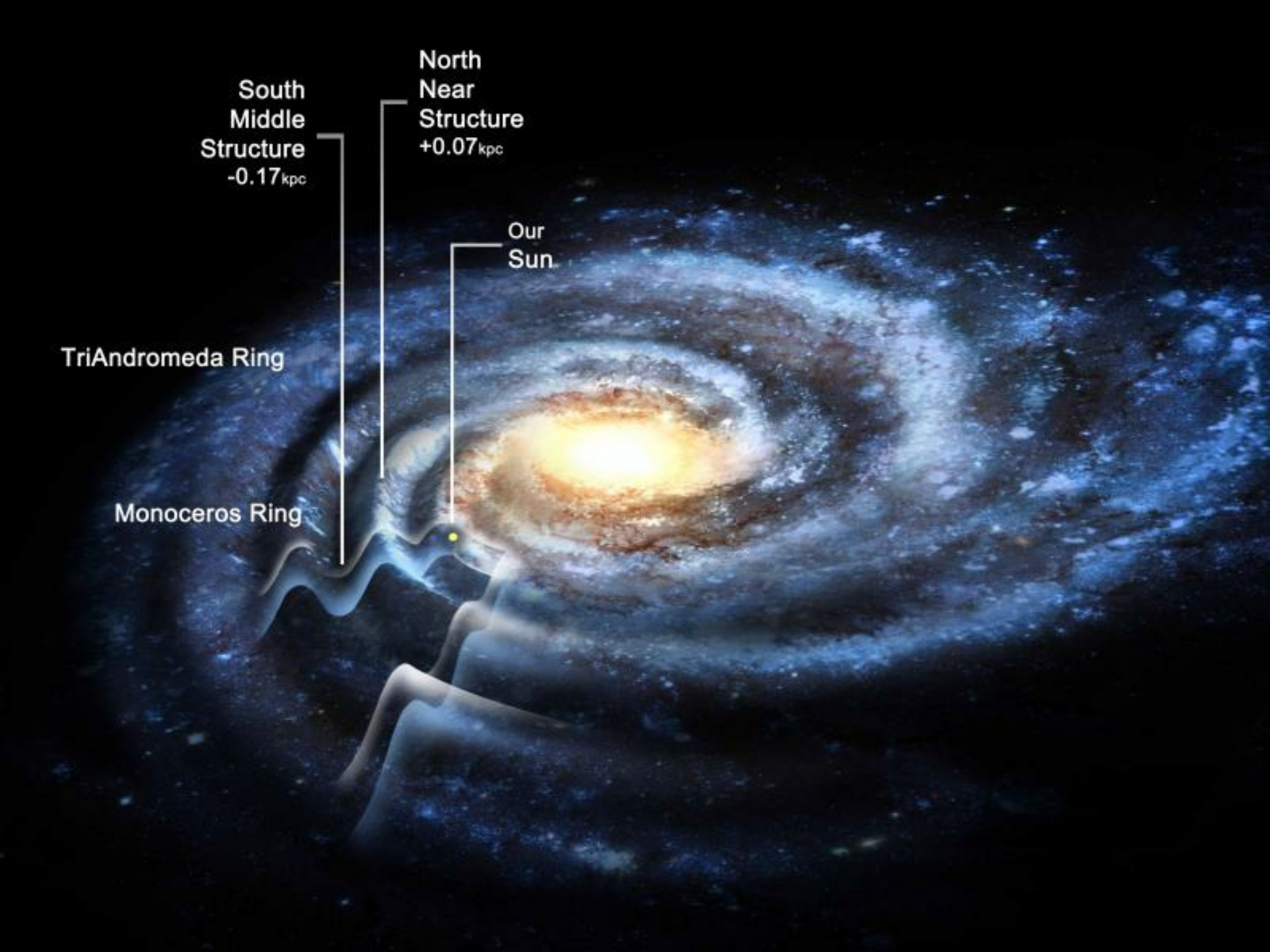}
\caption{Artist's conception showing 
corrugated ripples in the MW's disk \cite{kerr1957magellanic,alard2000flaring,xu2015rings}.
The vertical structure of the MW's disk is not well characterized except quite close to the Sun, 
with evidence for dynamical processes that couple 
vertical and planar motions \cite{Antoja2018youngperturbedMWdisk,Khanna2019ridgesarcheswaves}. 
Credit: Dana Berry/Rensselaer Polytechnic Institute\label{fig:ripples}}.
\end{center}
\end{figure}

In the outer reaches of the thin disk, beyond the solar radius at 8 kpc and extending to as far as 30 kpc, there is considerable evidence for warping, flaring, or more complex,  corrugated disturbances in the stars and gas as portrayed in Fig.~\ref{fig:ripples}  \cite{kerr1957magellanic,alard2000flaring,xu2015rings}. This is seen in both gas, via radio observations, and in the visible motion of stars.  As discussed below, this is indicative of a 
non-static potential \cite{GHY20}, which may be caused by recent interactions of our MW, in particular within the last Gyr, and ongoing. The primary culprits behind these perturbations are the LMC/SMC system as well as the massive Sagittarius dwarf system and its associated DM overdensities.

\begin{figure}[t]
\begin{center}
\includegraphics[width=0.85\textwidth]{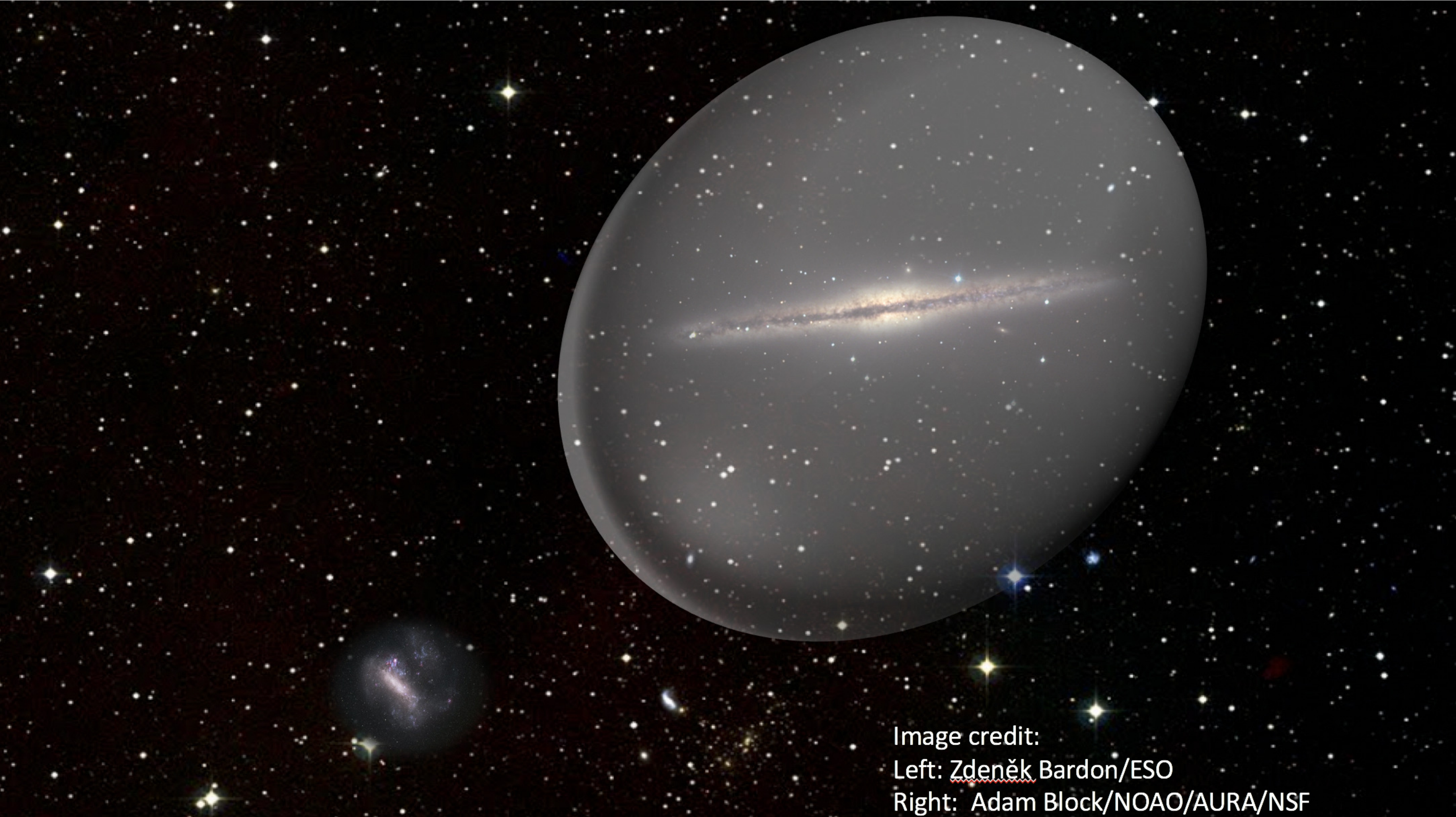}
\caption{Artist's conception showing the relative tilt of the dark halo surrounding the MW in the direction of the Large Magellanic Cloud system (lower left).  The MW and Magellanic clouds are separated by 50 kpc. Note that the halo actually extends well past 200 kpc in extent and is composed of clumps of unknown lumpiness. The MW's halo also completely envelopes the LMC/SMC and their own smaller dark matter halo. Credit: Austin Hinkel/University of Kentucky.
\label{fig:tiltedhalo}}
\end{center}
\end{figure}

Until recently the shape of the halo was assumed to be spherical or slightly oblate or prolate, with the halo axes lined up with the disk axes. Recent observations have updated this picture and provided substantial additional detail.   Though the outer {\bf stellar} halo of the MW is oblate, flattened with $c/a = 0.8$ \cite{Xue2015stellarhalo}
or flatter along the vertical axis, \cite{helmi2004dark} found evidence for a prolate rather than oblate DM dominated potential. N-body simulations of \cite{zentner2005prolatehalo} found prolateness of dark halos to be a common feature of large scale structure.
Detailed fits to the Sagittarius stream data, however, found that 
rotational 
symmetry about the $\hat{z}$axis 
did not hold \cite{law2010sagittarius} and in fact a triaxial halo, slightly misaligned, was necessary to fit the data.
More recent refinements to these fits by \cite{erkal2019LMCandorphan} found that the LMC/SMC system and its accompanying DM was a major perturber of the MW's disk and halo system. In fact, the mass of the LMC/SMC system may approach 25\% that of the MW, as reviewed in more detail below. This renders a perturbative expansion to the dynamics poorly behaved, and requires more detailed 3-D $N$-body simulations.   
A current view of the shape and orientation of our MW dark matter halo is of a tri-axial or tilted axisymmetric MW DM halo. This halo is mis-aligned with the stellar disk and has its long axis oriented in the direction of the LMC/SMC as shown in 
Fig.~\ref{fig:tiltedhalo} \cite{erkal2019LMCandorphan, GHY20}.

The ``clumpiness'' of the halo DM is 
very uncertain, and may be closely tied 
to the nature of DM itself.
Exploring the evidence for DM clustering on all scales from a few pc to a few kpc is 
an observational endeavor of much current focus.  Globular clusters on 10-50 pc scales and the solar system on scales of $10^{-4} \rm pc$ do not appear to have any significant DM and have measured mass-to-light ratios near unity (e.g. \cite{baumgardt2009masstolightglobular}) suggesting that all mass is accounted for by the visible starlight (or the Sun in the case of the solar system, though evidence for dark matter
may yet come from the outer reaches of the solar
system~\cite{Tremaine1990dmsolarsystem}).  
This is in contrast to dwarf galaxies on 500-1000 pc scales which clearly show evidence for very significant DM halos and mass-to-light ratios that exceed 100 in the most extreme cases \cite{faber1979masstolightgalaxies}.  The smallest scale 
on which DM is clumped may then lie somewhere in this $10-1000$ pc range.


\subsection{Rotation Curve \label{sec:MWrotcurve}}

\begin{figure}[t]
\begin{center}
\includegraphics[width=0.55\textwidth]{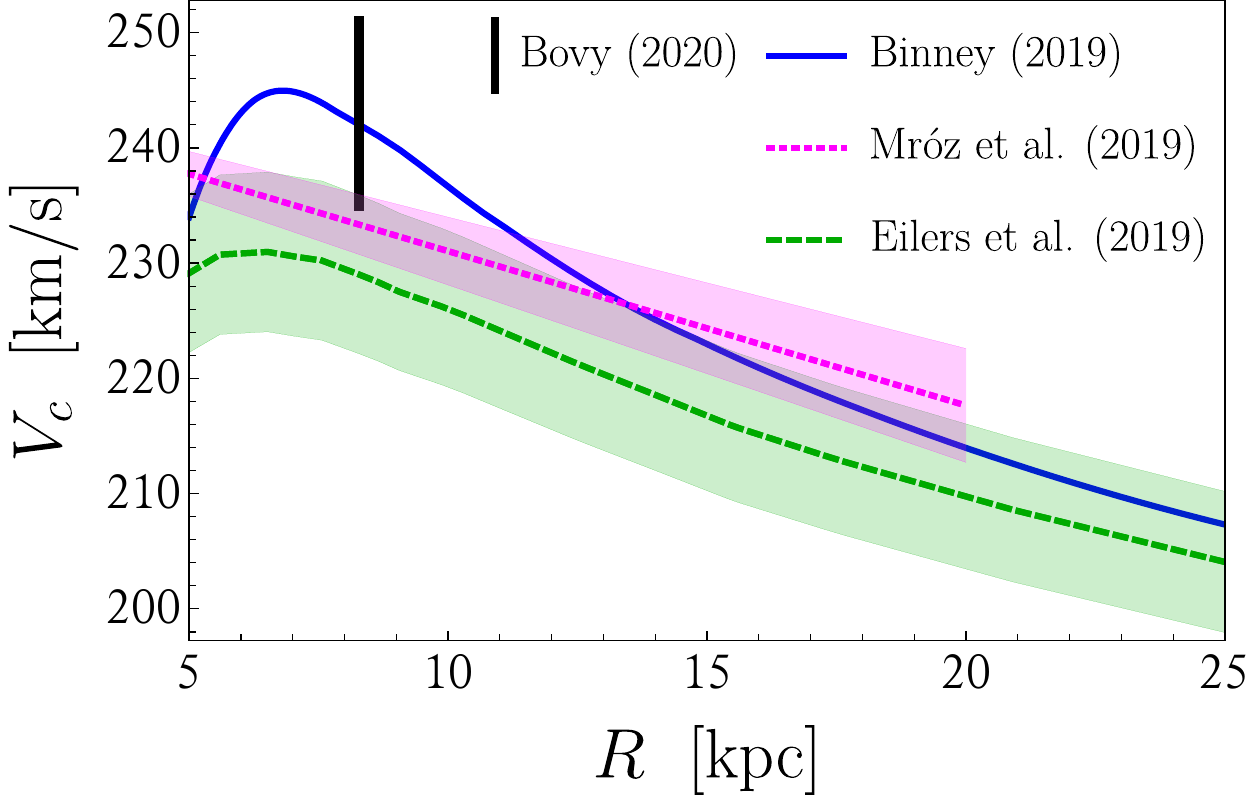}
\caption{Circular speed $V_c$ as a function of radius from the center of the MW, as determined from $f(\mathbf{J})$ 
modeling~\cite{binney2019modelling} and fits to 
measurements reported 
by \cite{eilers2019circular,mroz2019rotation}, 
as well as a value inferred from the measurement of
the solar system's acceleration~\cite{Bovy2020puremeasurement}. We include a constant 3\% error bar on the entire $V_c$ curve of Eilers et al.~\cite{eilers2019circular}, consistent with their systematic error assessment, 
and we simultaneously vary the best fit parameters of the linear model of Mr{\'o}z et al.~\cite{mroz2019rotation} by $1\sigma$. 
The spherically-aligned Jeans Anisotropic Method of \cite{Nitschai2021dynamicalmodelII} gives a result
intermediate to that of \cite{mroz2019rotation,eilers2019circular}.
\label{fig:vc_compare}}
\end{center}
\end{figure}

More insight can be gained into the distribution of DM in and around the stellar disks by examining our Galaxy's rotation curve; namely, the velocity at which a star on a circular orbit in the plane of the thin disk a given distance from the Galactic center rotates.  The GRAVITY collaboration's 
observations of the Galactic center have greatly improved 
the determination of the Sun-center 
distance~\cite{abuter2018detection}. This information, in concert with 
{\it Gaia} DR2 data, 
as well as with other optical and infrared surveys, has refined the rotation curve over the 
distance range $5 < R < 25 $ kpc, and yields 
$V_c = 229.0\pm 0.2 \>\rm km\> s^{-1}$, with a systematic uncertainty of ~3\%~\cite{eilers2019circular},
at the distance of the Sun to the Galactic Center
of $R_0=8.122 \pm 0.0.031$ kpc~\cite{abuter2018detection}.
Recent improvement in the 
treatment of optical aberrations have improved
the agreement between their earlier results~\cite{abuter2018detection,abuter2019geometric}
to yield $R_0=8.275\pm0.034$ kpc~\cite{Gravity2021improvedsundistance}, where we have 
combined statistical and systematic errors in 
quadrature throughout.
For reference, various other recent 
measurements and fits at $R_0$ give 
$V_c=242 \>\rm km\> s^{-1}$ \cite{binney2019modelling},
$V_c=233.6 \pm 2.6\>\rm km\> s^{-1}$ \cite{mroz2019rotation},
and $V_c=243\pm8  \>\rm km\> s^{-1}$ \cite{Bovy2020puremeasurement}.

We compare in Fig.~\ref{fig:vc_compare} the circular velocity curves obtained from the DF analysis based on 
$f(\mathbf{J})$ modeling and fits to data from the CGS survey,
which also yields the comparison with RAVE data shown
in Fig.~\ref{fig:cfRAVE}~\cite{binney2019modelling}, with
direct
determinations from observational data. In 
particular, we compare with analyses using observations of 
red-clump giants with {\it Gaia} DR2 and APOGEE~\cite{eilers2019circular}, which jointly fits these data to the parameters of an NFW DM halo under the assumption of axial symmetry,
and observations of nearby Cepheid variable stars~\cite{mroz2019rotation}.
We also compare to a direct measurement of 
the local acceleration of the solar 
system using the apparent proper motion of quasars
from {\it Gaia} Early Data Release 3 (EDR3) 
data~\cite{Bovy2020puremeasurement}, which 
yields fundamental Galactic parameters at the 
Sun's location. 
For the results of \cite{eilers2019circular}, we apply a 3\% error bar at all radii. For the result of \cite{mroz2019rotation} we depict their linear model including a prior on the Sun-center distance from \cite{abuter2019geometric}, simultaneously varying within $1\sigma$ the local value of $V_c$, the value of $R_0$, and the value of $dV_c/dR$. 
It is worth noting that the slopes of the two observational
analyses, over the region that they compare, 
are in good agreement with each other, yielding
$-1.7\pm 0.1 \pm 0.46\, {\rm km}\,{\rm s}^{-1}{\rm kpc}^{-1}$~\cite{eilers2019circular} and 
$-1.41\pm 0.11\, {\rm km}\, {\rm s}^{-1}{\rm kpc}^{-1}$~\cite{mroz2019rotation}. 
Each of these assessments is at odds with
older results that suggested the rotation curve
would be flat, see, e.g., \cite{reid2014trigonometric}; 
the local acceleration 
analysis of \cite{Bovy2020puremeasurement}
finds a comparable value to the most recent results, 
but with a much larger error, 
$-2 \pm 9 \, {\rm km}\,{\rm s}^{-1}{\rm kpc}^{-1}$~\cite{Bovy2020puremeasurement}.

An accurate rotation curve is important to constraining the amount and distribution of DM at these same radii near the Sun and in the outer parts of the disk. 
It is also pertinent to 
the assessment of other parameters, such as the
pattern speed of the Galactic bar.
For example, using the assessment of the Outer Lindblad Resonance (OLR) location from a distinct {\it Gaia} DR2 data sample~\cite{HGY20}, 
the rotation curve information determines the 
bar pattern speed~\cite{HInkel2020axialpatternspeed}. It is worth comparing the pattern speeds that result from different rotation curve information. The rotation curve 
result to which we have referred 
is much more precise than older 
studies~\cite{eilers2019circular}; for reference, 
we note earlier work which 
also employs HI data~\cite{huang2016MWrotationcurve},
making it quite distinct. 
Here, although 
$v_{c} = 240 \pm 6\>\rm km\> s^{-1}$~\cite{huang2016MWrotationcurve} is a bit bigger, 
the determined local radial derivative is also more negative, 
so that although the Eilers et al. result~\cite{eilers2019circular} gives
$49.3 \pm 2.2 \>\rm km\> s^{-1} kpc^{-1}$ for the 
pattern speed~\cite{HInkel2020axialpatternspeed}, the central value of the Huang et al. result~\cite{huang2016MWrotationcurve} evaluates 
to $50.7 \>\rm km \>s^{-1}\> kpc^{-1}$, within $1\sigma$
of the more precise determination. Thus 
reasonable consistency between the different rotation
curve assessments exists. 

The slope of the Galactic potential at a given radius translates directly to an estimate of the circular speed of the disk at that radius and thus, by inverting the relation, the MW's rotation curve is a sensitive probe of changes in the Galactic potential and ultimately of the underlying mass distribution.  Recent estimates of the rotation curve from \cite{eilers2019circular} can be closely compared with other data-based and theoretically driven estimates \cite{binney2019modelling} to challenge models containing such components as dark disks \cite{kramer2016darkdisks}.
We shall have more to say about the implications of the interpretation of the rotation curve for our understanding of the local density of dark matter in Sec.~\ref{sec:dm_phasespace}.

\subsection{Environment \label{sec:MWenvir}}

Within the local group of our MW, the LMC/SMC and Andromeda, many authors have remarked on a plane of satellites which appears to defy random infall of gas and dwarf galaxy satellites over cosmic time \cite{libeskind2015planeofsatellites}.  One explanation for this is that the alignment is guided by a DM filament, a component of a large scale structure
\cite{libeskind2015filaments}.  Structures on the largest scales ($> 50 \rm Mpc$)  of the so-called cosmic web consist of filaments of DM extending between large vertices of DM (where baryon-rich galaxy clusters collect). Individual field galaxies align along these filaments with a prolate dark halo configuration.  All these early simulations, however, assumed axisymmetry of the halo and alignment between the stellar disk axes and the dark halo axes.  Whether of not the existence of these coherent planes of satellites --- some of which also appear to show kinematic coherence --- are fully compatible with the predictions of CDM theory that predicts a more random or thermal build up of structure from smaller to larger scales remains uncertain \cite{mueller2021planeofsatellitesandcdm}.

On even larger scales, 
Galaxy clusters do appear to have common extended intra-cluster DM envelopes punctuated by sharp spikes of DM around each large cluster member galaxy, but the models are subject to some degeneracies \cite{Natarajan2017ClusterDarkMatterMap}.

\section{Probing the Milky Way at the Small-Scale Frontier}
\label{sec:smallscaleprobes}

Dark matter astrophysics has long been concerned with observational probes of the CDM paradigm. Studies of largest-scale structures are inevitably cosmological in nature. These cosmic tests range from the cosmic microwave background (CMB) radiation, which encompasses the entire visible Universe, to the baryon acoustic oscillation scale, which is visible in both the CMB and the distribution of galaxies from surveys of the low-redshift universe, to the behavior of galaxies in the immediate vicinity of the MW. Each of these is remarkably consistent with the CDM paradigm, which predicts that DM is organized in gravitationally self-bound structures termed ``halos''. The CDM prediction is that such structures are formed ``hierarchically'': small structures separate from the Hubble flow and collapse first, and larger halos are formed from successive collisions and mergers of these small initial objects. Partially merged subhalos that are at least partially self-gravitating may persist within a larger host halo for many orbits before being tidally disrupted and joining the larger virial distribution.

Notwithstanding the success of the CDM paradigm at cosmic scales, 
concerns at galactic and sub-galactic scales have existed for decades. These are commonly discussed as belonging to one of four particular problems: (i) the missing satellites problem, (ii) the too big to fail problem, (iii) the baryonic Tully-Fisher relation, and (iv) the core-cusp controversy. We will address the first of these in the context of the MW, and refer to other recent reviews on the remaining topics \cite{bullock2017challengeLCDM, buckley2018gravitational}.

Probes of DM halos currently span the approximate range $10^8-10^{15}\,\msun$. 
In this section, we will give an inventory of CDM structures within the MW, and briefly overview how they are affected by and arranged within the MW's gravitational potential. We will order this section roughly by 
size, beginning with the largest, most obviously apparent substructures of the MW with masses $\sim \mathcal O(10^{11})\, \msun$ and proceeding to lower masses and less prominent subsystems of mass $\lesssim \mathcal O(10^8)\, \msun$.

\subsection{The Large Magellanic Cloud}
\label{subsec:obs_LMC}

Our understanding of the LMC has evolved substantially in recent years \cite{vandermarel2006stars, kallivayalil2006lmc, besla2007lmcsmcfirstpassage, kallivayalil2013third, patel2016orbits}. Using data from the {\it Gaia} satellite \cite{brown2018gaia}, we have also gained significant insight into the influence 
of the LMC on the MW: the reflex motion of the MW in response to the gravitational influence of the LMC has now been detected \cite{laporte2018influence, erkal2019LMCandorphan, petersen2020MWreflexLMC, erkal2020MWsloshLMC}. Detailed studies of the interactions of the LMC with the MW halo and its satellites suggest that the LMC is relatively heavy, with a mass $M_{\rm LMC} \gtrsim 0.1 \times M_{\rm MW}$ \cite{besla2015magcloudmasses, erkal2019LMCandorphan, vasiliev2021heavyLMC, erkal2020MWsloshLMC, erkal2020MWmassLMC}. This is compatible with the results of a comparison of a census of LMC satellite members to $N$-body simulations \cite{kallivayalil2018magsatsgaia, patel2020orbitsmagsatsgaia, erkal2020LMCcensus}.

Of direct concern for improving our understanding of the MW is the growing appreciation that the LMC can directly and significantly influence the local phase space of DM particles \cite{besla2019highest}. This is possible due to both the high relative velocity of the centers of mass of the MW and the LMC as well as the reflex motion of particles in the solar neighborhood to the influence of the LMC, 
as can be probed through studies of 
axial symmetry breaking in the MW~\cite{GHY20,HGY20}.
Thus, our understanding of the MW is now sufficiently precise that further refining this understanding necessarily requires accounting for our largest satellites.

\subsection{Milky Way Satellite Galaxies}
\label{subsec:obs_satellite_galaxies}

\begin{figure}[t]
\begin{center}
\includegraphics[width=0.85\textwidth]{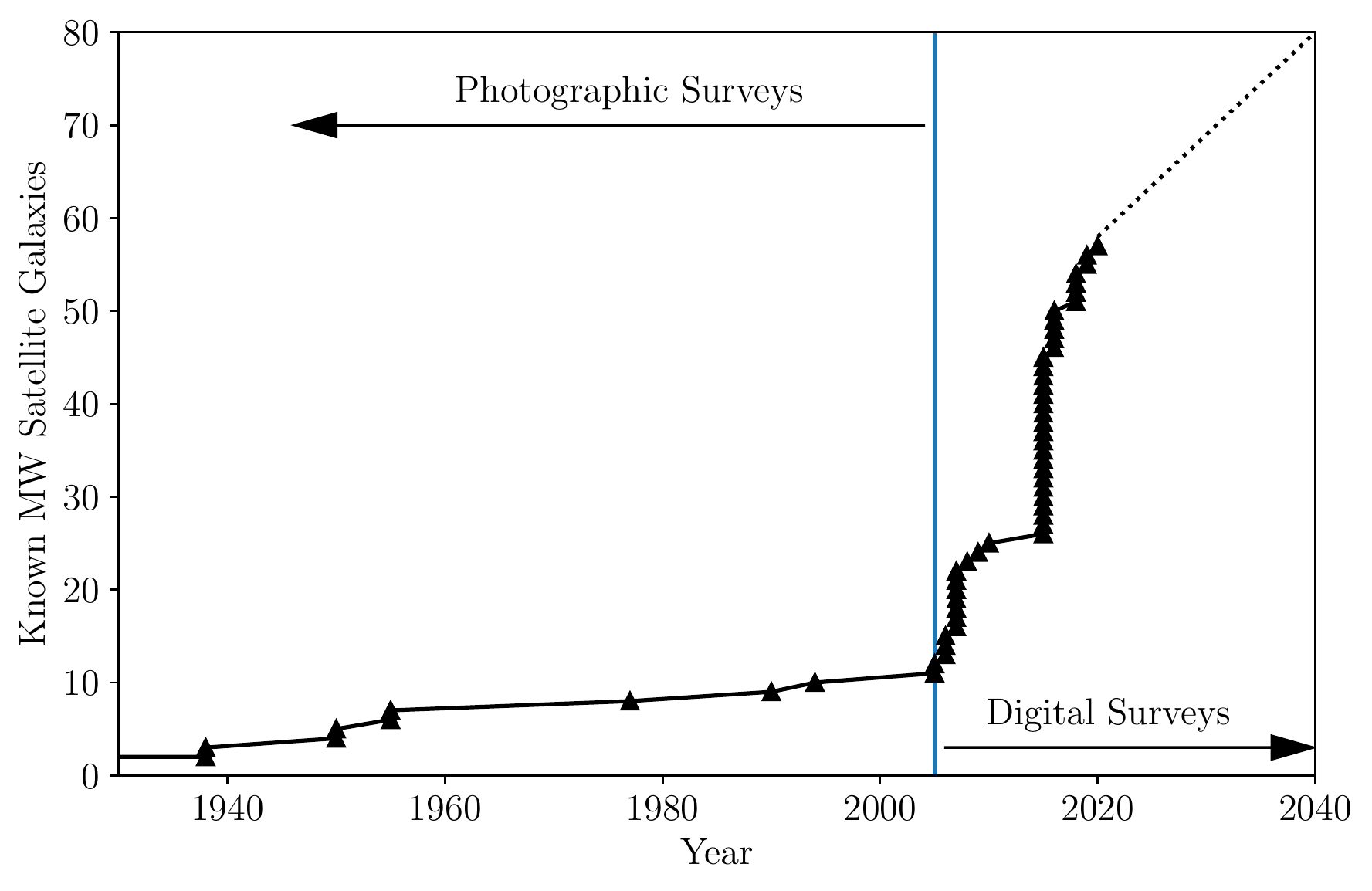} 
\caption{Known MW satellite galaxies as a function of time. The dotted line is a schematic projection for future surveys. Data courtesy of Joshua D.~Simon. \label{fig:simon_review_1}}
\end{center}
\end{figure}

The possibility that distant clusters of visible stars were island universes with their own history and identity independent of our immediate environment was originally suggested by Kant \cite{kant1755island, whitrow1967kant}. Almost a century ago, this hypothesis (and, by extension, the Copernican principle) was confirmed by Hubble \cite{hubble1922nebulae}. Some of these other galaxies of stars are now understood to be gravitationally bound to the MW. These smaller galaxies that are gravitationally bound to the MW are known as satellite galaxies. 

The essential foundation of any substantive understanding of these satellite galaxies is their enumeration. Small self-gravitating astronomical structures are classified in a variety of ways. One useful binary classifier of different such systems is whether or not their dynamics are determined by a significant DM component. Those with a large DM density and evidence of multiple epochs of star formation are commonly referred to as dwarf galaxies.
In the early years of this century, as numerical simulations improved, an apparent tension between the number of observed and predicted dwarf galaxies was noted by a number of authors (see {\it e.g.}~\cite{bullock2017challengeLCDM}). However, it is now believed that there is no missing satellites problem \cite{kim2018nomissingsatellites, bose2020abundancehistory}. The problem has been resolved by a number of factors. Chief among these have been recent discoveries of satellite galaxies around the MW \cite{drlica-wagner2015eightDESultrafaintdwarfs, bechtol2015eightDESdwarfs, koposov2015nineultrafaints}. Currently, almost 60 dwarf galaxies are known, a number that has changed by almost an order of magnitude in the past decade. For a recent review on the status of known satellites of the MW, see \cite{simon2019dwarfreview}, whose Fig.~1 we 
capture in Fig.~\ref{fig:simon_review_1}. This reveals the dramatic increase in known satellite galaxies as a function of time and technology. Predictions for the star-formation efficiency of these same small host environments have also been updated and improved \cite{tollerud2008luminositybias, hargis2014distributiondwarfs, jethwa2018halomasses, newton2018satellitequench, kim2018nomissingsatellites}, which has also helped to ameliorate the discrepancy.
Upcoming facilities such as the Rubin Observatory will extend the sensitivity to faint satellites even further \cite{drlica-Wagner2019lsstDM}, which will extend our ability to test the CDM paradigm to the frontier of fainter and smaller objects.

In a cosmic context, the number of small satellite galaxies probes the high-wavenumber tail of the matter power spectrum \cite{silk1968damping, peebles1970adiabatic, gott1975galaxyformation, bond1980lss, bond1983collisionlessdamping, bond1984cmbhighk, bond1987cmbstatistics}, where we note 
\cite{buckley2018gravitational, bullock2017challengeLCDM} for reviews. 
The power spectrum of matter at these large wavenumbers (corresponding to cosmologically small physical sizes) is in turn determined by the details of the DM of the universe. We make the connection between DM microphysics and galactic dynamics more explicit in Sec.~\ref{sec:dmcand}.

\subsection{Stellar Streams} 
\label{subsec:obs_stellar_streams}

Stellar streams are extended distributions of stars with similar kinematics and chemical composition. They are presumably formed from disruption of globular clusters and dwarf galaxies as those objects fall into the gravitational potential well of and virialize with their host halo. See \cite{newberg2016streamsbook} for a recent and comprehensive historical and methodological overview of stream finding in and around the MW.

\begin{figure}[t]
\begin{center}
\includegraphics[width=0.5\textwidth]{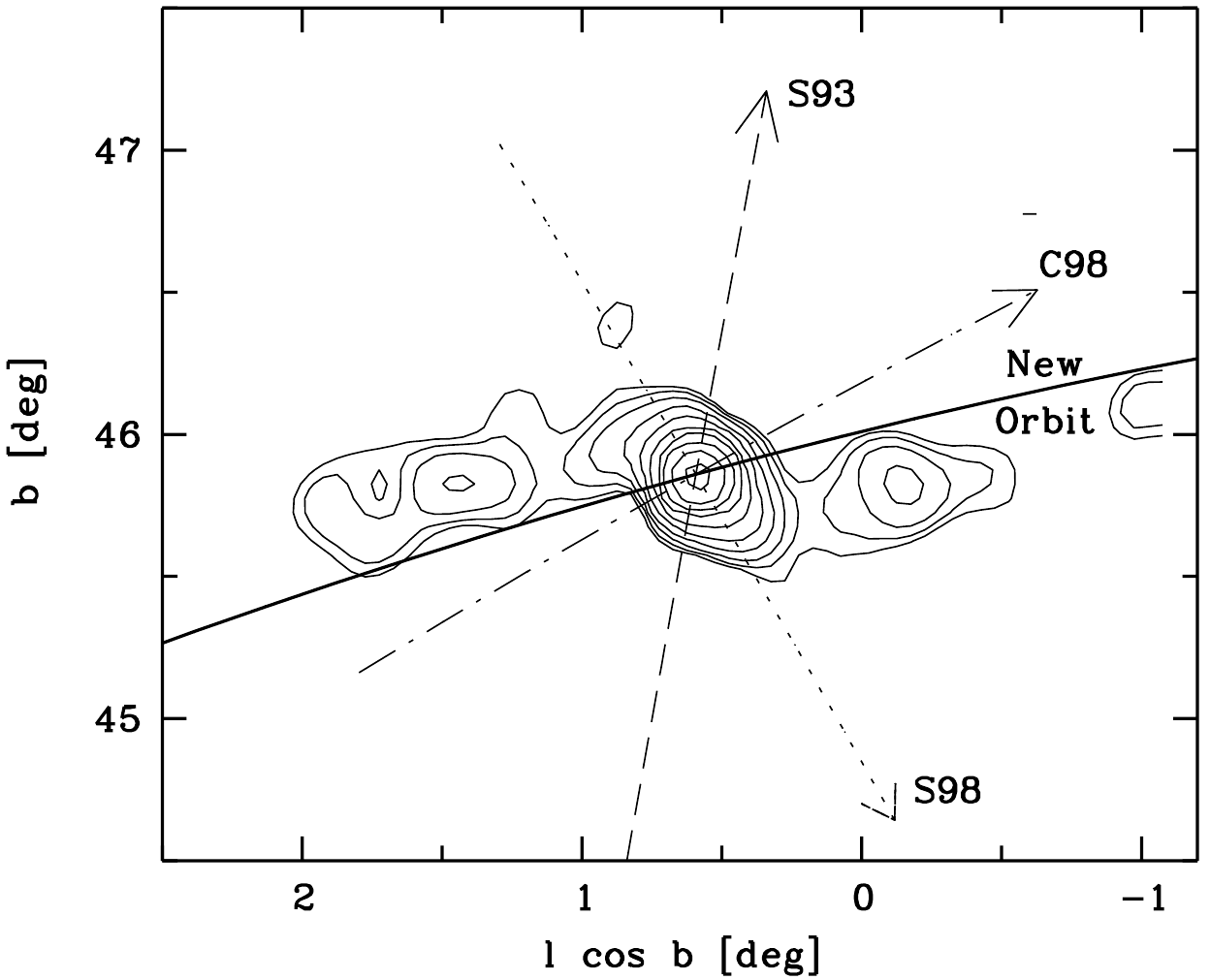}
\caption{The initial detection of tidal tails of the Pal5 stream. From \cite{odenkirchen2001pal5taildiscovery}$\copyright$AAS. Reproduced with permission. 
\label{fig:pal5_tail}}
\end{center}
\end{figure}

The discovery of tidal tails around Palomar-5 (Pal5), shown in Fig.~\ref{fig:pal5_tail} \cite{odenkirchen2001pal5taildiscovery}, was a landmark occasion for the study of stellar streams. The extended tidal tails of streams are in fact their essential distinguishing feature for tracing the interesting features of their host halo, including its assembly history \cite{johnston1996fossil, johnston1998streamshistory, dubinski1999streamsdarkpotential, bullock2005galaxyformationwithstreams, bell2008accretion} and its steady-state structure \cite{johnston1999streamspotential, johnston2005two, koposov2010mwpotentialfromgd1, law2010sagittarius, sanders2013astreampotential, sanders2013bstreampotential, bovy2014actionangle, sanderson2015actionangle,bonaca2014milkywaymass, gibbons2014skinnyMWsgr, price-whelan2014potential, bowden2015gd1stream, kupper2015streams, bovy2016streamsPal5GD1, Kamdar2021streamslifetime}. The tidal tails trace the trajectory of the disrupted progenitor along its original orbit, while in transverse directions the stream remains kinematically cold (unless perturbed by a massive satellite \cite{erkal2019LMCandorphan}). The phase space of the stream remains coherent, and the original characteristics of the progenitor can be inferred \cite{johnston1996fossil} as long as the stream is on a regular orbit \cite{knebe2005streamsmapping, price-whelan2016chaosstreams}.

\begin{figure}[p]
\begin{center}
\includegraphics[width=\textwidth]{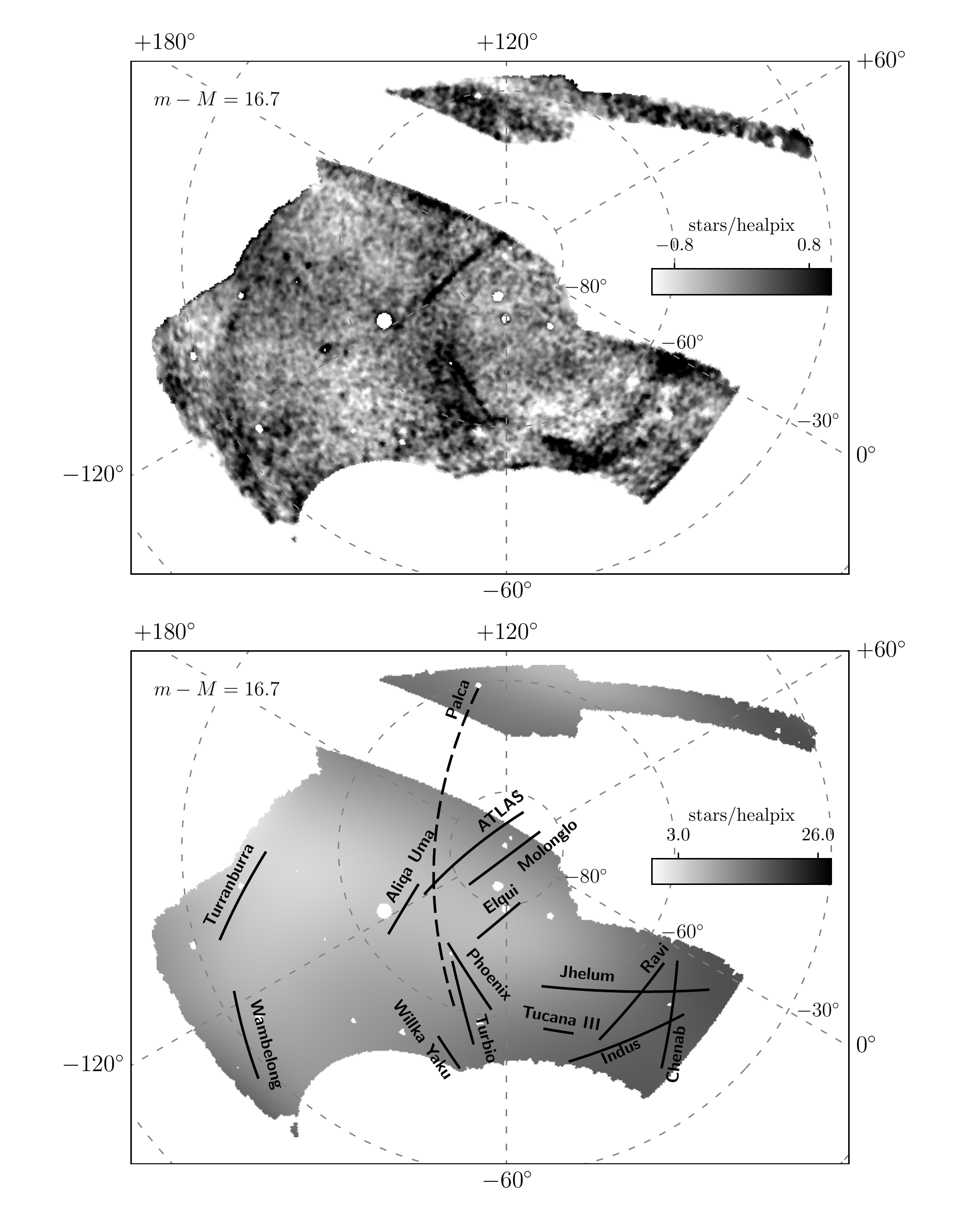}
\caption{Stellar streams of the MW as observed in the DES observational footprint. From 
\cite{shipp2018streamsfromdes}$\copyright$AAS. Reproduced with permission. 
\label{fig:des_streams}}
\end{center}
\end{figure}

There were 26 known streams circa 2016 \cite{grillmair2016chapterstreamsbook}, with 11 more discovered in DES data \cite{shipp2018streamsfromdes}. We show a recent collection of streams observed in the DES footprint in Fig.~\ref{fig:des_streams} \cite{shipp2018streamsfromdes}. At least three more streams have been discovered so far in {\it Gaia} data \cite{meingast2019pisceseridanusstream, ibata2019fimbithulstream, necib2020evidence}. Spectroscopic follow-up from the $S^5$ survey \cite{li2019s5survey} will further characterize these populations and provide additional clues as to their evolutionary history, as will complementary future surveys \cite{Martinez-Delgado2021streamsurvey}.

Streams carry with them a historical record of their gravitational interactions. The scars of this history will be particularly visible if the stream has been perturbed along its kinematically cold transverse directions by the influence of a massive perturber. Gaps and lumps transverse to the leading and trailing arms are entirely absent in a smooth background potential, but are induced by overdensities within the halo. For this reason, streams provide 
particular 
sensitivity to otherwise dark satellite members of their host halo \cite{johnston2002lumpy, ibata2002subhalosstreams, yoon2011satellitesfromstreams, carlberg2012subhalosstreams, erkal2015gapssubhalos, erkal2016gd1betterthanpal5}. Such substructures are predicted by the hierarchical assembly mechanism of DM halo formation, as discussed in Sec.~\ref{subsec:dmcand_hierarchical_structure}.
In Sec.~\ref{sec:dmcand} we shall have much more to say about how different theories of DM lead to different predictions for the structure and frequency of these perturbations.

\subsection{Patterns in Milky Way Field Stars}
\label{subsec:obs_patterns_in_stars}

All of astronomy hinges on the search for regularity in observations of different subsets of the Universe. Yet the detection of anomalies in the known positions of stars due to the gravitational influence of known objects is only one hundred years old. Finding {\it statistical} correlations in the 
positions, velocities, or accelerations of nearby stars are data- and computation-intensive challenges and thus particularly modern versions of this very old type of search.

Photometric gravitational lensing is a prominent
example of this type of statistical observational effort. Lensing detections of DM are thus far largely confined to extragalactic contexts, beyond the purview of this study, but we give a brief summary of aspects of the field here. We then extend some of the concepts from photometric gravitational lensing to related families of searches.

\begin{figure}[t]
\begin{center}
\includegraphics[width=\textwidth]{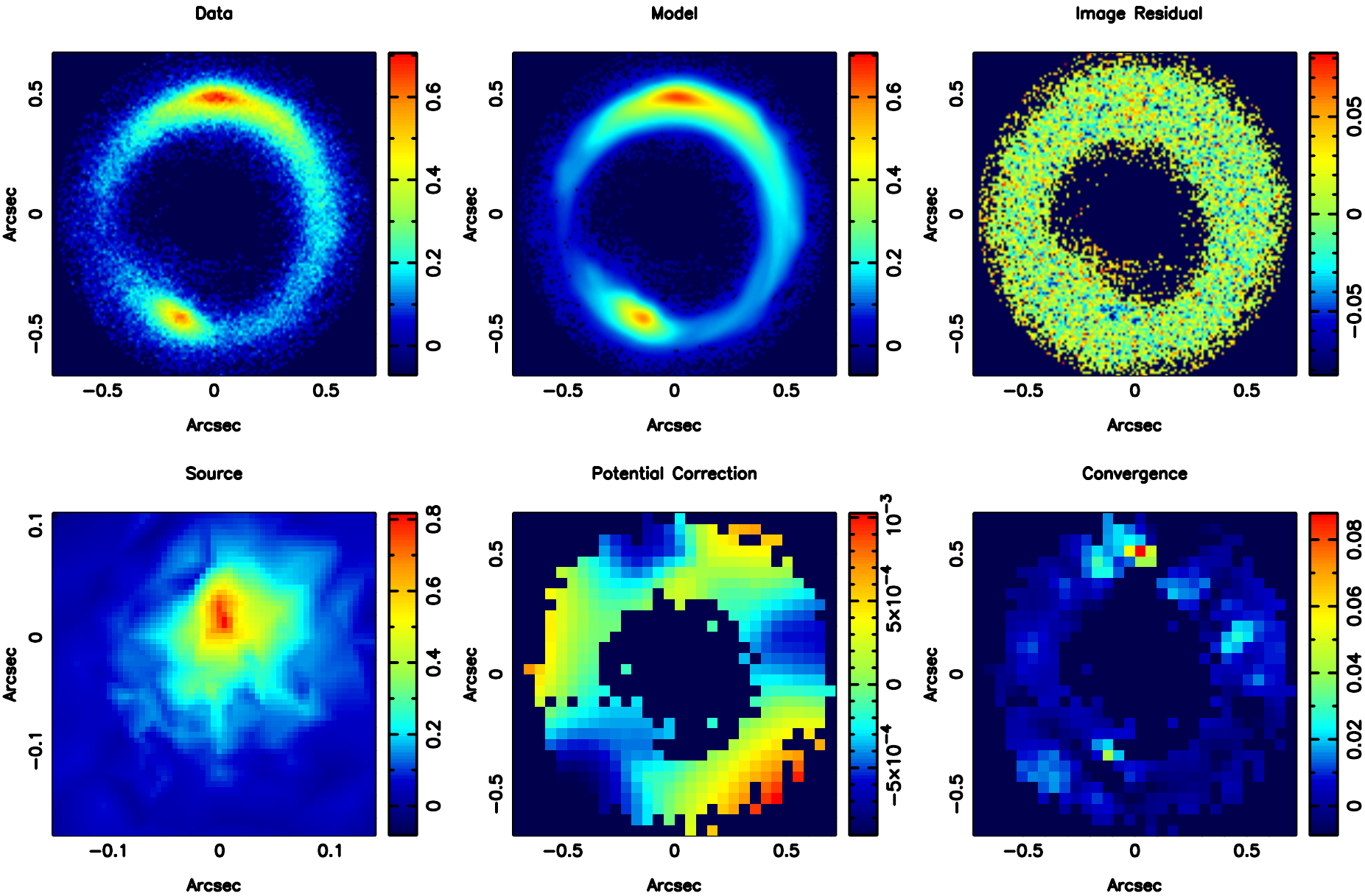}
\caption{A detection of dark matter substructure in a strongly gravitationally lensed image. Substructure is revealed in the lower right-hand panel. From \cite{vegetti2012subdetect},  reprinted by
permission from Springer Nature.
\label{fig:vegetti_lens}}
\end{center}
\end{figure}

First, we define photometric gravitational lensing to be the study of shear distortions in single images of background light from stars, quasars, and galaxies. This requires large photometric data sets and precise images. The lensing distortions 
are termed ``strong'' or ``weak'', depending on how severe the lensing is and over what angular range and across how many distant sources the lensing signal is correlated. Strong lensing is exemplified by the Einstein ring formed when a source, lens, and observer are exactly aligned. It is also possible to detect multiple images of the same object without seeing the extended arc of the Einstein ring. Anomalies in strongly lensed images can provide evidence of substructure in the lens \cite{mao1998lenssubstructure, hezaveh2016powerstronglensing}. Such studies have a number of exciting successes, as shown in Fig.~\ref{fig:vegetti_lens}, but are so far limited to cosmological distances \cite{vegetti2010subdetect, vegetti2012subdetect}. (Recently, it has been suggested that such lensing events are more numerous than expected in CDM \cite{meneghetti2020excesslenses}.)
Weak lensing in contrast is detected by statistical methods on large image sets \cite{kaiser1995weaklensing, kochanek2006lensingreview}. Transient effects like flux ratio anomalies and apparent magnification, which rely on relatively high cadence photometric observations, are 
termed ``weaker than weak'' lensing, because they do not lead to a change in the apparent position of the object, and are sometimes referred to as ``microlensing'' \cite{paczynski1986microlensingMWhalo, kochanek2006lensingreview}. (The even more subtle effect of a phase lag on the wavefront of a gamma ray burst, the so-called femtolensing, is unlikely to be observable after accounting for finite-source-size effects \cite{katz2018femtorevisited}.) Microlensing studies offer opportunities for measuring populations of particularly dense objects like planets and the very dense DM halos predicted in non-CDM cosmologies.
Gravitational lenses have been detected from radio to gamma-ray energies  \cite{schneider1992grlensbook, refregier1997xraylens, bartelmann2010lensreview, kneib2011lensreview, cheung2014gammaraylens}, using techniques spanning all of these methods. Due to the large, fluffy nature of halos in the CDM picture, a complete understanding of the building blocks of the MW halo is not likely to be completed in this way should the CDM hypothesis be correct. At this time, however, we can say that lensing provides evidence that is entirely compatible with the CDM picture of the MW halo. We defer to Sec.~\ref{sec:dmcand} the implications of lensing analyses for DM particle candidates in CDM and beyond.

Photometric lensing is but one technique to discover otherwise-invisible small-scale DM substructure. With the advent of extremely precise position (and thus parallax and proper motion) information from the {\it Gaia} satellite \cite{brown2016gaia}, studies of patterns in MW stars are no longer necessarily confined to photometric anomalies. For instance, these precise velocity data can be turned directly into a map of the overall halo potential \cite{bovy2010snapshot, green2020phasespace}. Alternately, astrometric or velocity-based lensing has recently been proposed. This requires studying perturbations to the precise {\it Gaia} astrometric solution. As with the photometric case, these perturbations can arise in the time-domain \cite{van2018halometry, mondino2020astrometricWL, vattis2020deepastrometry} or may be revealed through higher-point correlations across a broader statistical sample \cite{mishra2020power}. The hope is that these methods of astrometry can provide a census of dark compact objects. Similar to the case of photometric lensing, astrometric lensing is most promising for very dense subhalos.

In addition to measures of position- and velocity-space distortions, another frontier in the study of the halo of the MW 
is acceleration measurements \cite{Quercellini2008galacticgrav, ravi2018accelerometry,Silverwood2019stellaracc, Chakrabarti2020galacceleration,phillips2020pulsars,Chakrabarti2021massdenspulsaracc,buschmann2021potentialangular}. Following the trend identified in the case of velocity measurements, acceleration measurements can probe the local acceleration due to the overall DM halo, and thus finely map the local gravitational potential. Alternately, pulsar timing arrays that look for correlated phase lags in pulsar signals will be another important tool for understanding the substructure of the MW halo \cite{seto2007ptaPBH, baghram2011ptaprospects, dror2019ptasubhalos, ramani2020pta, Lee2021ptabayes}. These probes provide very exciting prospects for the next generation of studies of the immediate DM halo.

\section{Probes of Dark Matter Candidates via Milky Way Observations}
\label{sec:dmcand}

In general, the number of small DM halos is determined by the {\it primordial} distribution of DM at all distance scales. One convenient summary of the nature of the DM (but, as discussed below, not the unique one) is to characterize it as ``hot'' or ``cold'' depending on a single parameter: the {\it free-streaming scale} in the epochs when cosmic structures begin to form. Below this distance scale (or, equivalently, at larger wavenumber), density perturbations do not grow. A hot DM candidate is generally recognized as having been relativistic at the time of the formation of the density perturbations that characterize the CMB, with a characteristic free-streaming scale much larger than 10 kpc. A cold DM candidate will have a free-streaming scale much smaller than 10 kpc. (Naturally, warm DM interpolates between hot and cold DM and has a free-streaming scale of order 10 kpc.) This is often quantified by calculating the so-called DM transfer function, which is the ratio of the power spectrum of a given DM candidate to a cold DM particle with the same energy density. This transfer function will abruptly go to zero above the wavenumber of the free-streaming scale $k_{\rm fs}$. The vanishing of the transfer function implies the vanishing of the power spectrum, which in physical terms means the absence of structures with physical sizes below $\sim k_{\rm fs}^{-1}$.

Beyond the ``warmth'' of the DM, a full description of the small-scale distribution of DM requires knowledge of its power spectrum on all physical scales. However, calculating this quantity requires
detailed knowledge of 
the DM microphysics. For this reason, a precise treatment of the compatibility of DM candidates with cosmic and astrophysical observations requires a complete model of DM genesis and its interactions with itself and with the SM. 

In this section, we first give a brief overview of structure formation in CDM and non-CDM cosmologies. Then we summarize the status of several DM models that deviate from the CDM paradigm at small scales. We draw concrete conclusions about the properties of various DM particle candidates by connecting to the various observational probes discussed in Sec.~\ref{sec:smallscaleprobes}. We begin with a general discussion of structure formation in the conventional CDM paradigm and how this can be used to draw broad inferences about the nature of the DM. Then we extend our discussion by focusing on a handful of concrete models of DM-SM interactions.

\subsection{Hierarchical Structure Formation in the Milky Way and Beyond}
\label{subsec:dmcand_hierarchical_structure}

Observational diagnostics of the amount of DM in small scale structure come in many forms. At late cosmological times, these are often summarized in the halo mass function (HMF), or, in the case of a particular host galaxy such as the MW, the subhalo mass function (SHMF), which is the number of subhalos as a function of their mass, $dN_{\rm sh}/dM_{\rm sh}$. These functions are intimately related to the primordial DM density perturbations, but also encapsulate all the nonlinear gravitational, plasma, and potentially rich dark sector physics experienced by the structures after they are formed. Thus, the SHMF is a complicated function of inherent DM properties as well as of primordial cosmological information and more quotidian data like the host mass and environment.

By ``primordial'' we mean the data that set the initial boundary conditions for the era of linear structure growth. At the present time, our knowledge of the primordial characteristics of the Universe is limited to the information we can extract from the cosmic microwave background (CMB) radiation and indirect measurements of the era of Big Bang nucleosynthesis (BBN). Large-wavenumber perturbations in the CMB are relevant for structures of galactic size. By ``large-wavenumber'', we mean that these scales correspond to inverse length scales that are of order 10 kpc or shorter. These perturbations grow during the 
cosmic dark ages of the early universe 
into structures of a great range of density and a remarkable range of diversity. However, the number and the gross features of small DM halos ultimately depend on the distribution of the primordial DM density perturbations at large wavenumber \cite{bond1980lss, dodelson1993dodelsonwidrowneutrinos, viel2005wdmpower, schneider2012wdmpower, murgia2017wdmgeneral}.

The DM density perturbations at high wavenumber are in turn determined by the precise phase space distribution of the DM. This determines how the DM overdensities (and, eventually, the baryonic overdensities captured within) grow as a function of cosmic time \cite{boyanovsky2010smallscalepower}. The DM phase space distribution is itself determined by the DM particle properties: when and how it was produced and attained an appreciable cosmic abundance, what interactions it had after that time, and how and when it decoupled from those interactions \cite{buckley2018gravitational, nadler2020MWcensus}. For this reason, specific DM particle physics properties need to be specified in order to predict a particular distribution of DM on specific scales.

\begin{figure}[t]
\begin{center}
\includegraphics[width=0.485\textwidth]{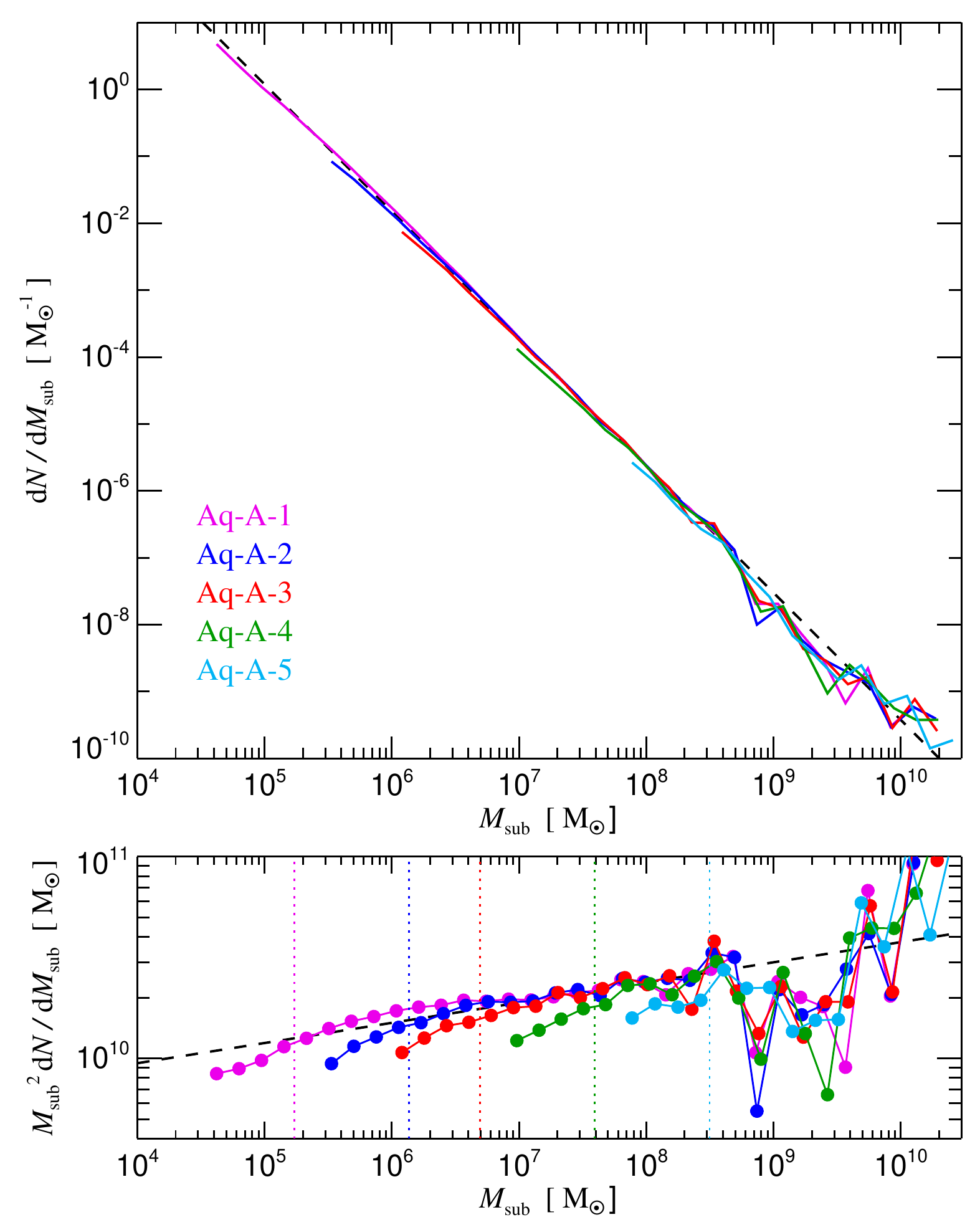}
\caption{The halo mass function in a cosmological CDM simulation. From \cite{springel2008aquarius}. \label{fig:hmf}}
\end{center}
\end{figure}

\begin{figure}[t]
\begin{center}
\includegraphics[height=2.6in]{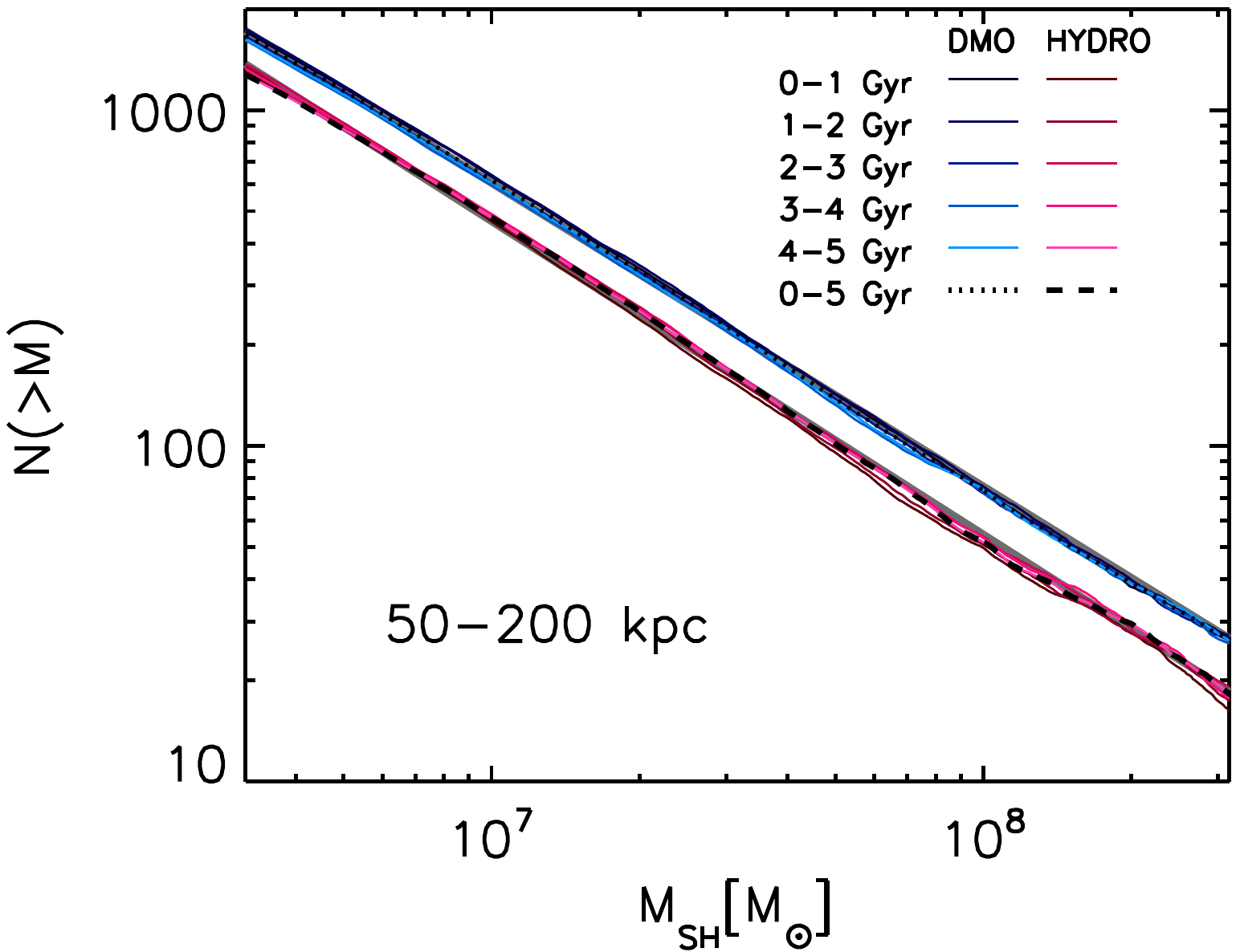}~~~~
\includegraphics[height=2.6in]{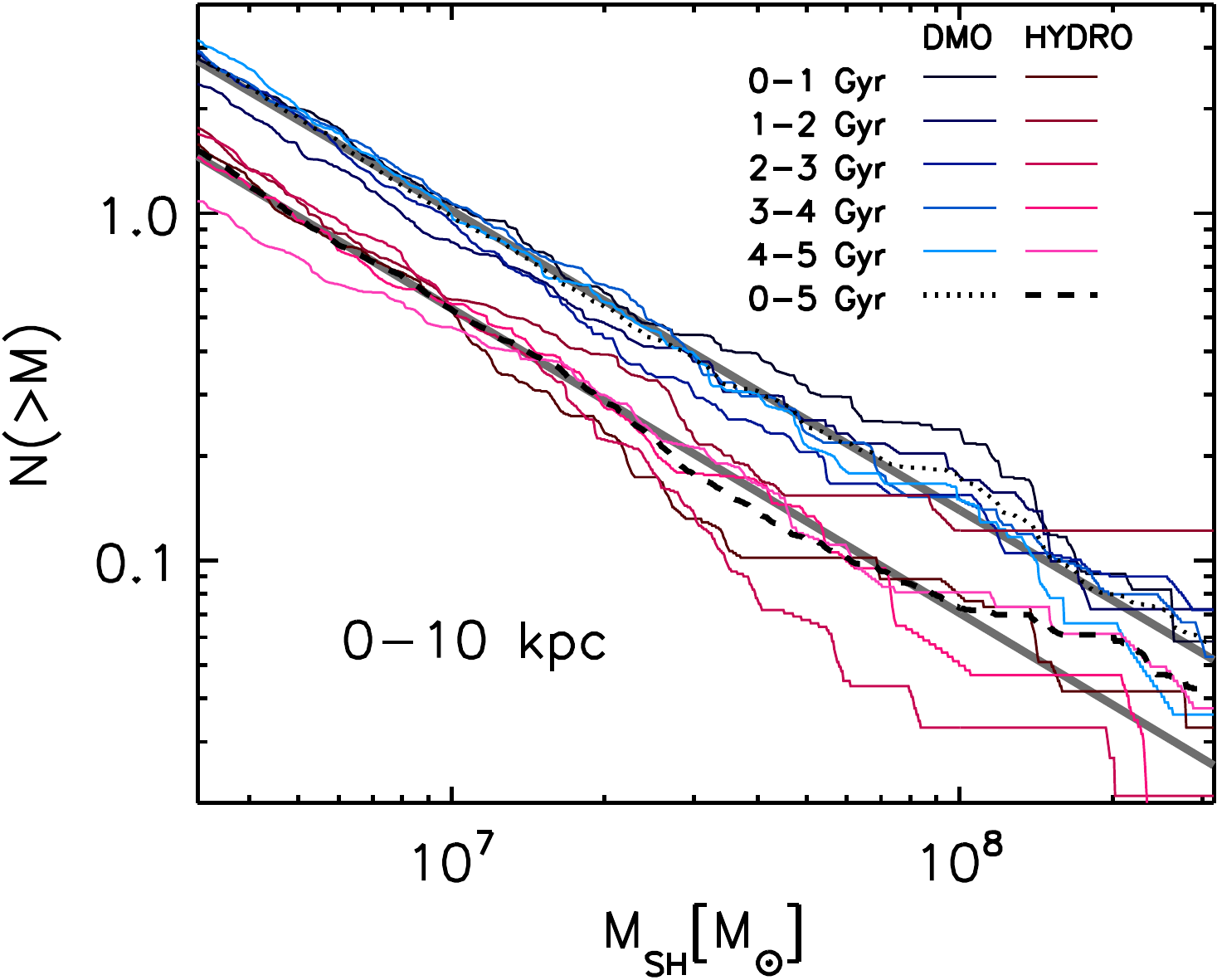}
\caption{The subhalo mass function in a cosmological CDM simulation. {\bf Left:} subhalos with large pericenter from their host halo. {\bf Right:} subhalos with small pericenter from their host halo. There is an increased scatter when tidal interactions are more important. From \cite{sawala2017mwsubstructure}. \label{fig:shmf}}
\end{center}
\end{figure}

The baseline model adopted for most simulations of cosmic structure evolution is that of 
cold and collisionless dark matter, the CDM paradigm. Roughly speaking, CDM is expected to have a hierarchical distribution of structures, wherein small structures form first and merge to form large structures at lower redshift. Cosmological simulations reveal a HMF 
$dN_{\rm h}/dM_{\rm h} \propto M_{\rm h}^{-\alpha}$ for some positive constant $\alpha \lesssim 2$ \cite{springel2008aquarius}; this is illustrated in Fig.~\ref{fig:hmf}, where the dashed line compares simulation results against the model $dN_{\rm h}/dM_{\rm h} \propto M_{\rm h}^{-1.9}$. This scaling with halo mass appears to be largely robust against tidal effects from interactions with the host halo, aside from an increased scatter in the number of halos \cite{sawala2017mwsubstructure}, which we provide an example of in Fig.~\ref{fig:shmf}. As discussed in more detail in this section, any deviation from the cold, collisionless, low-density, weakly coupled paradigm will result in a departure from (and generally a suppression of \cite{bose2017wdmSHMF}) the HMF expected within CDM. Often, these departures will have characteristic features that reveal further details of the DM. In this way, an understanding of the SHMF of the MW is a sensitive method for investigating the essential characteristics of the DM.

\begin{figure}[t!]
\begin{center}
\includegraphics[width=\textwidth]{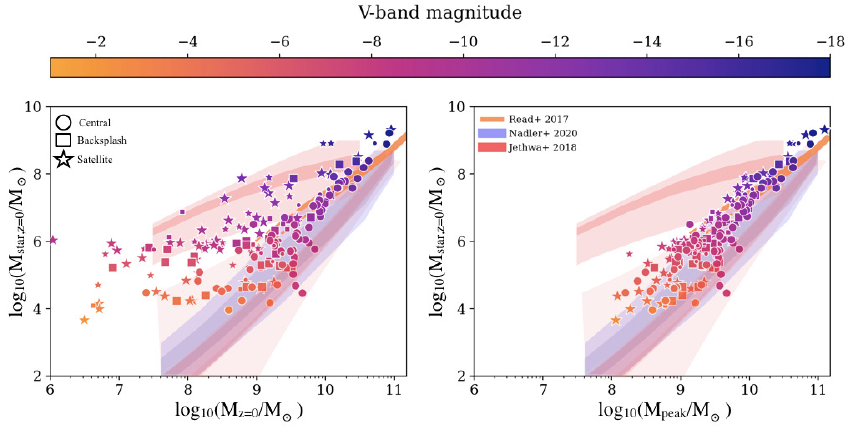}
\caption{The stellar-mass/halo-mass relation as a function of subhalo mass at {\bf (left)} $z=0$ and {\bf(right)} at peak mass before infall. Subhalos with peak mass before infall of $\gtrsim 10^8\,\msun$ efficiently form stars. Shaded bands are theoretical predictions without tidal mass loss. Points are simulated galaxies. From \cite{munshi2021UFDscatter}. \label{fig:smhm}}
\end{center}
\end{figure}

A direct probe of the SHMF for halo masses above $\sim 10^8\, \msun$ comes from counting dwarf galaxies within the MW. Above this characteristic mass scale, halos are expected to become efficient at encouraging star formation \cite{hoeft2006dwarfgalaxiesSFR, okamoto2008galaxySFRUV, sawala2015lowmassSFR, munshi2019UFDsimSFH, munshi2021UFDscatter}. Critically, it is the {\it peak} halo mass, rather than the present-day mass, which may have decreased due to disruption effects during infall, that controls this efficiency \cite{munshi2019UFDsimSFH, munshi2021UFDscatter}. 
The efficiency of star formation is the key parameter making such halos detectable, so we illustrate the correlation between the stellar mass and the halo mass  in Fig.~\ref{fig:smhm}. We conclude that halos are amenable to searches for visible self-gravitating structures if they have masses $M \gtrsim 10^8 \,\msun$ or a characteristic velocity dispersion greater than $ \sigma_v \sim \sqrt{2GM/R} \sim 10$ km/s, assuming that $R\sim 10$ kpc defines the size of small subhalos. (Smaller, denser, structures that are dominated by their baryonic gravitational potential such as globular clusters, nuclear star clusters, and giant molecular clouds are therefore not as informative as to the nature of the DM.) As discussed above in Sec.~\ref{subsec:obs_satellite_galaxies}, our census of satellite galaxies of the MW has undergone a substantial 
growth in the recent past. 
The consensus is now that the MW resides in a DM halo with a typical population of subhalos \cite{kim2018nomissingsatellites}. This constrains major deviations from the 
CDM paradigm.

Because of these recent advances, the frontier of the search for the DM is now in the search for dark structures below the characteristic mass scale of star formation, $M \lesssim 10^8\, \msun$. This requires new searches that extend observations from the direct search for luminous satellites to indirect methods that are sensitive to entirely dark substructures of the MW.

One method for looking for dark substructures are studies of lensing, as discussed in Sec.~\ref{subsec:obs_patterns_in_stars}. This is a promising route for determining the mass function of cosmological CDM halos \cite{zackrisson2010lensingCDM}. The amount of anomalous flux ratios observed in samples of gravitational lenses is consistent with the amount of substructure predicted in CDM cosmologies \cite{dalal2002lensingCDMsubstructure, vegetti2010subdetect}. Direct searches for microlensing of stars in the MW and Andromeda galaxies and the Magellanic Clouds by the OGLE \cite{udalski1992ogle, udalski2015ogleIV}, MACHO \cite{alcock2001macho}, EROS \cite{tisserand2007eros}, and Subaru HSC \cite{niikura2019hsc} collaborations confirm the CDM halo picture. The low rate of lensing events observed by these collaborations are compatible with the expectation of fluffy and low density CDM subhalos, and thus are primarily used to rule out other candidates \cite{smyth2019lensingPBH, croon2020genlensingDM, croon2020lensingHSCDM}. Combining lensing with complementary information allows for even stronger constraints on all models \cite{Nadler2021unifiedanalysis}.

Searches for patterns in other measurements of MW stars provide another handle for studying the particle nature of DM. 
The prospects for a positive detection of local substructure with pulsar timing arrays depends on the DM theory. Though the possibility of finding the dense subhalos predicted in non-CDM cosmologies is promising on the timescale of decades \cite{dror2019ptasubhalos, ramani2020pta}, it is less promising in the case of fluffy CDM halos \cite{lee2020ptadetectability}. So far, other pattern-based studies mentioned in Sec.~\ref{subsec:obs_patterns_in_stars} are also not competitive with lensing probes.

\begin{figure}[t]
\begin{center}
\includegraphics[width=\textwidth]{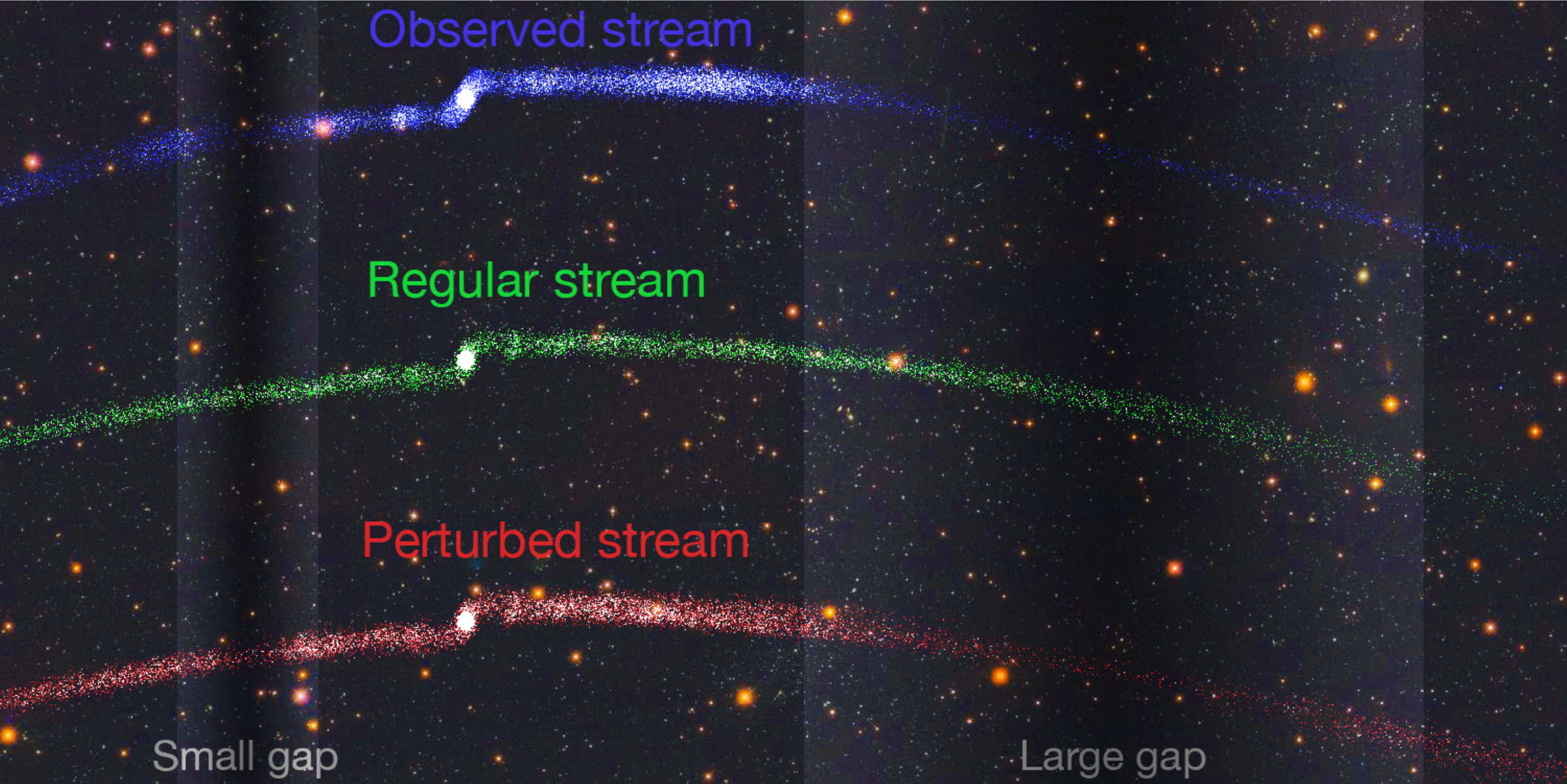}
\caption{Comparison of observation and simulations of the Pal 5 stream with and without the gravitational influence of massive satellites of the simulated host. The body of the remaining Pal 5 progenitor is the white oval, left of center in each panel.  Tidal stream stars escape the progenitor at two Roche-lobe overflow points, leading to a kink in the complete stream. From \cite{erkal2017sharperpal5}. \label{fig:sim_pal5}}
\end{center}
\end{figure}

Stellar streams, surveyed in Sec.~\ref{subsec:obs_stellar_streams}, offer another promising probe of the structure of the MW. Of particular interest is their ability to probe the nature of the smallest building blocks of the MW dark matter halo \cite{choi2007satellitesandstreams, yoon2011satellitesfromstreams}. Because of the small number and the unique nature of these streams, they are studied on an individual (rather than an ensemble) basis. We discuss two particularly well-known streams here.

The ``poster child'' stream Palomar 5 (Pal5) \cite{kupper2015streams} is known to have a significant interaction with the MW bar \cite{pearson2017Pal5bar, banik2019pal5baryons}, presenting an irreducible background to searches for DM substructure in this system. We show updated observations of Pal5 compared with a set of simulated versions of a Pal5-like stream with and without simulated perturbers in Fig.~\ref{fig:sim_pal5} \cite{erkal2017sharperpal5}: observations are reproduced at the top of the figure in blue; a simulation of an unperturbed, or ``regular'', stream which has undergone tidal disruption by its host galaxy, but has experienced no other external perturbative interactions, is shown in green; and a simulated stream with two interactions, evidenced by two gaps on the left (leading) and right (trailing) arm, is shown in red. We note that very recent simulations have also raised the possibility that mass segregation, which concentrates many black holes each of 3 $\msun$ or more in the central regions of Pal 5, can have a significant effect on its evolution \cite{gieles2021bhspal5}.

\begin{figure}[t]
\begin{center}
\includegraphics[width=\textwidth]{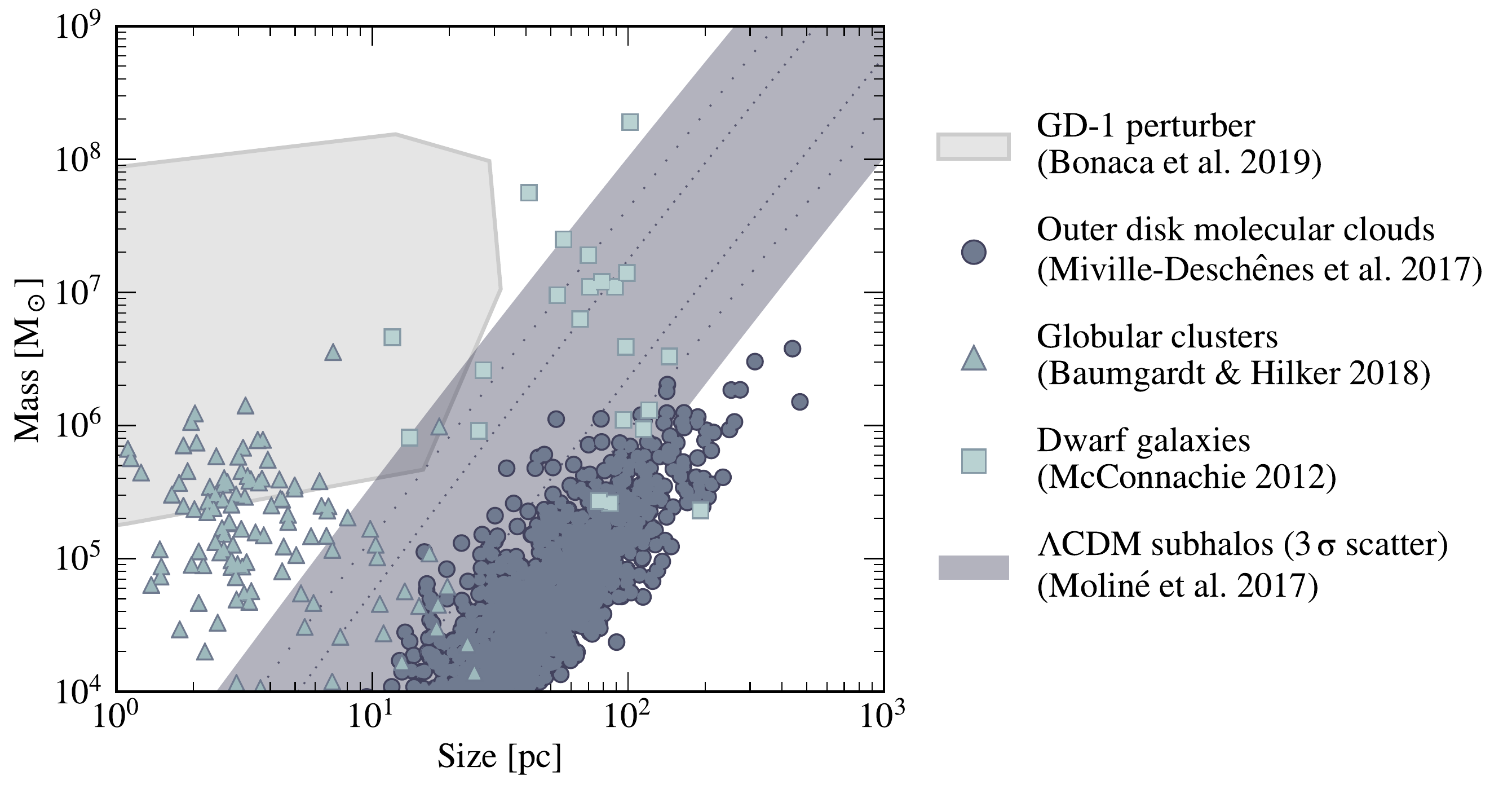}
\caption{Inferred mass and size of the perturber of GD-1 compared to known constituents of the MW, both dark and luminous. From \cite{bonaca2019spurgapGD-1}$\copyright$AAS. Reproduced with permission.  \label{fig:bonaca_gd1_pert_features}}
\end{center}
\end{figure}

The GD-1 stream is expected to be more robust against these confounding baryonic effects \cite{amorisco2016gd1betterthanpal5, erkal2016gd1betterthanpal5} due to its large pericenter, retrograde orbit, and large inferred separation from known baryonic substructure \cite{bonaca2019spurgapGD-1}, and is thus a promising target for detecting DM substructure \cite{banik2018streamsDM}. Recently, observations of GD-1 have been used to probe the SHMF of the DM of the MW \cite{bonaca2019spurgapGD-1, banik2019streamsDM, banik2021streamsDM}. These studies indicate that the GD-1 stream most likely was perturbed by at least one dense, 
massive object --- and it is unlikely to have been 
the Sagittarius dwarf remnant~\cite{banik2019streamsDM}.
The rate of encounters of a stream on a GD-1-like orbit with a substructure of the virialized MW dark matter halo is expected to be large enough to account for the observed perturbation of GD-1 \cite{erkal2016gd1betterthanpal5, bonaca2019spurgapGD-1}. Following the logic laid out above, this rate is evidence that the subhalo mass function must not overly be suppressed in the mass range $10^6-10^8 \,\msun$ \cite{banik2019streamsDM, banik2021streamsDM}. This provides evidence in favor of cold, rather than warm, DM, and has been used to constrain the mass of thermal DM to be larger than 4.6 keV \cite{banik2021streamsDM}. On the other hand, the perturber may be somewhat more dense than expected in a conventional CDM-like scenario \cite{bonaca2019spurgapGD-1}, potentially pointing the way to new physics, as shown in Fig.~\ref{fig:bonaca_gd1_pert_features}. However, if the interaction with the subhalo remnant is taken into account, $N$-body simulations suggest that a kinematically warm DM population may actually be preferred \cite{malhan2021DMaccretedstreams}. Studies of streams are still a relatively young subject, and will benefit from further exploration of potentially confounding physical effects.

In conclusion, studies of MW satellites and their invisible brethren are powerful but still-developing probes of the nature of the hierarchical structure formation paradigm. Counts of visible satellites match qualitative expectations from simulations, thus providing compelling evidence that the DM in our galaxy is formed roughly hierarchically at least down to subhalo masses $M \sim 10^8\, \msun$. The stellar stream GD-1 provides further suggestive hints that hierarchical structure formation continues at least one order of magnitude lower in halo mass, and streams in general will be a compelling testbed for future advances in understanding the nature of the MW's assembly and current constitution.

\subsection{
Nearly Thermal Dark Matter
}
In the classification scheme adopted above, DM is characterized as being either cold, warm, or hot. This scheme relies fundamentally on the assumption that DM has reached thermal equilibrium at some point in its cosmological evolution. A stronger assumption that is often implicitly adopted is that the DM was in fact in {\it chemical} equilibrium {\it with the 
SM 
thermal bath}, as in the canonical weakly interacting massive particle (WIMP) scenario, where we use chemical equilibrium to refer to the equilibration of annihilation processes. We explore departures from both of these assumptions in this subsection, though for now we will still find it useful to compare the DM phase space distribution to the thermal SM
phase space.

One possible departure from the conventional scenario is that DM could have attained thermal equilibrium and have a thermal phase space distribution $f_{\rm DM}({\bf v},z) \propto \exp(-3m_{\rm DM}v^2/2T_{\rm DM}(z))$, but the equilibrium temperature $T_{\rm DM}(z)$ describes only its own 
``dark sector''. In order for the DM particles to shed their energy and entropy, they must be able to annihilate away some of its equilibrium abundance, except for a one-parameter family of particles with 
a temperature $\xi = T_{\rm DM}/T_{\rm SM} \simeq (T_{\rm mr}/m_{\rm DM})^{1/3}$, and with the requirement $m\gtrsim 530$ eV in order to become nonrelativistic at sufficiently high temperatures to match the high-wavenumber modes of the CMB. Suffice to say, such a scenario is sufficiently fine-tuned as to be unappealing. Instead, it is conventional to assume that a secluded dark sector has a light partner particle into which the DM particle can annihilate, thereby satisfying constraints on the relic density.

A related alternative is that the dark sector has a secluded number-changing self interaction. Such a DM candidate cannibalizes itself to keep warm \cite{carlson1992cannibalDM}. This extra self-generated warmth slows the redshifting of energy and suppresses the growth rate of density perturbations \cite{carlson1992cannibalDM, machacek1994cannibalperturbations} to a degree that is ruled out by current observations \cite{buen-Abad2018cannibalgrowth} unless the interactions couple very late and change the density negligibly \cite{heimersheim2020cannibalS8}.

A DM candidate that differs in certain key respects from a simple thermal relic is sterile neutrino DM \cite{dodelson1993dodelsonwidrowneutrinos, shi1998shifullersterileneutrino}. This particle never reaches full thermal equilibrium with the SM 
bath. Rather than freeze out like a canonical WIMP, so
that its relic density is determined once its interactions
are no longer strong enough to maintain thermal equilibrium, such particles may ``freeze in'' over a long period of time or may attain the correct energy density during a brief period of resonance production (due, {\it e.g.}, to a large lepton asymmetry). The sterile neutrino is constrained by MW satellite counts in much the same way as in the strongly interacting scenario discussed below in the context of dark matter with a large scattering cross section with SM material: too light of a sterile neutrino will have too large of a free-streaming length, and will suppress the formation of MW satellites below a characteristic mass scale \cite{nadler2020MWcensus}. We show these bounds in Fig.~\ref{fig:dmconstraints_SHMF}.

\begin{figure}[t]
\begin{center}
\includegraphics[width=0.55\textwidth]{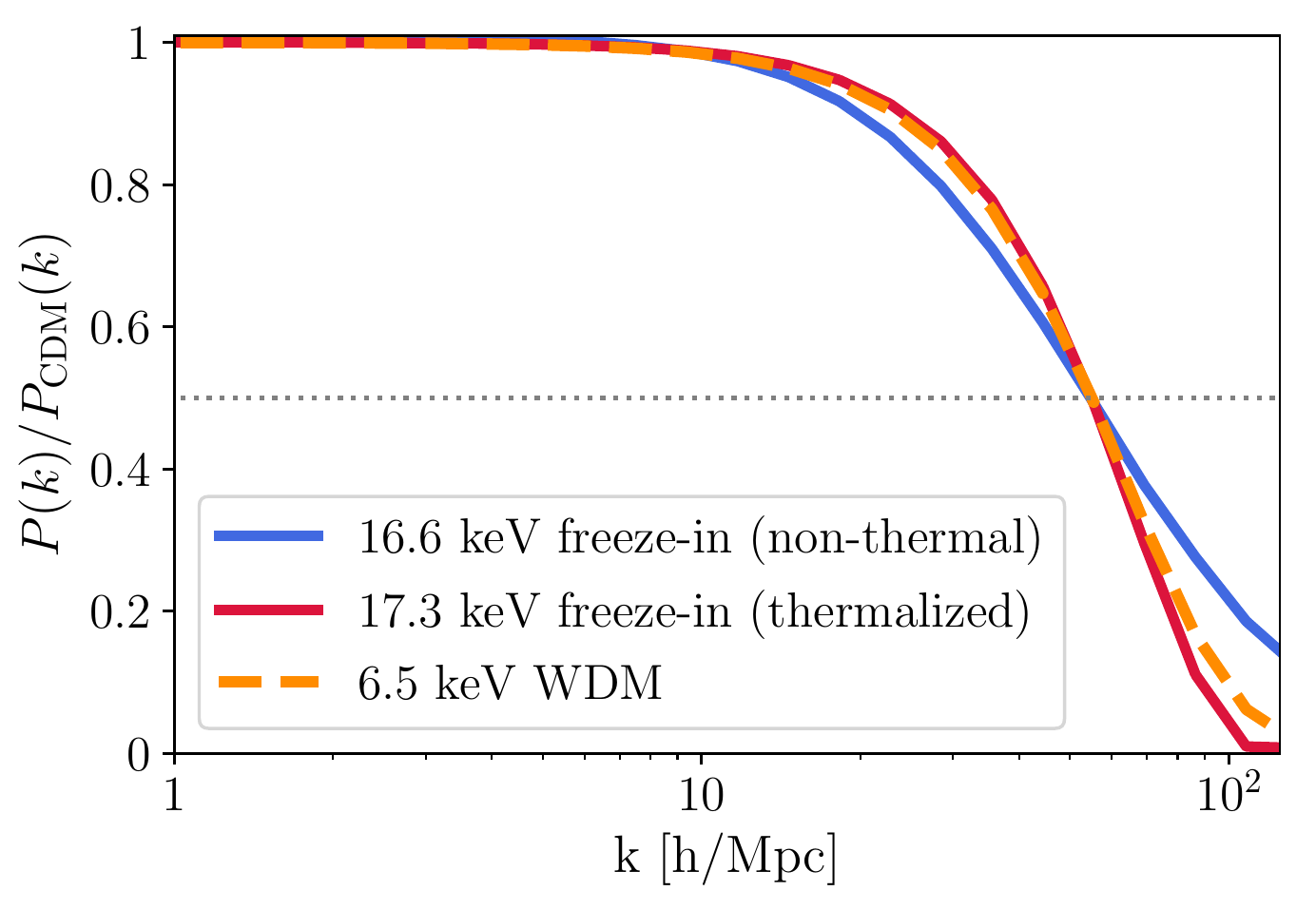}
\caption{Transfer function for density perturbations of a dark matter particle that is ``frozen-in'' through a coupling with a light kinetically mixed dark photon compared to thermalized dark matter of different masses, from \cite{dvorkin2020plasmafreezein2}. \label{fig:dmconstraints_SHMF_2}}
\end{center}
\end{figure}

The sterile neutrino freeze-in process has been generalized in a number of ways since \cite{Hall2009freezein, Krnjaic2017freezeall, Evans2019leakin}. Recently, the calculation of the necessary parameters to achieve the correct DM energy density has been performed in the context of DM that interacts with the SM 
through a very light kinetically mixed dark photon \cite{dvorkin2019plasmafreezein1, dvorkin2020plasmafreezein2}. This model, like the sterile neutrino, is kinematically colder than a completely thermalized particle of the same mass, but with a very long high-velocity tail. Both of these novel features must be accounted for when calculating the expected subhalo abundance. We show the transfer function of this DM candidate in Fig.~\ref{fig:dmconstraints_SHMF_2}.

\subsection{Extremely Massive and Ultralight Dark Matter Candidates}

Extremely massive and ultralight DM candidates differ from the preceding cases by having no notion of temperature. Instead, these candidates are completely athermal, and derive their energy density from a novel mechanism unrelated to the thermal energy of the SM 
or dark sector bath. See \cite{Lin2019tasi} for a recent overview of model possibilities.

By extremely heavy DM candidates, we mean DM composed of particles that are so massive that their individual nature becomes apparent. In some sense, the particulate 
nature of DM structures is then probed directly, instead of the averaged thermodynamic quantities that we typically consider, such as the local and cosmic density. The archetypal heavy DM candidate is the primordial black hole (PBH) \cite{Carr1974pbh}, but composite extended structures such as nuclear-like dark many body states can also grow to become extremely massive in the early universe \cite{Wise2014boundstates1, Wise2014boundstates2, Hardy2014boundstates1, Hardy2015boundstates2, Gresham2017nuclear1, Gresham2017nuclear2, Gresham2018nuclear3}, with unique observational signatures  \cite{Nussinov2018blobs, Grabowska2018blobs, Montero-Camacho2019pbh}.

On the other hand, ultralight DM candidates are those whose mass is so low that they form extremely high-occupancy states, and thus behave more like a classical wave than like a classical particle: their de Broglie wavelength $\lambda_{\rm dB}$, which is inversely proportional to their mass and their characteristic velocity dispersion, exceeds their interparticle spacing in the MW and its satellites when $m_{\rm DM} \lesssim 100$ eV. For a recent review, see \cite{Hui2021wavereview}. Such particles will induce a diminished dark matter transfer function above a characteristic wavenumber set by their mass because their quantum nature prevents them from collapsing on scales smaller than their de Broglie wavelength. This is macroscopically important when that wavelength matches the $\sim 10$ kpc  dwarf galaxy length scale mentioned above. By assuming a characteristic velocity dispersion of $\bar v\sim$10km/s, one can calculate that ultralight DM will match the dwarf galaxy length scale at currently probed sizes of $\gtrsim 10$ kpc if the dark matter mass is less than $m_{\rm DM} \sim 10^{-22}$ eV, since  $\lambda_{\rm dB} = (m_{\rm DM} \bar v)^{-1} \simeq 10 {\rm kpc} \times (m_{\rm DM}/10^{-22}{\rm eV})^{-1}$ \cite{Hu2000fdm}. Dark matter of mass well below this value will not ``fit'' into dwarf galaxies, inhibiting their growth and unacceptably suppressing the subhalo mass function in conflict with observations \cite{Hui2016ultralight, schutz2020fdmSHMF, Benito2020ultralightsubstructures}. This may be extended to larger and smaller systems with current and future data \cite{Blas2020ultralightpulsars, Diehl2021ultralightdwarfs}.

\begin{figure}[t]
\begin{center}
\includegraphics[width=0.985\textwidth]{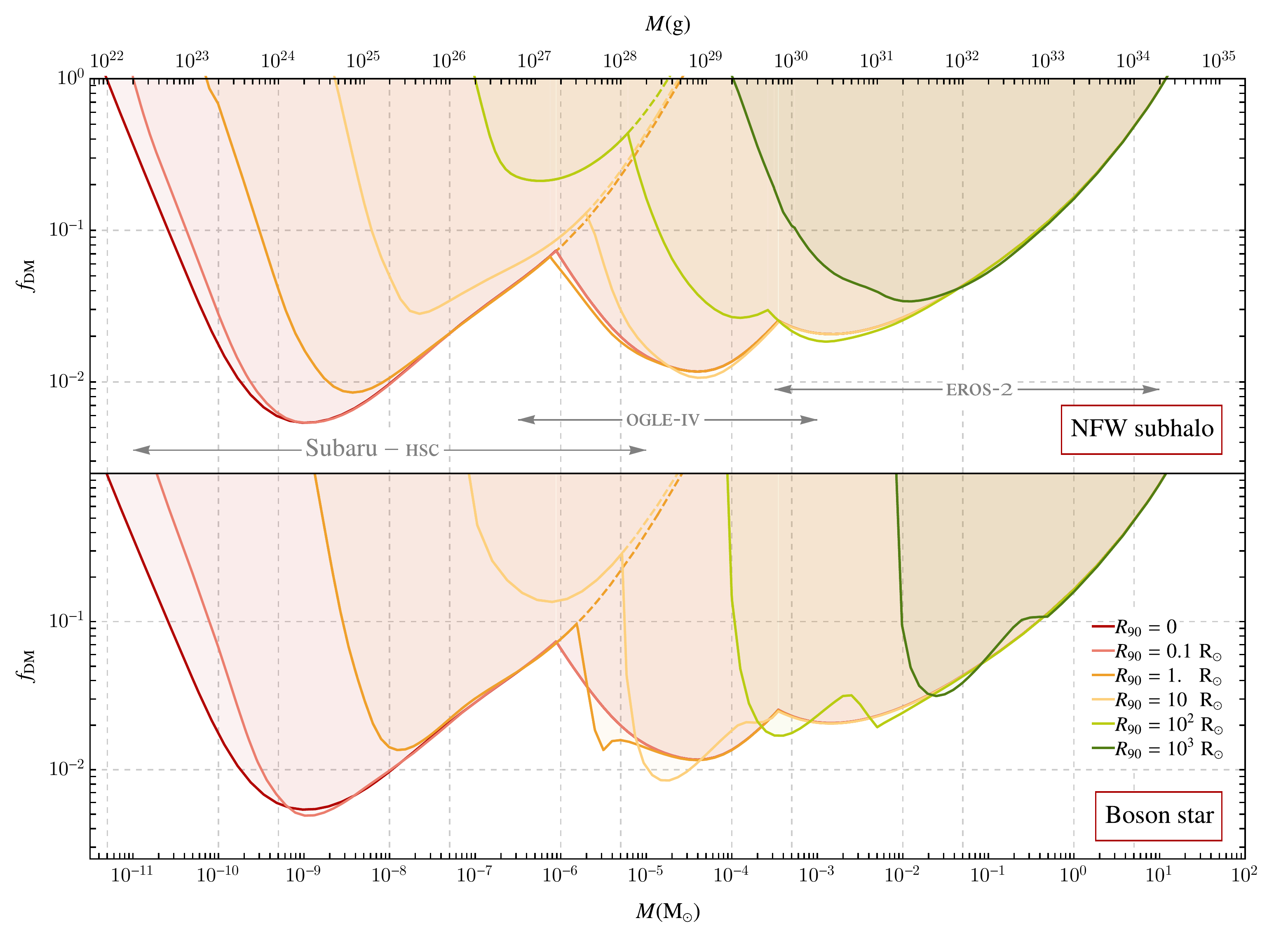}
\caption{Constraints on dense halo objects formed from 
extremely massive or ultralight dark matter candidates.
Reprinted with
permission from \cite{croon2020genlensingDM, croon2020lensingHSCDM}.
Copyright 2020 by the American Physical Society. 
\label{fig:dmconstraints_lensing}}
\end{center}
\end{figure}

In the context of the MW, there is an interesting and perhaps surprising convergence between ultralight and extremely massive DM particles. This happens because ultralight particles typically have attractive self-interactions that cause them to form ultra-dense agglomerations over cosmic timescales \cite{Semikoz1994bosecond, Guth2014axioncorrelation, Levkov2018gravbose, Hertzberg2020bosestars1, Kirkpatrick2020axiontime, Hertzberg2020bosestars2}. Thus in both cases, these DM candidates are probed by lensing searches. Treating the heavy objects as point lenses \cite{smyth2019lensingPBH} (as appropriate for a PBH) or extended lenses \cite{croon2020genlensingDM, croon2020lensingHSCDM} (as appropriate for a composite object of self-interacting particles) changes the constraints
somewhat. 
We show results assuming either an NFW or a boson-star-like density profile 
in Fig.~\ref{fig:dmconstraints_lensing}.

Compact objects such as PBHs also have the ability to dynamically disrupt observed stellar systems \cite{Bahcall1985machodisk, Yoo2004machohalo, Brandt2016machostars}. Fokker-Planck simulations of dwarf galaxies exclude compact objects of mass $M_{\rm CO}$ from constituting a fraction $f \simeq (M_{\rm CO}/10\, \msun)^{-1}$ of the total DM in the MW \cite{Zhu2018fokkerplanckmacho, Stegmann2020fokkerplanckmacho}. Simulations of wide binaries in the MW stellar halo lead to similar bounds with 
slightly lower masses \cite{Monroy-Rodriguez2014machorevisited}. See \cite{Green2020PBHreview} for a more detailed discussion and for additional non-dynamical constraints on compact objects. The convergence in physical effects 
noted above in the case of lensing also carries over to dynamical considerations, and has been used to place strong constraints on ultralight DM \cite{Marsh2019eriIIultralight}.

\subsection{Interactions of Dark Matter with Standard Model Matter}
Dark matter that interacts so strongly with the SM 
that it cannot penetrate the Earth and reach underground DM direct detection experiments (sometimes referred to as IDM) is probed in a multitude of ways. We set aside cosmological and direct probes of such DM models, and focus here on the implications for the subhalo abundance of the MW.

\begin{figure}[t]
\begin{center}
\includegraphics[width=0.485\textwidth]{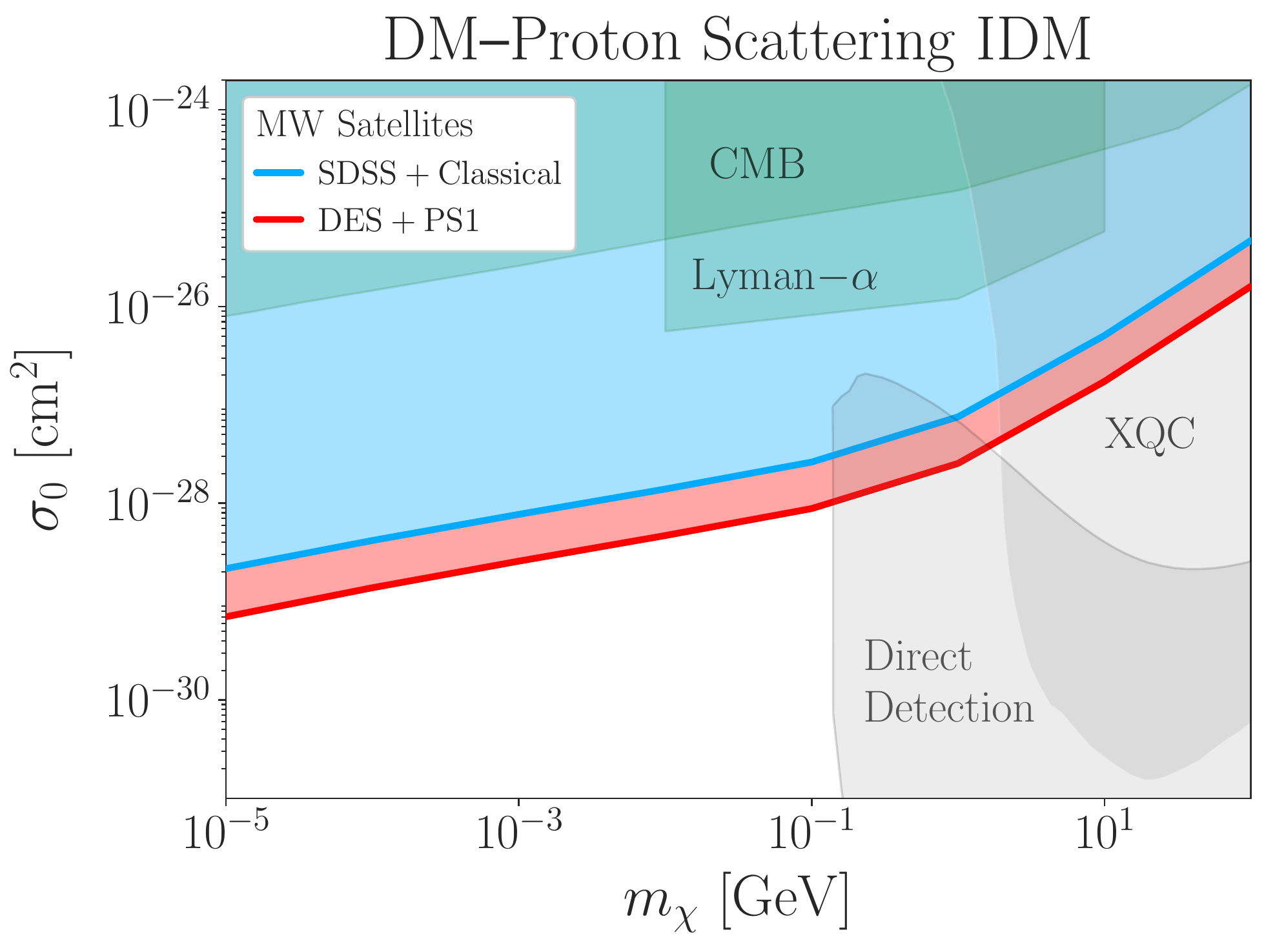}~~~~
\includegraphics[width=0.485\textwidth]{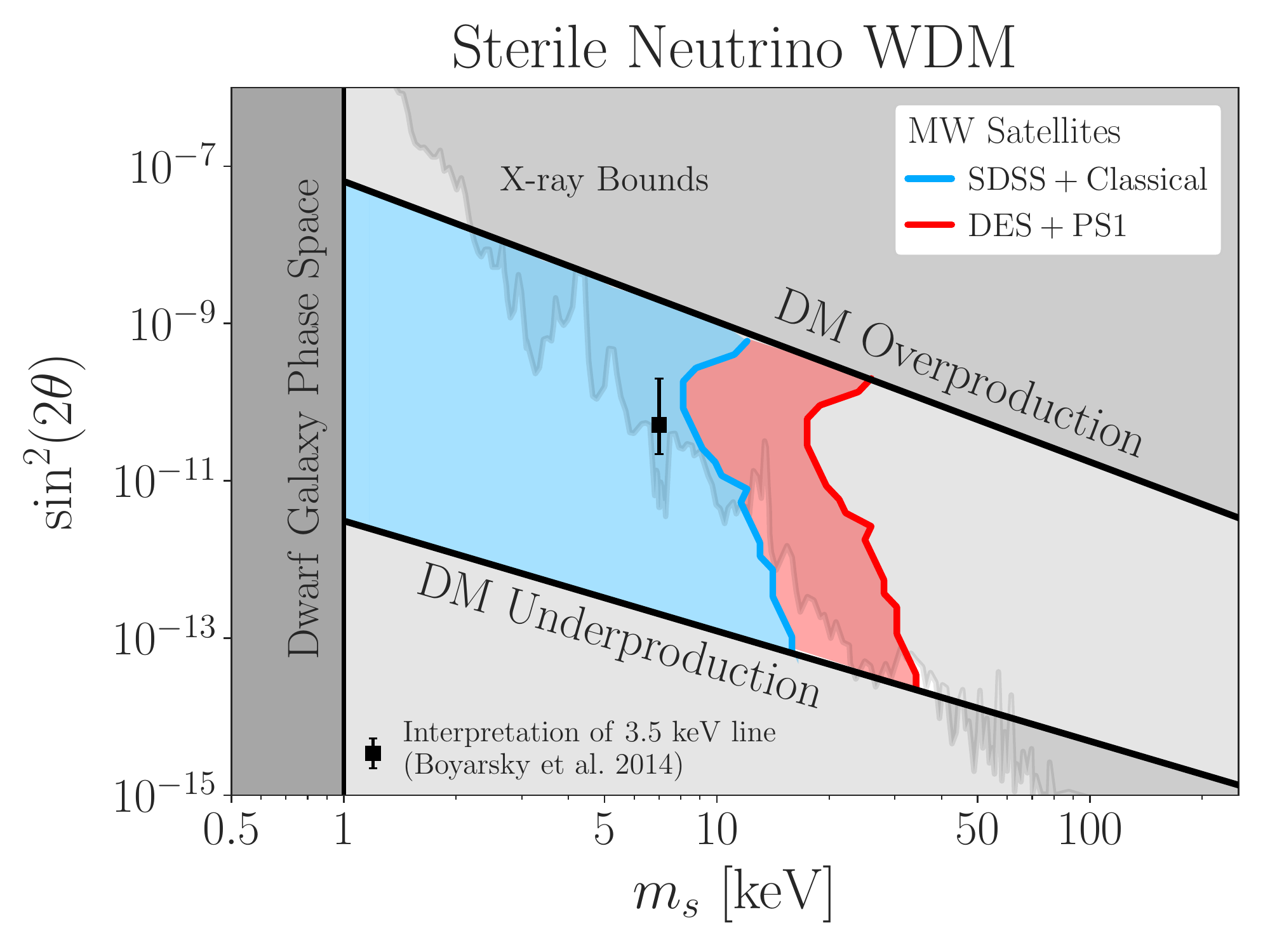}
\caption{Constraints on non-CDM dark matter candidates reproduced from \cite{nadler2020MWcensus}. {\bf Left:} constraint on the mass and elastic scattering cross section between a dark matter particle and the proton. {\bf Right:} constraint on the mass and mixing angle of a sterile neutrino. \label{fig:dmconstraints_SHMF}}
\end{center}
\end{figure}

These strong interactions will couple the DM fluid to the baryon fluid at the time of the formation of the CMB, at a time when the baryons are being prevented from falling into overdense halos by the pressure of the hot and abundant photons. This means that the subhalo abundance is suppressed in these scenarios \cite{nadler2019SHMFdmconstraints, nadler2020MWcensus, Maamari2020SHMFdmconstraints}. The severity of the suppression depends on the temperature of the decoupling of the DM and SM  
fluids. The scattering cross section of such a DM candidate with baryons is constrained by comparing calculations against the current measurements of the SHMF, in analogy with the SIDM case above. We show recent constraints on this type of model in Fig.~\ref{fig:dmconstraints_SHMF}. In addition, much of the parameter space of these models are subject to stringent constraints from their contributions to the energy density of the Universe during the epoch of BBN \cite{Krnjaic2020BBN}.

\subsection{Self-Interacting Dark Matter}


Self-interacting DM (SIDM) was initially proposed as a solution to both the core-cusp problem and a seeming dearth of small MW satellites \cite{spergel1999spergelsteinhardtSIDM}. We will not further address the inner profiles of MW satellites except to say that in SIDM scenarios these problems and their resolutions are inherently linked.

The SIDM power spectrum features a high-$k$ cutoff that is similar to that of warm DM \cite{huo2017sidmMPSCMB, sameie2018sidmhaloabundance}, although SIDM can additionally feature dark acoustic oscillations \cite{cyrRacine2012atomicDM, cyr-Racine2013daoconstraints, egana-Ugrinovic2021sidmevolution} that can serve as a powerful distinguishing marker of additional model complexity. However, the additional complexity of the self-interacting dark sector means that any prediction of high-$k$ power is dependent on a large number of model parameters. These can be summarized in terms of a 
dark kinetic decoupling temperature, analogous to the temperature at which the SM 
particles decouple from the photon bath and begin to form structures  \cite{huo2017sidmMPSCMB, sameie2018sidmhaloabundance, egana-Ugrinovic2021sidmevolution}. If this temperature happens to be close to the corresponding temperature at which high-$k$ SM 
modes decouple from the photon bath, $z_{\rm kd,\,SM}^{\text{(high-k)}} \simeq 4$ keV as discussed above, then the deviation from the CDM power spectrum can be visible in field galaxy counts.

Understanding the SHMF of SIDM involves physics that is both unique to the subhalo and physics that connects the subhalo to its host. For instance, the gravothermal evolution of the self-gravitating SIDM halo, and the eventual gravothermal catastrophe befalling all self-gravitating systems \cite{balberg2002gravocatastrophe1, balberg2002gravocatastrophe2, loeb2011SIDMcores, essig2019dissSIDM}, is a necessary ingredient for drawing conclusions about the nature and the density of observed SIDM halos \cite{nishikawa2020SIDMcollapse, correa2021vdSIDMshmf, yang2021sidmsmallscale}.
When considering SIDM subhalos in the MW, tidal interactions can play a critical role in interpreting the SHMF \cite{zavala2019SIDMshmf, nishikawa2020SIDMcollapse, correa2021vdSIDMshmf}. These factors, sometimes mitigating and sometimes compounding the effects of the self-interactions, must be taken into account when drawing inference about DM self interactions.

\subsection{Dark Matter with Inelastic Transitions}
Dark matter that can interact via inelastic transitions, either through internal excitation  \cite{TuckerSmith2001inelasticDM} or light-particle emission \cite{Fan2013dissipativeDM}, has a very different phenomenology than DM that only interacts elastically. This type of DM can form structures very similar to those formed by the SM,
ranging from the length scales of acoustic oscillations in the early universe \cite{cyr-Racine2013daoconstraints} to structures of the size of the MW disk \cite{Fan2013dddm, Fan2013ddu} and below \cite{egana-Ugrinovic2021sidmevolution, Curtin2019mirrorstars1, Curtin2019mirrorstars2, hippert2021mirrorNS}. In fact, such dissipation can be the key to forming the extremely massive composite DM candidates discussed in the previous subsection \cite{Wise2014boundstates1, Wise2014boundstates2, Hardy2014boundstates1, Hardy2015boundstates2, Gresham2017nuclear1, Gresham2017nuclear2, Gresham2018nuclear3}.

Most interesting for studies of galactic dynamics is DM equipped with a dissipative force. This can form a dark disk, which may be coincident with or at least concentric with the MW stellar disk. This impacts the matter surface density observed by stars in the local neighborhood, and can thus have an observable effect on their phase space distribution \cite{Fan2013dddm, Fan2013ddu, schutz2017disklimit, buch2018disklimit}. Inferences on the dark disk density using observations from the {\it Gaia} satellite have thus far been limited to the 
equilibrium case \cite{deSalas2020darkdensitydisk}, which, as argued above, needs to be improved upon given our updated knowledge of the dynamics of the MW. We discuss inference of the local DM energy density more below in Sec.~\ref{sec:dm_density}.

\section{Probes of Change} 
\label{sec:Change}

Our ability to ascertain {\it change} in the MW --- 
beyond the birth and death of single stars --- has
come as quite a surprise. Certainly, the DF formalism 
we have outlined in Sec.~\ref{sec:Prologue} is
constructed to address a system in steady state. 
Although the existence of the MW's spiral arms and 
the discovery of the Galactic bar speak to non-steady-state
effects, it has been thought that these effects could be
accommodated with only small adjustments 
of the DF formalism, despite the fact that the spiral arms signal
a spiral distortion in the Galaxy's gravitational field. 
This engenders radial mixing of the gas and stellar disks, 
with stellar distributions in metallicity and age 
giving observational support to such effects 
--- yet the overall angular momentum distribution may be largely undisturbed~\cite{sellwood2002radialmixing}.
Considerable evidence for imperfections throughout
the Galactic disk also exists, however: it is 
warped and flared in HI gas \cite{levine2006vertical, kalberla2007dark} and 
in stars \cite{alard2000flaring, ferguson2017milky}, and we
 have already noted the striking evidence for the latter 
from three-dimensional maps of Cepheids~\cite{chen2019intuitive,skowron2019three}. 
Rings \cite{newberg2002ghost, morganson2016mapping} and ripples \cite{pricewhelan2015reinterpretation, xu2015rings} 
in regions farther from the Sun, where the disk
is relatively thin, have also been noted. 

We believe that observations of
wave-like asymmetries near the Sun's location 
in main-sequence stars 
from the 
SDSS~\cite{widrow2012galactoseismology, yanny2013stellar}, 
in vertical velocities of red-clump stars from the RAVE survey~\cite{williams2013wobblyRAVE}, 
as well as from {\it Gaia} 
DR2 \cite{bennett2018vertical},  
speak to a sea change, revealing the existence of
non-steady-state effects in the solar neighborhood. 
Evidence for axial-symmetry breaking of 
out-of-plane main-sequence 
stars in the north with SDSS 
has also been observed~\cite{ferguson2017milky}. 
The astrometry of 
{\it Gaia} DR2~\cite{prusti2016gaia, brown2018gaia}
has greatly enriched these studies. For example, the 
striking snail shell and ridge correlations within the 
position and velocity components of the DF 
\cite{Antoja2018youngperturbedMWdisk} have also been 
discovered, revealing 
axially asymmetric and presumably 
non-steady-state behavior. This, 
as they note~\cite{Antoja2018youngperturbedMWdisk}, is attributable to the existence of the Galactic bar, spiral arms, as well as of other, external 
perturbations.

\begin{figure}[tb]
\begin{center}
\subfloat[]{\includegraphics[scale=0.75]{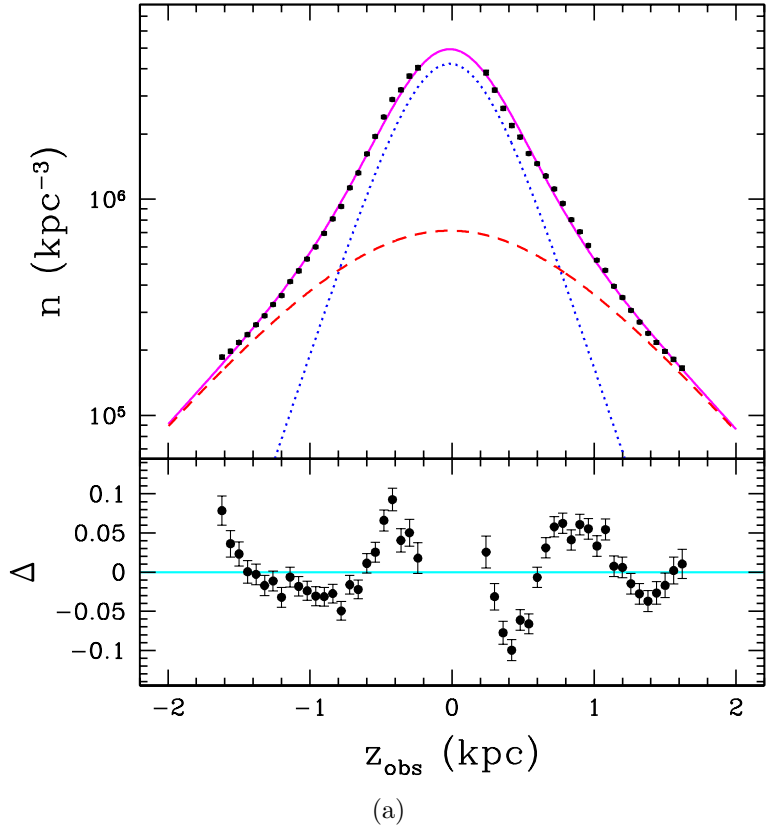}}
\caption{Estimated number density $n$ as a function
of observed distance $z_{\rm obs}$ from the Sun (top), using 
photometric parallaxes and SDSS observations 
of K and M dwarfs within an in-plane distance of 
1 kpc, with the black 
points showing the observed star counts, including an
overall normalization factor. 
The magenta curve is a model fit to those points, assuming no North-South breaking, and  
the contributions from the thin (blue dots) and
thick (red dashes) disks as well. The bottom panel 
shows $\Delta\equiv \rm (data-model)/model$, revealing
a wave-like vertical asymmetry, north and south. 
From \cite{widrow2012galactoseismology}$\copyright$AAS. Reproduced with permission. 
\label{fig:NSwavefirst}}
\end{center}
\end{figure}

In what follows we discuss probes of 
non-steady-state effects broadly, considering
first evidence from north-south symmetry breaking
and phase-space correlations, before turning to 
the study of broken stellar streams and intruder
stellar populations. The latter two --- and
probably the last three --- probes are
of such complexity that they serve as 
{\it prima facie} evidence for non-steady-state
effects. In the case of north-south symmetry
breaking, we note that a study of north-south
and axial symmetry breaking can framed to show
concretely that the MW is not in steady state, 
following the theorems we have discussed
in Sec.~\ref{sec:Prologue}~\cite{GHY20,HGY20}. 
The particular origins of these various 
symmetry-breaking and phase-space correlation
effects are not well-established, but 
we review various proposed explanations --- all 
of which involve external perturbations. 

\subsection{North-South Symmetry Breaking}

\begin{figure}[tb]
\begin{center}
\subfloat[]{\epsfig{file=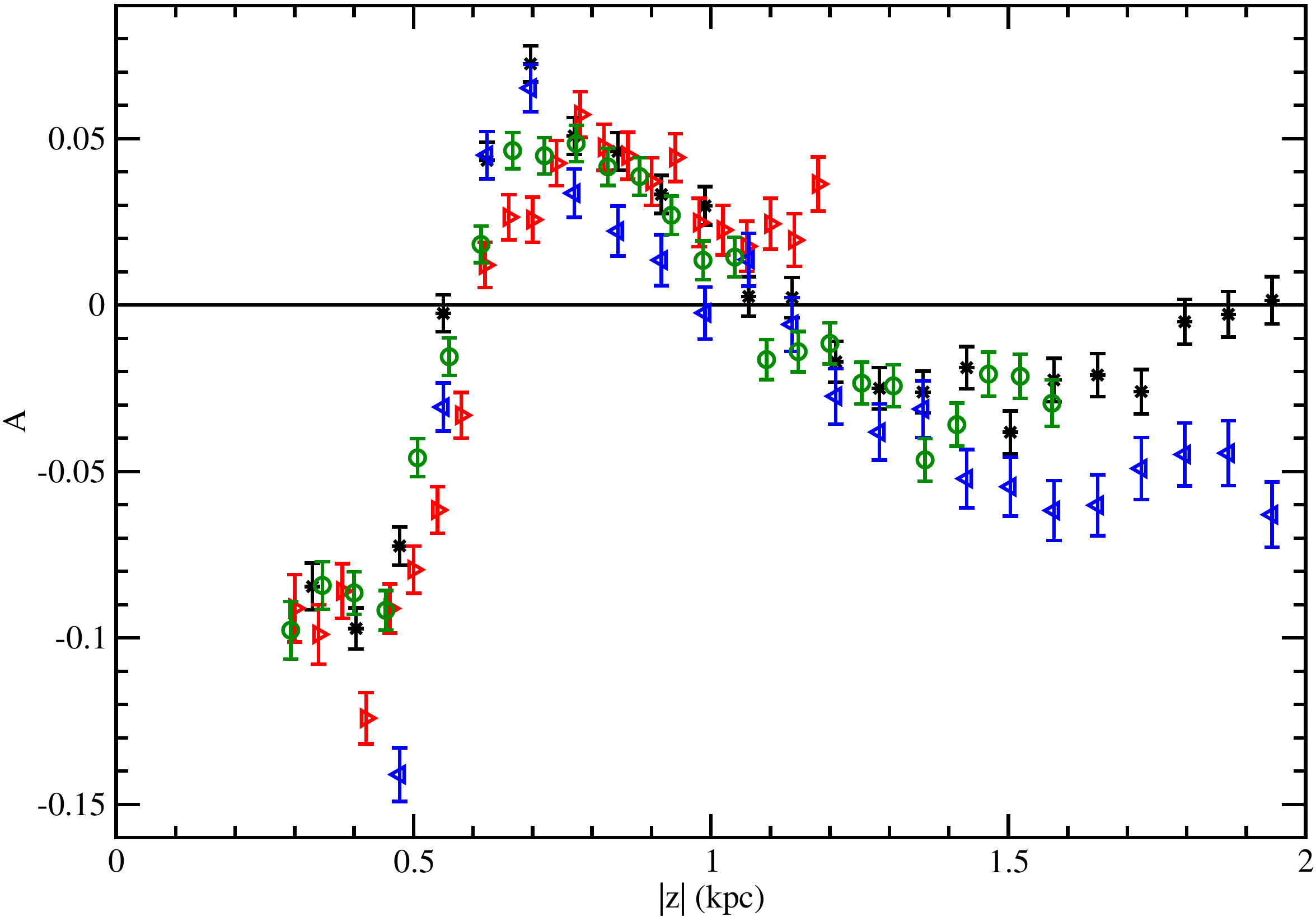,scale=0.5}}
\caption{The 
asymmetry in star counts, north and south, 
with height $|z|$ from the Galactic mid-plane, 
for different bands in $(g-i)_0$ color. 
Note $1.8 < (g-i)_0 < 2.4$ (black), 
$0.95 < (g-i)_0 < 1.8$ (blue), $0.95 < (g-i)_0 < 1.8$ (red), 
and the 
green points employ 
a distance relationship
based on $(r-i)_0$ color. From 
\cite{yanny2013stellar}$\copyright$AAS. Reproduced with permission. 
\label{fig:NSwave}}
\end{center}
\end{figure}

We show the first observation of a vertical
asymmetry in stellar number counts, from \cite{widrow2012galactoseismology}, in Fig.~\ref{fig:NSwavefirst}, with distances
computed using a photometric parallax relation. 
A follow-up analysis  
confirms this
result, as shown in Fig.~\ref{fig:NSwave}~\cite{yanny2013stellar}. 
The north-south {\it asymmetry} 
is defined as 
\begin{equation}
    A(|z|)\equiv \frac{n(z>0) -n(z<0) }{n(z>0) + n(z<0)}\,,
\end{equation}
where $n(z)$ are the
stellar number counts north $(z>0)$ and south $(z<0)$, 
measured from the Galactic mid-plane. 
The insensitivity of
the observed vertical asymmetry to stellar selection 
suggests that it is indeed a density wave. 
We also note 
results from the RAVE velocity survey
that show evidence for vertical ringing in $V_z$ of 
stars~\cite{williams2013wobblyRAVE} at similar distances to those 
studied in 
\cite{yanny2013stellar}, as well as 
an observed vertical wave in mean 
metallicity \cite{an2019asymmetric}, inferred from 
SDSS photometry, with features similar to the observed density wave.

\begin{figure}[t]
\begin{center}
 \subfloat[]{\includegraphics[scale=0.55]{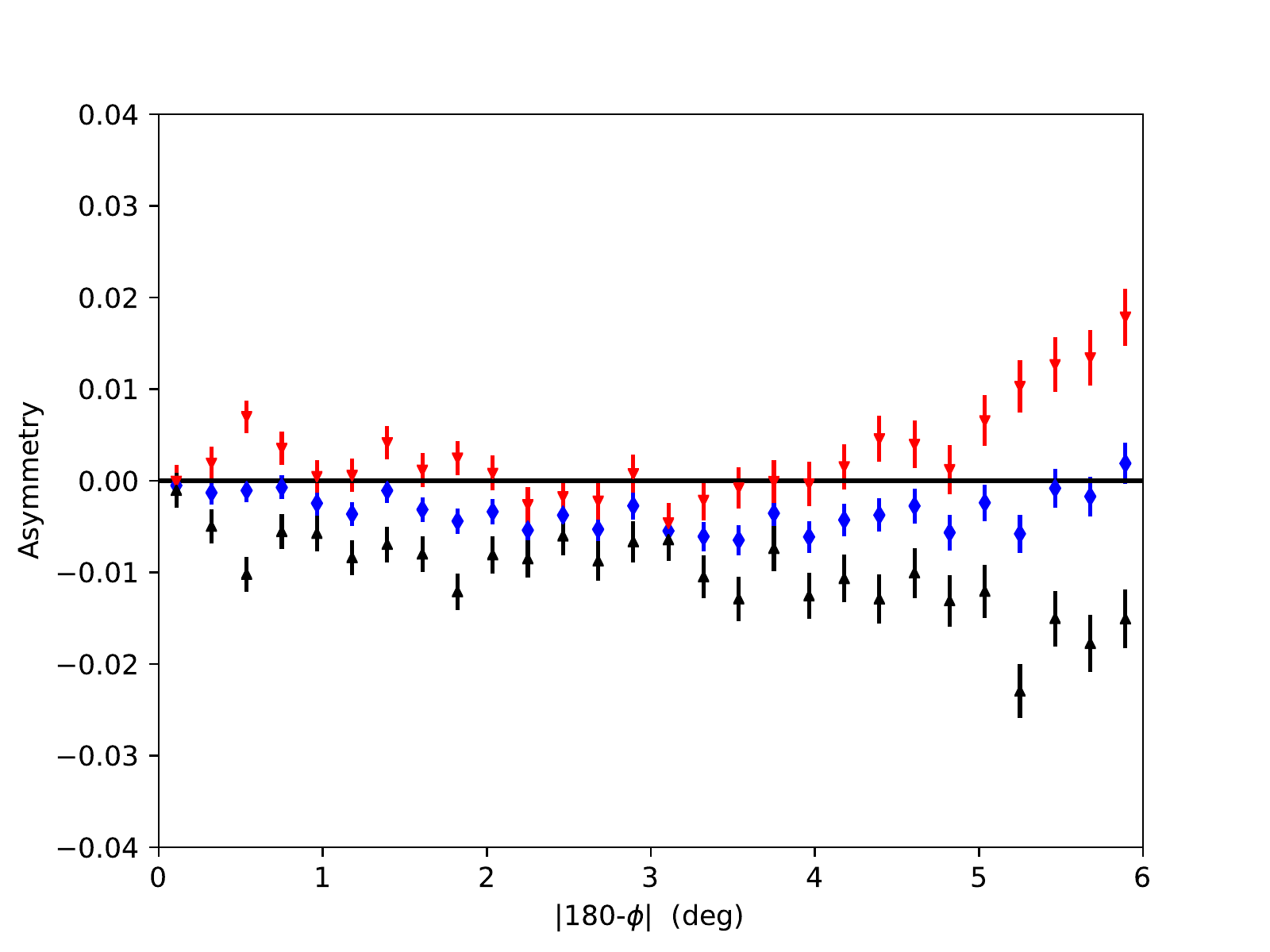}}
 \subfloat[]{\includegraphics[scale=0.55]{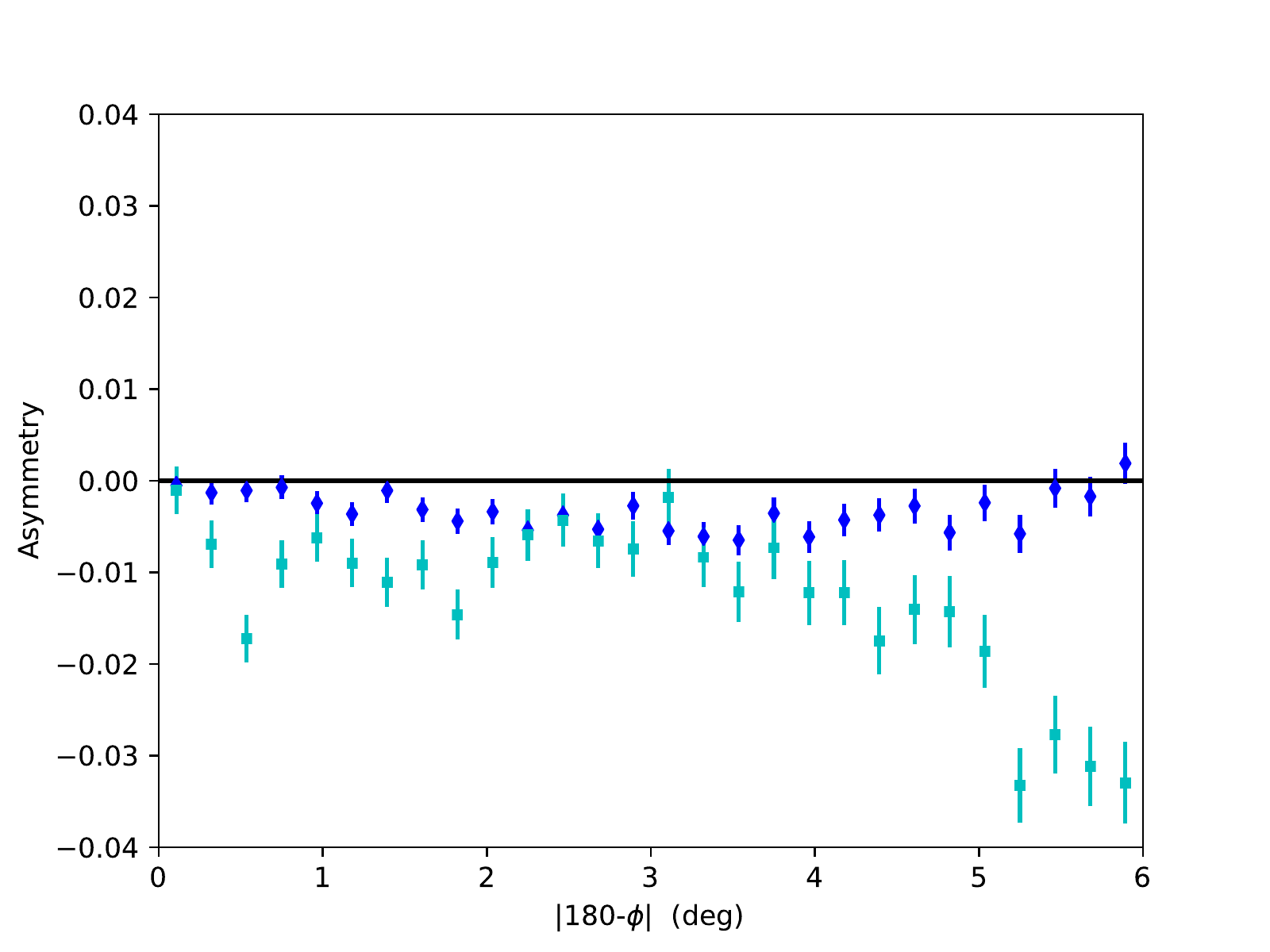}}
\caption{
The axial asymmetry $A(\phi)$ with $\phi$, for
selections north (N) and
south (S), (a) red (S), black (N), and blue (N+S), (b) 
we compare $A(\phi)$
in the  N+S sample with the difference of $A(\phi)$ 
in the north and {\bf $A(\phi)$} in the south (N-S) (squares). From \cite{GHY20}$\copyright$AAS. Reproduced with permission. 
\label{fig:axialasym}}
\end{center}
\end{figure}

To probe the possible 
origin of the vertical symmetries we have noted, 
we turn to 
a combined analysis of axial and north/south 
symmetry breaking~\cite{GHY20}, using a 
sample of 14 million {\it Gaia} DR2 stars
within $3\,\rm kpc$ of
the Sun's location, 
carefully selected for sensitive studies 
of symmetry breaking~\cite{HGY20}. 
The axial asymmetry $A(\phi)$ is determined
by counting stars on either side of the
anti-center line 
in the Galactocentric longitude $\phi=180^\circ$,
computing
\begin{equation}
    A(\phi) = 
    \frac{n_L(\phi) - n_R(\phi)}{n_L(\phi) + n_R(\phi)} \,,
\end{equation}
where $n_L(\phi)$ and $n_R(\phi)$ are defined as the number
of stars, left and right, of the anti-center line. 
This observable probes the vertical component of 
the angular momentum as an integral of motion; that is, 
if $A(\phi)=0$, it is conserved, as per our discussion
of Noether's theorem~\cite{noether1918} in Sec.~\ref{sec:Prologue}.
The result of this evaluation 
for stars with $z>0$ (north N), $z<0$ (south S), and
for all $z$ (N+S) is shown in Fig.~\ref{fig:axialasym}. 
Since $A(\phi)\ne 0$ in all cases, axial symmetry is
broken, but it is also apparent that 
$A(\phi)|_N - A(\phi)|_S \gg A(\phi)|_{N+S}$. 
This pattern of symmetry breaking, as
per our discussion of \cite{an2017reflection} in 
Sec.~\ref{sec:Prologue}, can only result if our
large sample of stars is not in steady state. 
Ergo the Galaxy near the Sun's location 
is not in gravitational equilibrium. 
This outcome supports an external perturbation
origin for the vertical asymmetries, and for
the novel phase space structures we consider
in the next subsection.

We first pause to consider the north and south pattern of the asymmetries shown
in Fig.~\ref{fig:axialasym}a. 
These results in the solar neighborhood can be compared 
to the axial asymmetries expected 
from the distorted halo shapes determined from the analysis
of peculiar velocities of stars in the Orphan
stream~\cite{erkal2019LMCandorphan}. The outcome 
suggests that the halo of the MW is prolate in shape and tilted in the direction of the LMC/SMC system~\cite{GHY20},
as illustrated in Fig.~\ref{fig:tiltedhalo}. 
The prolateness of the halo is distinguished by comparing subtle differences in star counts in the northern vs. southern hemispheres, building 
on the result of \cite{erkal2019LMCandorphan}. 
Including the effects of the Sagittarius dwarf as well 
appears to adjust the picture to favor an oblate
and possibly radius-dependent geometry~\cite{Vasiliev2021Tangoforthree},
and their results also support a large mass for
the LMC, in agreement with \cite{erkal2019LMCandorphan}.
On the other hand, 
earlier 
studies of flaring HI gas in the outer galaxy 
support a prolate DM 
distribution \cite{banerjee2011progressively}; these authors note that a prolate halo can support
long-lived warps \cite{ideta2000time}, which would help to explain why they are commonly seen  \cite{banerjee2011progressively}.

In other galaxies, 
\cite{Xu2020prolateoblatelight} shows 
that one can determine whether external galaxies are more prolate or oblate in their stellar distributions by observing their stellar density profiles 
as another check on N-body models of DM if 
the DM DF reflects that of light.
Whether or not this is so has
not yet been established.

\subsection{Phase-Space Correlations}

\begin{figure}[t]
\begin{center}
\subfloat[]{\includegraphics[
width=\textwidth]{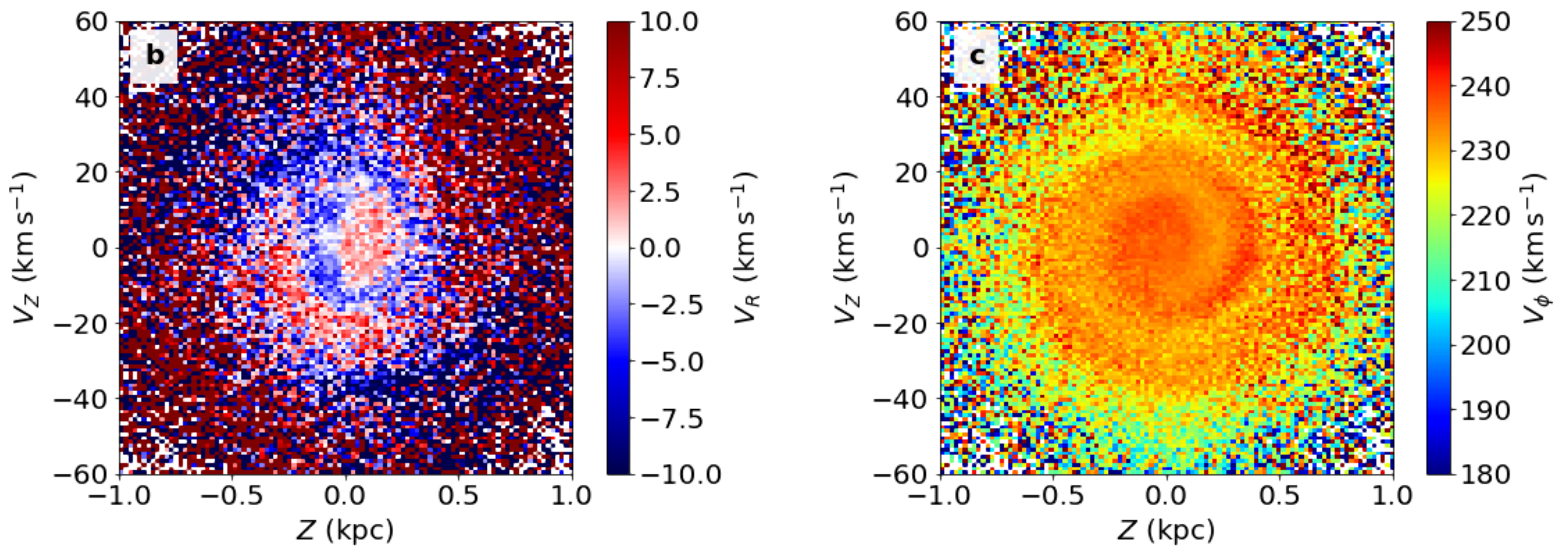}}
\caption{
Distribution of stars located $8.24 < R < 8.44 \,\rm kpc$ 
from {\it Gaia} DR2 data 
in the vertical position-velocity 
($Z-V_Z$) plane colored 
as a function of median $V_R$ (left) and
$V_\phi$ (right) in bins of $\Delta Z=0.02 \,\rm kpc$
and $\Delta V_Z = 1 \,\rm km s^{-1}$. From 
\cite{Antoja2018youngperturbedMWdisk}, reprinted by
permission from Springer Nature.
\label{fig:gaiasnail}}
\end{center}
\end{figure}

We show the novel phase-space correlations 
of the so-called {\it Gaia} snail in 
Fig.~\ref{fig:gaiasnail}. 
This structure is 
apparent in a vertical velocity-position phase space diagram. It 
indicates strong evidence for recent and on-going disturbances of the disk of the MW by interloper structures such as the Sagittarius dwarf or other massive structures with a significant vertical component to their motion.  

Thus it would seem that the 
vertical asymmetries in the stellar density 
may indeed be due to an ancient 
impact, possibly by the Sagittarius dwarf galaxy \cite{widrow2012galactoseismology}. Support for the impact hypothesis comes
from numerical simulations~\cite{purcell2011sagittarius, gomez2012vertical}. The novel phase-space structures noted by \cite{Antoja2018youngperturbedMWdisk,li2021verticalphasemixing} also offer support to the impact hypothesis, as such features had been predicted as a consequence \cite{purcell2011sagittarius, gomez2012signatures, d2016excitation, fux2001order}.
We refer to 
\cite{binney2018gaiasnail,blandhawthorn2021snail} 
for a review of the state of observations of phase-space correlations and the theory behind the {\it Gaia} snail.   Note that assigning responsibility for the snail to one particular dwarf such as Sagittarius may be problematic \cite{bennett2021snailisnotsgr}, and that models continue to evolve and benefit from additional data. 

Recently, too, the discovery of stars with non-prograde kinematics in the disk has led to determination
of a previously unidentified ancient impact, 
from {\it Gaia}-Enceladus (or the {\it Gaia}-Sausage) in the inner halo \cite{helmi2018merger,belokurov2018co}, which we discuss
further later. 
Finally, \cite{koppelman2018one} have noted significant merger debris, and streams, in the halo, which are also 
an expected consequence of ancient impacts.

\subsection{Fitting Broken streams and the Galactic potential shape}

Detailed observations of tidal stellar streams in the halo of our MW are potentially strong probes of the distribution of matter, including the DM, on scales of 20-100 kpc, as overviewed in Sec.~\ref{subsec:obs_stellar_streams}.  Combining observations of stellar position, proper motion, radial velocity, distance estimates and density of stars along halo stellar streams is potentially an extremely strong probe of the Galactic potential at radii inside the streams' orbit.  There are at least two known cases where segments of streams originally identified as independent are now thought to be originate from the same stellar cluster. The Orphan stream (above the Galactic plane) and the Chenab stellar stream (below the plane) have been observed to share close orbital parameters and in fact may have a common origin \cite{koposov2019piercing}. Models are being undertaken to show how an interaction with the LMC at some time in the past can explain a prominent feature in the stream \cite{erkal2019LMCandorphan}.  More recently the ATLAS and Aliqa Uma stream segments appear to share a common origin \cite{Li2020ATLASAliqaUma}. The nearby halo streams of GD-1 and Khir also potentially share a common origin \cite{Malhan2019gd1andkshir}.
The full potential of using broken streams to constrain the mass of the LMC are still limited by the precision of the available data, with error on distances being the limiting factor.  
A stream which covers a wide range in distance between its apo- and peri-galacticons can be used to put a constraint on the mass within --- such as has been done with the Orphan stream \cite{newberg2010fitorphanstream}.

Aside from broken streams, at least one stream is observed to have a kink along its length. This is thought to be evidence of an interaction with an otherwise unseen massive perturber \cite{bonaca2019spurgapGD-1, banik2021streamsDM}. The implications of this interaction for theories of DM have been discussed in more detail in Sec.~\ref{sec:dmcand}.  A calculation by 
\cite{erkal2016gd1betterthanpal5}  shows how the presence or absence of gaps in the density of stars along a stellar stream may be used in a more general way to constrain the mass and number density of DM blobs floating through the halo.   

\subsection{Intruder Stellar Populations \label{sec:intruders}}

{\it Gaia} has contributed to our understanding of the history of how the Galaxy has built up over time, as recently reviewed by \cite{helmi2020reviewaraasequoia}.  The largest past merger has been that of the {\it Gaia}-Enceladus-Sausage, though recent work suggests that this is in fact two separate mergers. The smaller of these nearly simultaneous mergers, dubbed the Sequoia merger \cite{myeong2019sausagedistinctfromsequoia}, is distinguished by having released stars on kinematically distinct retrograde orbits. The Sequoia merger may possibly have left a dwarf galaxy nucleus remnant today in the form of the so-called $\omega$-Cen globular cluster, though 
it is likely not a true globular cluster due to its multiple stellar populations. The case for two ancient mergers has also been made by \cite{Evans2020sausageminussequoia}.  The metallicities and kinematics of these merger remnants are present today in the disk of the MW.

Whether or not smaller merger remnants in the disk can be isolated is an on-going topic of research with an ever-growing list of techniques.
For instance, \cite{Ostdiek2020cataloging, necib2020chasing, necib2020evidence} find evidence for stars in a nearby stream, Nyx, using, in part, machine learning techniques, and \cite{malhan2018streamfinder} are able to disentangle populations of stream stars embedded within the disk of relatively low contrast, again taking advantage of the excellent {\it Gaia} dataset. Determining the limits of sensitivity of these new techniques is the subject of ongoing work, and 
caution is advised in identifying 
intruder populations via any 
single technique.
If possible, using multiple methods, such as 
chemical signatures and 
kinematic markers, is 
necessary to draw inference on the nature of the 
possible merger remnants, as 
carried out by a recent analysis of Nyx candidate stars with the GALAH survey \cite{zucker2021nyxnotintruder}.

\section{Implications for the Local Dark Matter Phase Space Distribution} 
\label{sec:dm_phasespace}
The local DM phase space distribution, namely, 
the distribution function, $f_{\rm DM} (\mathbf{x},\mathbf{v},t | R_\odot)$, at the Sun's 
location, reflects 
all of the environmental and evolutionary 
factors we have discussed thus far. This object is defined as the one-body distribution function, and is well-posed regardless of whether the assumptions that would lead it to be a solution of the collisionless Boltzmann equation, as we discuss in Sec.~\ref{sec:Prologue}, are fulfilled.
Here we provide a summary of what is known about this
elusive object,
noting first that it is the 
local DM mass density,  
$\rho_{\rm DM}(R_\odot) = M_{\rm MW} \int d^3\mathbf{v} f_{\rm DM} (\mathbf{x},\mathbf{v},t | R_\odot)$, 
where $M_{\rm MW}$ is the total mass of the MW, 
and the 
local DM velocity distribution, 
$f_{\rm DM}(\vec v|R_\odot)
= \int d^3\mathbf{x} f_{\rm DM} (\mathbf{x},\mathbf{v},t | R_\odot)$, that are of greatest interest to DM 
direction detection searches, as we have highlighted
in Sec.~\ref{sec:Prologue}.
We will first treat the total local DM density, $\rho_{\rm DM}(R_\odot)$. Then we will discuss the local DM velocity distribution, $f_{\rm DM}(\mathbf{v}|R_\odot)$, including possible contributions from partially mixed phase-space structures.

\subsection{The Local Dark Matter Density}
\label{sec:dm_density}

\begin{figure}[t]
\begin{center}
\includegraphics[width=0.55\textwidth]{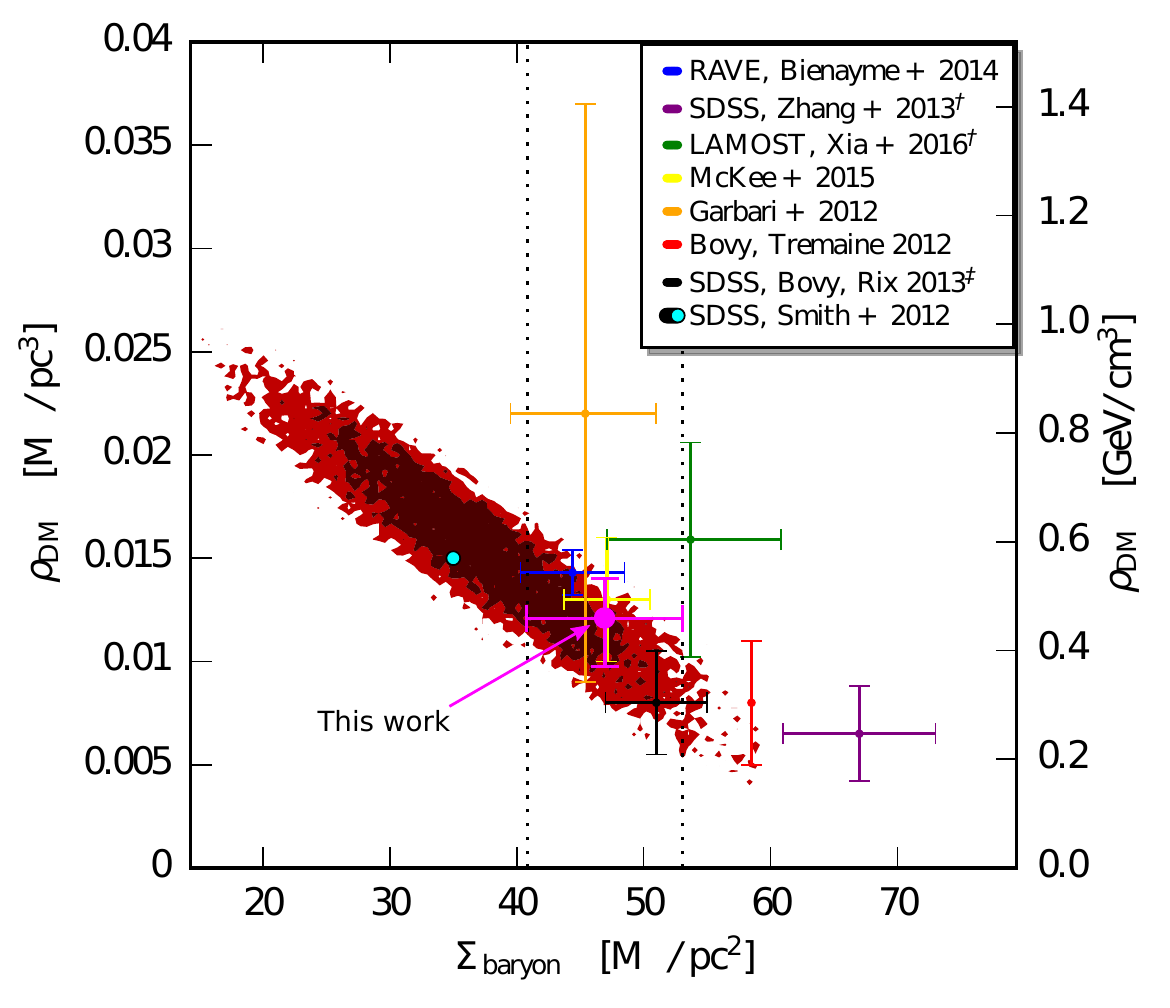}
\caption{
Illustration of the degeneracy in the
determination of the local dark matter density and the baryonic
surface density, comparing the two-dimensional marginalised
posterior of \cite{Sivertsson2018localDMgdwarfs},
with the results of particular groups. 
From \cite{Sivertsson2018localDMgdwarfs}, and
we refer to that reference for all details. 
\label{fig:density_anticorrelation}}
\end{center}
\end{figure}

A long-standing problem has been 
the determination of the matter density in the vicinity of the Sun~\cite{kapteyn1922first}, as inferred from the measured kinematics of the local stars~\cite{oort1932force}. In this so-called Oort problem, a sample of stars, assumed to be
in a gravitationally relaxed, or steady, state 
is used to trace the local gravitational potential and to infer the local matter density.  
We note Refs.~\cite{binney2008GD,Read2014localdmreview} for 
a detailed account of earlier work. Since such dynamical 
methods yield the total matter density, a 
careful accounting of ordinary, or baryonic, matter 
must be made simultaneously, because it is their difference that 
gives the local dark matter density. 
A local census of baryonic matter
in the solar neighborhood, including stars, brown dwarfs, 
and gas, gives $0.0889 \pm 0.007 M_\odot \rm \> pc^{-3} = 3.4 \pm 0.3 \, GeV \> cm^{-3}$~\cite{schutz2017disklimit},
consistent with the assessment of $0.09 \pm 0.01 M_\odot \rm \> pc^{-3}$
in \cite{binney2008GD}, even if their detailed 
contents differ. The dynamical mass estimate has
been assessed to exceed this by 10\%~\cite{binney2008GD}, 
and indeed 
$0.3\, \rm GeV\> cm^{-3}$ has traditionally been estimated to be the local dark matter density 
in the 
SHM~\cite{drukier1986SHM}, which we have reviewed
in Sec.~\ref{sec:Prologue}.
Recently, progress has been made 
through the simultaneous, but separate,
analysis of the baryonic and dark matter 
contributions in an integrated Jeans equation 
analysis~\cite{Sivertsson2018localDMgdwarfs}. 
Although the stellar tracers of the gravitational potential are 
certainly blind to the distinction between visible and dark matter, 
the degeneracy between these two forms of 
matter is broken if one works a few vertical 
scale heights above the Galactic plane~\cite{bahcall1984selfconsistent}, because there
the contribution would be mostly dark matter. 
The outcome of this analysis is shown in 
Fig.~\ref{fig:density_anticorrelation}; it 
can be seen that the apparent discrepancies 
between groups are ameliorated 
once the baryonic-dark matter
degeneracy is taken into account~\cite{Sivertsson2018localDMgdwarfs}. 
We also note the outcome from a Jeans analysis of 
{\it Gaia} EDR3 and APOGEE data, 
$\rho_{\rm DM} (R_\odot) =(8.92 \pm 0.56\, ({\rm sys}))\times 10^{-3} M_{\odot} {\rm pc}^{-3} 
\,(0.339\pm 0.022 \,(\rm sys)\, GeV cm^{-3})$ \cite{Nitschai2021dynamicalmodelII}, 
as well as a determination from precision binary pulsar timing measurements, 
$\rho_{\rm DM} (R_\odot) = -0.004{\stackrel{{+0.05}}{{}_{-0.02}}} M_{\odot} {\rm pc}^{-3}$ \cite{Chakrabarti2021massdenspulsaracc}, once the
baryon density is removed \cite{McKee2015DMsolar}, with 
a Jeans analysis of {\it Gaia} DR1 data finding local DM densities compatible 
with either result depending on the stellar tracer population chosen~\cite{schutz2017disklimit}. 


The appearance of vertical oscillations in 
the stellar number counts and velocity distributions of the 
MW~\cite{widrow2012galactoseismology,yanny2013stellar} and
variations in the effective vertical height of
the Sun across the 
Galactic plane~\cite{ferguson2017milky}, speaking
to warping in the disk, strongly suggests 
the existence of non-steady-state effects, 
for which we have reviewed definitive 
evidence in Sec.~\ref{sec:Change}.
Thus the explicit time-dependence of the 
DF must be taken into account, 
as in Refs.~\cite{banik2017localdensity,Sivertsson2018localDMgdwarfs}.
It is important to emphasize that suitable 
DFs have to be simultaneous solutions of the 
collisionless Boltzmann and Poisson equations, 
as discussed in Sec.~\ref{sec:Prologue}. 
This work was completed before the release of
the {\it Gaia} DR2 data, and with the discovery
of striking axial symmetry breaking features, such 
as the {\it Gaia} 
Sausage~\cite{Antoja2018youngperturbedMWdisk},
it is apparent that 
terms neglected in previous analyses can be significant. 
Indeed, larger variations in the dark matter density
have been found, 
even varying up to a 
factor of 2 larger than the SHM result~\cite{evans2018dmhalosausage}.
For reviews of systematic uncertainties and different approaches to this problem, see \cite{Read2014localdmreview, desalas2020dmlocaldensity}.

We note that increasing the local dark matter density 
should not shift the normalization of the rotation curve (i.e., the plot of $V_c$ versus galactocentric distance, as shown in Fig.~\ref{fig:vc_compare} and discussed in Sec.~\ref{sec:MWrotcurve}) by a large amount. This is true because most of the enclosed mass at the solar circle, and thus the largest contribution to the circular velocity at $R \approx R_\odot$, is due to the density of baryonic matter (stars). This latter mass density is not perfectly known, and extractions of its value are anticorrelated with the local DM density \cite{Sivertsson2018localDMgdwarfs}, as shown in Fig.~\ref{fig:density_anticorrelation}. On the other hand, if the density of luminous stars is well constrained, then small dips and wiggles in the rotation curve can be used to map out a 
residual 
dark component with some accuracy.
Certainly any axial asymmetries there, or 
in the local stars~\cite{GHY20,HGY20} could be 
used to constrain the axial-symmetry-breaking terms
in a Jeans analysis, leading to improved  
assessments of $\rho_{\rm DM}(R_\odot)$.

While data so far have focused nearly entirely on stellar positions and velocities (i.e., the 0th and 1st derivatives of motion), over time, {\it Gaia} data and upcoming data from pulsar timing arrays are of sufficient accuracy that accelerations in star motions may be eventually be able to be used to constrain the MW's potential \cite{phillips2020pulsars, buschmann2021potentialangular}, as discussed in Sec.~\ref{subsec:obs_patterns_in_stars}.

\subsection{The Local Dark Matter Velocity Distribution}
\label{sec:MW_fv}
The local DM velocity distribution $f_{\rm DM}({\bf v}|R_\odot)$ is important chiefly for interpreting experimental results related to the direct detection (DD) of DM, although we note that {\it comparisons} between such experiments are nevertheless possible without knowledge of this distribution \cite{Fox2010integratingout, Anderson2015haloindDD}. Our knowledge of $f_{\rm DM}({\bf v}|R_\odot)$ is informed by our knowledge of global properties of the MW, which allow simulators to identify MW analogues in cosmological $N$-body simulations of DM structure formation \cite{Vogelsberger2009phasespaceDD, Ling2010nbodyDMsim, Kuhlen2010DDnonmaxwellian, Mao2013nbodyDMfv, Butsky2016nihaofv, Bozorgnia2016simMWDD}. These global properties have been discussed in Sec.~\ref{sec:Parameters}.

When constructing a semi-empirical dark matter velocity distribution suitable for interpreting dark matter direct detection data, it is particularly important to account for known local structures, such as the Sagittarius dwarf \cite{Purcell2012DDsgrstream}. In the limit of a large number of DM events in a near-future DD experiment, it may conversely be possible to extract this information from the recoil spectra \cite{Lee2012constrainDMvdistrib, Kavanagh2016reconstructingfv}.

The task of identifying kinematically distinct components of the stellar halo and converting these to weighted contributions to the DM halo has undergone a rapid and substantial evolution in the last several years \cite{Herzog-Arbeitman2018metalpoor1, Herzog-Arbeitman2018metalpoor1, Bozorgnia2019correlationDMstellar, OHare2018dmhurricaneS1, evans2018dmhalosausage, Necib2019inferredsubstructure, Necib2019firelight, OHare2020substructureDMGaia}. A consensus has roughly emerged that distinct phase-space substructures can account for at most $\lesssim 20\%$ of the DM velocity distribution \cite{myeong2018sausage, evans2018dmhalosausage, OHare2020substructureDMGaia}, and departures from the SHM 
given in Eq.~\eqref{eq:isothermal-MBfv}, which assumes a Maxwellian velocity distribution for 100\% of the DM halo, 
as we have discussed in Sec.~\ref{sec:Prologue},
are correspondingly small \cite{evans2018dmhalosausage, Bozorgnia2019correlationDMstellar, Bozorgnia2020radiallyanisotropic, OHare2020substructureDMGaia, Callingham2020orbitalcontracted}.

\begin{figure}[t]
\begin{center}
\includegraphics[width=0.55\textwidth]{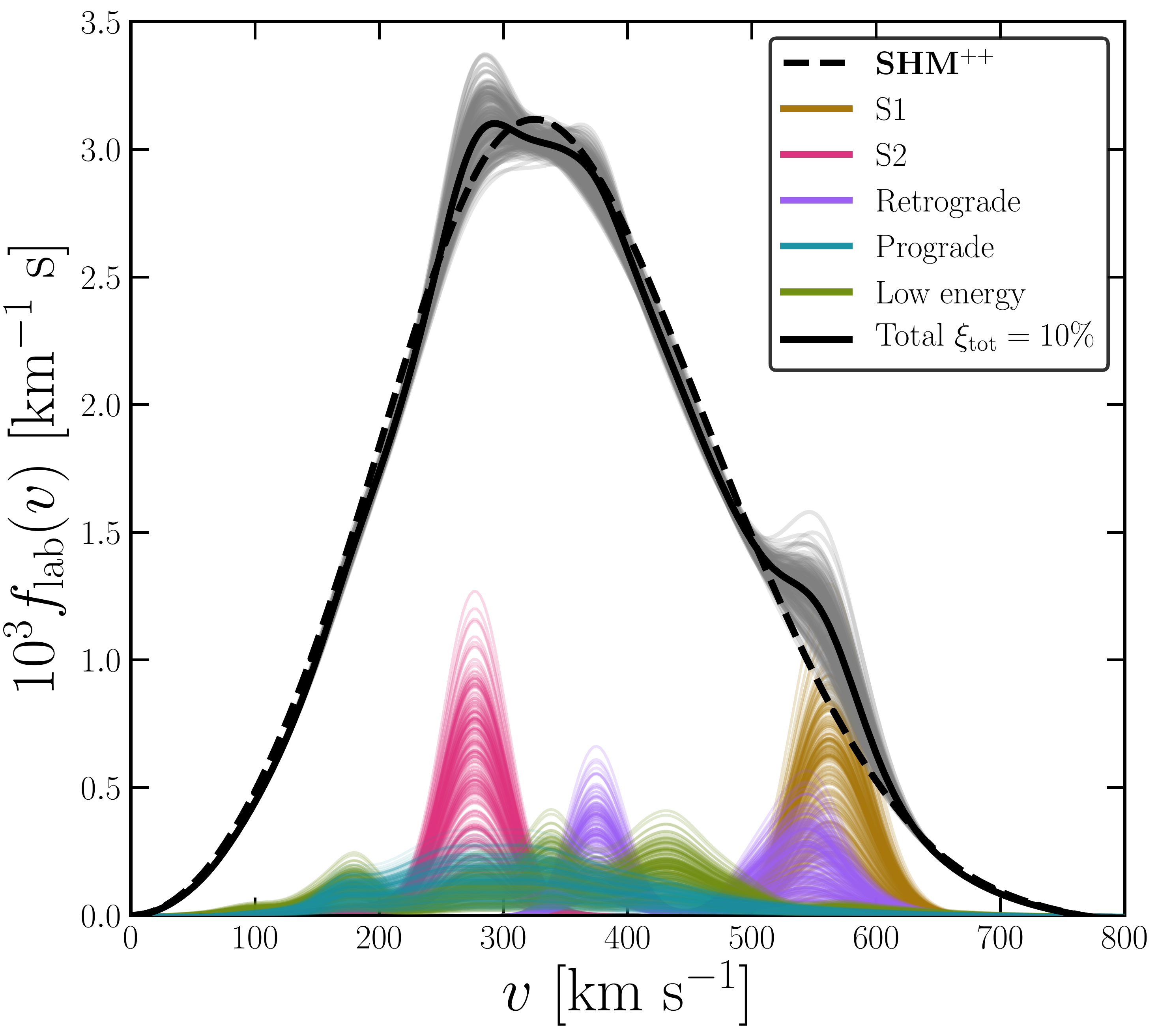}
\caption{Possible MW velocity distributions as inferred from substructures in {\it Gaia} data consistent with other global constraints on the MW DM halo~\cite{OHare2020substructureDMGaia}. Reprinted with
permission from \cite{OHare2020substructureDMGaia}. 
Copyright 2020 by the American Physical Society. 
\label{fig:ohare_fv_Shards}}
\end{center}
\end{figure}

The fact that these departures are small does not mean they cannot be quantified. A velocity distribution that is more realistic than the SHM has been developed recently. The SHM${}^{++}$ \cite{evans2018dmhalosausage} is not strictly isotropic and isothermal. It accounts for the {\it Gaia}-Enceladus merger by adding to the SHM a new anisotropic component governed by a single parameter $\beta=1-(\sigma_\theta^2+\sigma_\phi^2)/2\sigma_r^2$, where $\sigma_i$ are the velocity dispersions in the spherical galactocentric reference frame:
\begin{equation}
    f_a({\bf v}) \propto \exp \left( -\frac{v_r^2}{2 \sigma_r^2}-\frac{v_\theta^2}{2 \sigma_\theta^2}-\frac{v_\phi^2}{2 \sigma_\phi^2} \right), \quad \sigma_r= \frac{3v_0^2}{2(3-2\beta)}, \quad \sigma_\theta=\sigma_\phi= \frac{3(1-\beta)v_0^2}{2(3-2\beta)}.
\end{equation}
The parameter $v_0 = \sqrt2 \sigma_v$ is the speed of local standard of rest, mentioned above, which determines the characteristics of the bulk halo in the SHM${}^{++}$, $f_{\rm SHM} \propto \exp(-v^2/v_0^2)$. The total velocity distribution according to the SHM${}^{++}$ also requires a relative normalization $\eta$ to relate the relative contribution of the anisotropic component: $f({\bf v}) = (1-\eta) f_{\rm SHM}({\bf v}) + \eta f_a({\bf v})$ \cite{evans2018dmhalosausage}. The parameter $\eta \sim 20\%$ ensures that the contributions from this anisotropic, phase-space unmixed, component are not dominant \cite{OHare2020substructureDMGaia}.

The SHM${}^{++}$ does not account for smaller structures, which may yet be energetically important. For example, counter-rotating structures can have noticeable effects for dark matter particles with kinetic energies near experimental thresholds, and thus these structures are especially important for the original goal of interpreting DM DD experiments \cite{evans2018dmhalosausage, Ibarra2019substructure, OHare2020substructureDMGaia}. Such small counter-rotating 
structures have been identified using a number of methods \cite{Ostdiek2020cataloging, necib2020chasing, necib2020evidence, OHare2020substructureDMGaia}. We show in Fig.~\ref{fig:ohare_fv_Shards} a range of possible MW velocity distributions including these yet smaller structures as cataloged by \cite{OHare2020substructureDMGaia}. We compare these semi-empirical $f(\bf v)$ distributions to the SHM${}^{++}$, which itself differs from the SHM.

\section{Summary and Future Prospects} 
\label{sec:Summary}

In this review we have considered how precision 
astrometry, particularly from the {\it Gaia} space 
telescope, has enriched our ability to study the MW, 
particularly in regards to its dynamical and DM aspects. 
In particular, we have reviewed our current understanding of the 
DM phenomenon as illuminated by the stellar halo of the Milky Way. Starting from very general principles, we have discussed the structure and contents of the MW and how the DM is distributed within it. 

We have discussed what we know observationally about the MW halo and the subhalos within it. We have considered concrete models covering the range from a weakly (or sub-weakly) interacting fermion (WIMP), to 
an ultralight 
boson with a super-saturated phase space density, to a non-luminous massive compact object, such as a rock, planet, or black hole (MACHO).
We have recapped studies of the impacts of such DM candidates on a variety of stellar systems spanning a huge range of masses.

Cosmological data, primarily astrophysical in nature, show that DM has an overall mass content within the observable universe of about five times that of the known baryonic matter (stars, gas, and dust). This DM is largely cold and collisionless (unlike a gas with collisions) and is non-luminous (does not interact electromagnetically). No clear resolution to the question of the fundamental nature of the DM has manifested itself despite nearly 90 years of observations since the existence of DM was first proposed. Here, we have emphasized the essential features of what we consider to be a promising route to identifying the DM: using observations of motions and distributions of stars in and around the MW to constrain what we know about how DM is clumped. A minimum clumping scale, if observed, is inversely related to the mass scale of a DM particle.  Studies of star kinematics in globular clusters and dwarf galaxies suggest a minimum clumping scale could exist at the threshold of our experimental and theoretical sensitivity, in the range $10-1000$ pc, and future work should be able to search for and refine such a scale, if it exists uniquely.

Studying stellar motions, especially the dispersions of motions, illuminate the total mass distribution of the MW, because the virial theorem dictates that these motions are determined by the entire mass enclosed by their orbit, including the invisible DM. In this review, we have reviewed the theoretical foundation of such searches, and we have gone beyond studies of stellar motion to consider the study of number counts of stars in balanced volumes of space.  We have shown that such counts are a powerful probe of the symmetry of the underlying matter distribution.  Any observed breaking of such symmetries, even at the 1\% level, can lead to important insights about the distribution and extent of unseen DM on sub-galactic scales.  Upcoming refinements to the already essential {\it Gaia} data on MW stellar kinematics, combined with photometric stellar population information from optical surveys at the Rubin observatory and from other complementary surveys, and even multimessenger studies of 
compact astrophysical objects, 
will continue to provide datasets capable of further constraining DM properties.
The future of the study of the DM halo of the MW promises to continue to grow ever more illuminated by these studies.

\section*{Acknowledgments}
SG and SDM thank the Aspen Center for Physics, which is supported by National Science Foundation grant PHY-1607611, and the organizers of ``A Rainbow of Dark Sectors'' for (virtual) hospitality while this work was completed. 
SDM thanks Nikita Blinov, Djuna Croon, Matthew Lewandowski, Annika H.~G.~Peter, Katelin Schutz, Josh Simon, and W.~L.~Kimmy Wu for helpful discussions. 
SG and BY thank Austin Hinkel for collaborative discussions.
SG acknowledges partial support from the U.S.
Department of Energy under contract DE-FG02-96ER40989. Fermilab is operated by Fermi Research Alliance, LLC under Contract No. DE-AC02-07CH11359 with the United States Department of Energy.
 
\typeout{}
\bibliography{main_15jun21_final}
\bibliographystyle{doiplain}

\end{document}